\documentclass[]{WileyASNA-v1}

\usepackage[utf8]{inputenc}
\usepackage{calc}
\usepackage{anyfontsize}
\usepackage{hyperref}
\usepackage{endnotes}

\newlength{\okinalen}
\newcommand{\okina}{\hbox to.666\okinalen{\hss`\hss}}

\articletype{Article Type}

\received{1 March 2021}
\revised{1 April 2021}
\accepted{26 April 2021}

\begin{document}

\title{Gaia Search for stellar Companions of TESS Objects of Interest II}

\author[1]{M. Mugrauer}

\author[1]{K.-U. Michel}

\authormark{Mugrauer \& Michel}

\address[1]{Astrophysikalisches Institut und Universit\"{a}ts-Sternwarte Jena}

\corres{M. Mugrauer, Astrophysikalisches Institut und Universit\"{a}ts-Sternwarte Jena, Schillerg\"{a}{\ss}chen 2, D-07745 Jena, Germany.\newline \email{markus@astro.uni-jena.de} \thanks{This research is based on observations made with the NASA/ESA Hubble Space Telescope, obtained from the Space Telescope Science Institute, which is operated by the Association of Universities for Research in Astronomy, under NASA contract NAS 5–26555. These observations are associated with the program 14260.}}

\abstract{We present the latest results of our ongoing multiplicity study of (Community) TESS Objects of Interest, using astro- and photometric data from the ESA-Gaia mission, to detect stellar companions of these stars and to characterize their properties. In total, 113 binary, 5 hierarchical triple star systems, as well as one quadruple system were detected among 585 targets surveyed, which are all located at distances closer than about 500\,pc around the Sun. As proven with their accurate Gaia EDR3 astrometry the companions and the targets are located at the same distance and share a common proper motion, as it is expected for components of gravitationally bound stellar systems. The companions exhibit masses in the range between about 0.09\,$M_\odot$ and 4.5\,$M_\odot$ and are most frequently found in the mass range between 0.15 and 0.6\,$M_{\odot}$. The companions are separated from the targets by about 120 up to 9500\,au and their frequency is the highest and constant within about 500\,au while it continually decreases for larger separations. Beside mainly early to mid M dwarfs, also 5 white dwarf companions were identified in this survey, whose true nature was revealed by their photometric properties.}



\maketitle

\section{Introduction}

In 2020 we have initiated a new survey at the Astrophysical Institute and University Observatory Jena with the goal to explore the multiplicity of (Community) TESS Objects of Interest (CTOIs), i.e. stars, which are photometrically monitored by the Transiting Exoplanet Survey Satellite \citep[TESS, ][]{ricker2015} and exhibit promising dips in their light curves, that could be caused by potential exoplanets, which revolve around these stars.

In our survey stellar companions of (C)TOIs are detected and their properties are determined with astro- and photo\-metry, originally taken from the 2nd data release \citep[Gaia DR2 from hereon, ][]{gaiadr2} of the ESA-Gaia mission. The first results of the survey were presented by \cite{mugrauer2020}, who have already explored the multiplicity of about 1400 (C)TOIs, which were all listed in the (C)TOI release of the \verb"Exoplanet" \verb"Follow-up" \verb"Observing" \verb"Program" for TESS (ExoFOP-TESS)\footnote{Online available at:\newline\url{https://exofop.ipac.caltech.edu/tess/view_toi.php}\newline\url{https://exofop.ipac.caltech.edu/tess/view_ctoi.php}} by the end of May 2020. In the meantime, several of these (C)TOIs, which were revealed as members of multiple star systems in the course of our survey, could already be confirmed by follow-up observations to be exoplanet host stars, e.g. TOI\,451, 1098, 1259, and 1333 \citep{newton2021, tofflemire2021, martin2021, rodriguez2021}.

Due to the successful execution of the TESS mission and the photometric analysis of its data the number of (C)TOIs and hence the number of targets of our survey is continuously growing. Since the end of May 2020 many hundreds of new (C)TOIs have been announced by the ExoFOP-TESS and we could already investigate the multiplicity of these stars, which is presented in this paper.

In the following section we describe in detail the properties of the selected targets, as well as the search for companions around these stars. In section 3 we present all (C)TOIs with detected companions and characterize the properties of these stellar systems. A summary of the current status of our survey and an outlook for the project are given in the last section of this paper.

\section{Search for stellar companions of (C)TOIs by exploring the Gaia EDR3}

In contrast to \cite{mugrauer2020} the search for companions, presented here, uses astro- and photometric data from the early version of the 3rd data release \citep[Gaia EDR3 from hereon, ][]{gaiaedr3} of the ESA Gaia mission, which was just recently published on 3 December 2020. The Gaia EDR3 is based on data, which could be collected with the instruments of the ESA-Gaia satellite during the first 34 months of its mission. This data release contains astrometric solutions, i.e. positions ($\alpha$, $\delta$), parallaxes $\pi$, and proper motions ($\mu_{\alpha}cos(\delta)$, $\mu_{\delta}$) of about 1.5 billion sources down to a limiting magnitude of 21\,mag in the G-band, i.e. white light observations, taking advantage of the full spectral response of the utilized CCD-detectors. Parallaxes are determined with an uncertainty in the range of about 0.02 milliarcsec (mas) for bright ($G<15$\,mag, with a lower magnitude limit of $G\sim1.7$\,mag) up to 0.5\,mas for faint ($G=20$\,mag) detected sources. Proper motions are measured with an uncertainty of about 0.02\,mas/yr for bright objects, which increases up to 0.6\,mas/yr at $G=20$\,mag. In addition, for all sources their G-band magnitude is given with a photometric precision in the range between about 0.3\,millimagnitude (mmag) for the brightest and 6\,mmag for faint sources.

In the survey, presented here, stellar companions of the investigated (C)TOIs are identified at first as sources, which are located at the same distances as the targets, and secondly share a common proper motion with these stars. In order to clearly detect co-moving companions and prove their equidistance with the (C)TOIs, only sources are taken into account in this survey, which are listed in the Gaia EDR3 and exhibit significant measurements of their parallaxes ($\pi/\sigma(\pi) > 3$) and proper motions ($\mu/\sigma(\mu) > 3$). Thereby sources with a negative parallax are neglected.

As this survey was originally based on data of the Gaia DR2, which exhibits a typical parallax uncertainty of 0.7\,mas for faint sources down to $G = 20$\,mag, the survey is constrained to (C)TOIs within 500\,pc around the Sun (i.e. $\pi > 2$\,mas), to assure $\pi/\sigma(\pi) > 3$ even for the faintest detectable companions. This distance constraint is slightly relaxed to $\pi + 3\sigma(\pi)>2$\,mas, i.e. taking into account also the parallax uncertainty of the (C)TOIs. Although from now on data from the Gaia EDR3 are used, which has smaller parallax uncertainties, we will keep for the survey the chosen distance constraint for continuity reasons.

Between the end of May and the beginning of December 2020, in total 585 stars were added to the (C)TOI release of the ExoFOP-TESS, which fulfil this distance constraint, and therefore were selected as new targets of our survey. The target selection did not take into account (C)TOIs with dips in their light curves, which could already be classified as false positive detections by follow-up observations, carried out in the course of the ExoFOP-TESS. Furthermore, (C)TOIs which are confirmed exoplanet host stars, whose multiplicity was already studied with Gaia data by \cite{mugrauer2019} or \cite{michel2021}, were excluded as targets as well.

\begin{figure*}
\resizebox{\hsize}{!}{\includegraphics{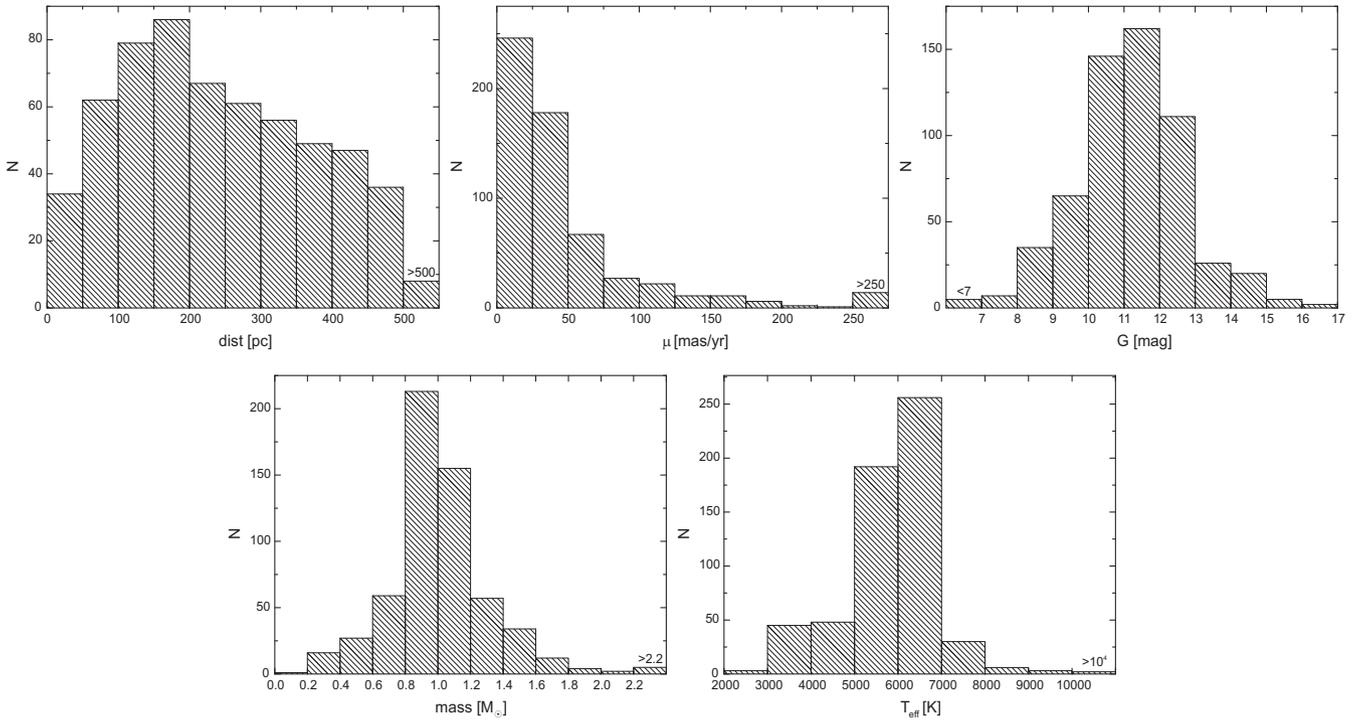}}\caption{The histograms of the individual properties of all targets.}\label{HIST_TARGETS}
\end{figure*}

The histograms of the properties of all selected targets are summarized in Fig.\,\ref{HIST_TARGETS}. The distances ($dist$) and the total proper motions ($\mu$) of the targets were derived with their accurate Gaia EDR3 parallaxes ($dist[\rm{pc}]=1000/\pi[\rm{mas}]$) and proper motions in right ascension and declination. The G-band magnitudes of all targets were taken from the Gaia EDR3, their masses and effective temperatures ($T_{\rm eff}$) from the Starhorse Catalog \citep[SHC from hereon,][]{anders2019}, respectively.

The targets are located at distances between about 10 up to 550\,pc and exhibit proper motions in the range between about 1 up to 1650\,mas/yr, G-band magnitudes from 4.7 to 17\,mag, effective temperatures from about 2900 up to 14200\,K, and masses, which range between about 0.2 and 4.8\,$M_{\odot}$.

According to the cumulative distribution functions of the individual properties, the targets are most frequently located at distances between about 100 and 250\,pc and have proper motions in the range between about 10 and 30\,mas/yr, as well as G-band magnitudes from $G=10$ to 13\,mag. The targets are mainly solar like stars with masses in the range between 0.9 and 1.2\,$M_{\odot}$. This population also emerges in the $T_{\rm eff}$ distribution of the targets at intermediate temperatures of about 5900 and 6400\,K. In addition, another but fainter pile-up of targets is evident in this distribution at lower effective temperatures between about 3000 and 4900\,K, which is the early K to mid M dwarf population.

As defined and described in \cite{mugrauer2020} our survey is limited to companions with projected separations up to 10000\,au, which guarantees an effective companion search on one side but also detects the vast majority of all wide companions of the selected targets. This results in an angular search radius for companions around the targets of $r [\rm{arcsec}] = 10 \pi[\rm{mas}]$, with $\pi$ the Gaia EDR3 parallaxes of the (C)TOIs.

All sources, listed in the Gaia EDR3, which are located within the used search radius around the targets and exhibit significant parallaxes and proper motions are considered as companion candidates. In total, 36132 such objects were detected around 518 targets, investigated in the course of this survey. The companionship of all these candidates was tested based on their accurate Gaia EDR3 astrometry and that of the associated (C)TOIs, exactly following the procedure, as described in \cite{mugrauer2020}. The vast majority of these sources\linebreak ($>99.7$\,\%) could be excluded as companions, as they do not share a common proper motion with the (C)TOIs and/or are not located at the same distances as  these stars. In contrast, for 119 candidates the companionship to the (C)TOIs could clearly be proven with their accurate Gaia EDR3 astrometry. The properties of these companions and of the associated (C)TOIs are described in detail in the next section of this paper.

\section{(C)TOIs and their detected stellar companions}

\begin{figure}
\resizebox{\hsize}{!}{\includegraphics{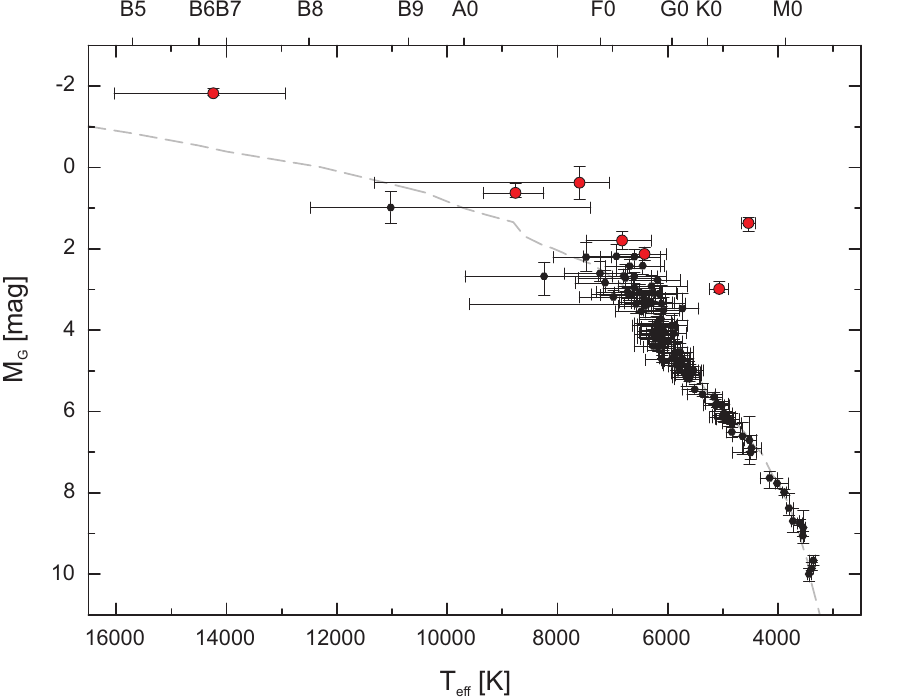}}\caption{The $T_{\rm eff}$-$M_{\rm G}$ diagram of all (C)TOIs with detected companions, presented here. The main sequence is shown as grey dashed line. (C)TOIs, listed in the SHC with surface gravities $\log(g[\rm{cm/s^{-2}}])\lesssim3.8$, are illustrated as red circles, those with larger surface gravities with black circles, respectively.}\label{HRDCTOIS}
\end{figure}

The masses, effective temperatures, and absolute G-band magnitudes of the (C)TOIs with detected companions, presented here, are all listed in the SHC and we plot these stars in the $T_{\rm eff}$-$M_{\rm G}$ diagram, which is shown in Fig.\,\ref{HRDCTOIS}. For comparison we plot in this diagram the main sequence from \cite{pecaut2013}\footnote{Online available at: \url{http://www.pas.rochester.edu/~emamajek/EEM_dwarf_UBVIJHK_colors_Teff.txt}}.

The vast majority of all targets with detected companions are main-sequence stars. Few (C)TOIs are located (significantly) above the main sequence and all of these stars exhibit surface gravities $\log(g[\rm{cm/s^{-2}}])\lesssim3.8$, as listed in the SHC, hence they are classified as (sub)giants.

\begin{figure}
\includegraphics[width=\linewidth]{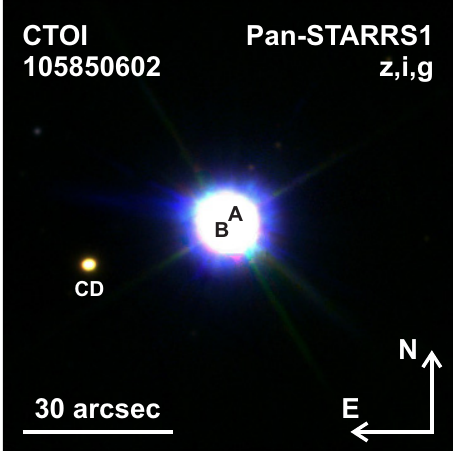}\caption{Color(RGB)-composit image of the quadruple system CTOI\,105850602\,AB+CD, made of z-, i-, and g-band images, taken by the Panoramic Survey Telescope and Rapid Response System (Pan-STARRS). The close binary companion (CD), whose components are separated from each other by only about 0.7\,arcsec, is not resolved by Pan-STARRS but it appears clearly elongated in this image, different to other point like sources detected around the star.}\label{PIC_Quad}
\end{figure}

The parallaxes, proper motions, apparent G-band magnitudes, and extinction estimates of the (C)TOIs and their companions, detected in this survey, are summarized in Tab.\,\ref{TAB_COMP_ASTROPHOTO}, which lists in total, 107 binary, and 5 hierarchical triple star systems. In the case of CTOI\,105850602 (alias HD\,146759), which is shown in Fig.\,\ref{PIC_Quad}, beside its close binary companion CTOI\,105850602\,CD, located at an angular separation of about 29\,arcsec ($\sim$\,3600\,au of projected separation), whose two components are resolved by Gaia, the star also exhibits a close companion-candidate ($\rho \sim 0.3$\,arcsec), whose relative astrometry is listed for two observing epochs in the Washington Double Star Catalog \citep[WDS from hereon,][]{mason2001}. The companion was observed in 2008 at $\rho=0.3$\,arcsec \& $PA=136^\circ$, and in 2014 at $\rho=0.3$\,arcsec \& $PA=132^\circ$, respectively. Adopting that this candidate is a non-moving background star, in 2014 we would have expected the object to be located at $\rho=0.35$\,arcsec \& $PA=96.4\,^\circ$, based on its 1st epoch astrometry and the accurate Gaia EDR3 parallax and proper motion of the CTOI. Because the companion exhibits a constant angular separation in both observing epochs and in particular its position angle does not decrease by about $40\,^\circ$ between both obser\-ving epochs, as expected for a non-moving background source, we conclude that this candidate is a co-moving companion of CTOI\,105850602. Therefore, this CTOI is actually the primary component of a stellar quadruple system.

For all detected companions we determined their angular separation ($\rho$) and position angle ($PA$) to the associated (C)TOIs, using the accurate Gaia EDR3 astrometry of the individual objects. The obtained relative astrometry of the companions is listed in Tab.\,\ref{TAB_COMP_RELASTRO}, together with its uncertainty, which remains below about 1\,mas in angular separation, and 0.03\,$^{\circ}$ in position angle, respectively.

The parallax difference $\Delta \pi$ between the (C)TOIs and their companions together with its significance $sig\text{-}\Delta\pi$ was calculated (in addition also by taking into account the astrometric excess noise of the individual objects) and is summarized in Tab.\,\ref{TAB_COMP_RELASTRO}. In the same table for each companion its differential proper motion $\mu_{\rm rel}$ relative to the associated (C)TOI is listed with its significance, as well as its $cpm$-$index$\footnote{The degree of common proper motion of a detected companion with the associated (C)TOI is characterized by its common proper motion (cpm) index, as defined in \cite{mugrauer2020}.}.

The parallaxes of the individual components of the stellar systems, presented here, do not significantly differ from each other ($sig\text{-}\Delta\pi < 3$), when the astrometric excess noise is taken into account. This clearly proves the equidistance of the detected companions with the (C)TOIs, as expected for components of physically associated stellar systems. Furthermore, the vast majority of the detected companions (more than 96\,\% of all) exhibit a $cpm\text{-}index> 10$ and all companions reach a $cpm\text{-}index > 5$. Hence, the detected companions and the associated (C)TOIs clearly form common proper motion pairs, as expected for gravitationally bound stellar systems.

\begin{figure}
\resizebox{\hsize}{!}{\includegraphics{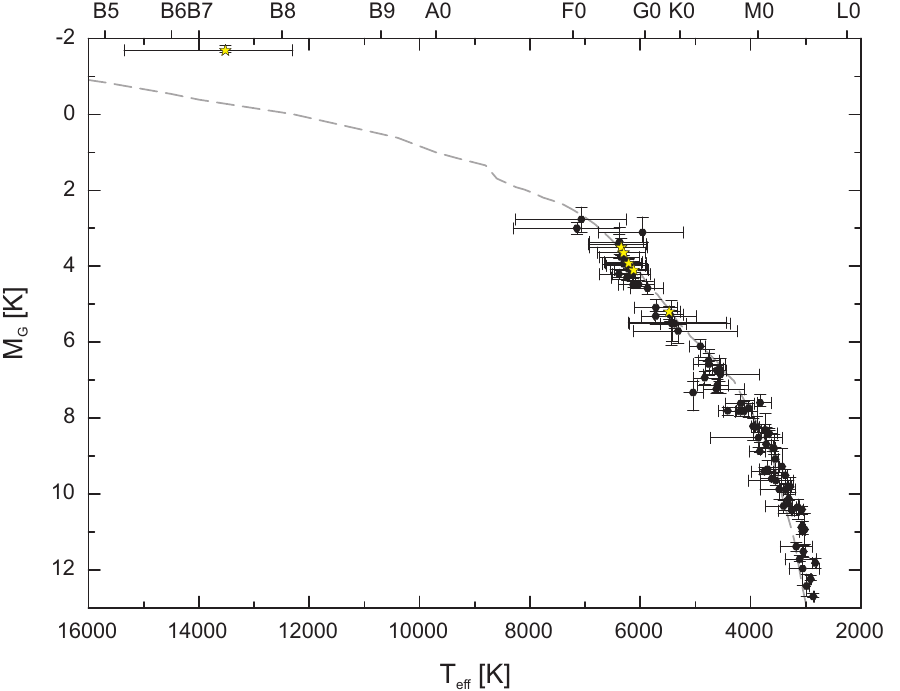}}\caption{This $T_{\rm eff}$-$M_{\rm G}$ diagram shows all detected companions, whose effective temperatures are either listed in the SHC or for which Apsis-Priam temperature estimates are available. Companions, which are the primary components of their stellar systems, are illustrated as yellow star symbols. The main sequence is plotted as dashed grey line for comparison.}\label{HRDCOMPS}
\end{figure}

\begin{table*}[h!]
\caption{The photometry of the five white dwarf companions, detected in this survey. For each companion we list the color difference $\Delta(B_{\rm P}-R_{\rm P})$, and the G-band magnitude difference $\Delta G$ to the associated (C)TOI, its apparent $(B_{\rm P}-R_{\rm P})$ color, as well as its derived intrinsic color $(B_{\rm P}-R_{\rm P})_{0}$, and effective temperature $T_{\rm eff}$.}\label{TAB_WDS_PROPS}
\centering
\begin{tabular}{lccccc}
\hline
Companion          & $\Delta(B_{\rm P}-R_{\rm P})$ $[$mag$]$ & $\Delta G$ $[$mag$]$  & $\left(B_{\rm P}-R_{\rm P}\right)$ $[$mag$]$ & $(B_{\rm P}-R_{\rm P})_{0}$ $[$mag$]$   & $T_{\rm eff}$$[$K$]$\\
\hline
CTOI\,253040591\,B & $-0.2954 \pm 0.0454$            & $8.6180 \pm 0.0042$   & $0.5392 \pm 0.3322$                  & $~~~0.3672_{-0.3555}^{+0.3378}$ & $~7263_{-1141}^{+1932}$\\
CTOI\,341411516\,B & $-0.3460 \pm 0.0714$            & $8.6006 \pm 0.0053$   & $0.4841 \pm 0.0698$                  & $~~~0.4724_{-0.0859}^{+0.0704}$ & $~6879_{-230}^{+309}$\\
CTOI\,369376388\,C & $-1.3829 \pm 0.0120$            & $5.1889 \pm 0.0040$   & $0.1757 \pm 0.0088$                  & $-0.1438_{-0.1141}^{+0.1745}$   & $10916_{-1841}^{+1038}$\\
TOI\,2092\,B       & $-0.2398 \pm 0.2533$            & $8.3105 \pm 0.0096$   & $0.6278 \pm 0.2532$                  & $~~~0.5768_{-0.2599}^{+0.2548}$ & $~6542_{-820}^{+974}$\\
TOI\,2127\,B       & $-0.3113 \pm 0.1537$            & $6.5909 \pm 0.0064$   & $0.9283 \pm 0.1537$                  & $~~~0.8203_{-0.1594}^{+0.1637}$ & $~5785_{-513}^{+485}$\\
\hline
\end{tabular}
\end{table*}
The equatorial coordinates, as well as the derived absolute G-band magnitudes, projected separations, masses, and effective temperatures of all detected companions are summarized in Tab.\,\ref{TAB_COMP_PROPS}.

The absolute G-band magnitudes of the companions are taken from the SHC if available, or were derived with their apparent G-band photometry and the parallaxes of the (C)TOIs from the Gaia EDR3, as well as the Apsis-Priam G-band extinction estimates, listed in the Gaia DR2. Thereby, always the extinction estimates of the companions if available, otherwise those of the (C)TOIs were used. For systems with no G-band extinction estimates, listed for any of their components, we have used the extinction estimates of the (C)TOIs from the SHC if available or from the Starhorse catalogue for 5 surveys \citep{queiroz2020}, indicated with the \texttt{SHC} and \texttt{SHC5} flag in Tab.\,\ref{TAB_COMP_ASTROPHOTO}, respectively. Thereby, V-band extinctions given in these catalogues were transformed to the G-band using the relation $A_{\rm G}/A_{\rm V}=0.77$, as determined by \cite{mugrauer2019}.

The projected separations of all companions were derived from their angular separations to the associated (C)TOIs and the parallaxes of these stars.

The masses and effective temperatures of all detected companions, presented here, including their uncertainties, are taken from the SHC if available, which applies to about 70\,\% of all companions. In Fig.\,\ref{HRDCOMPS} we plot these companions in a $T_{\rm eff}$-$M_{\rm G}$ diagram, together with the companions for which Apsis-Priam estimates of their effective temperatures are available\footnote{Following the recommendation of \cite{andrae2018} Apsis-Priam temperature estimates are only used in this survey if their flags are equal to $\texttt{1A000E}$ with $\texttt{A}$ and $\texttt{E}$ that can have any value.}, indicated by the $\texttt{PRI}$ flag in Tab.\,\ref{TAB_COMP_PROPS}. Except for the brightest and hottest companion, which is located slightly above the main sequence and exhibits a low surface gravity ($\log(\rm{g[cm/s^{-2}}])\lesssim3.8$), i.e. this companion is a subgiant, the photometry of the majority of all detected companions is well consistent with that expected for main-sequence stars.

For two companions, namely TOI\,2001\,B and TOI\,2115\,B no G-band photometry is listed neither in the Gaia EDR3 nor DR2, hence the properties of these companions could not be determined (indicated with the flag $\texttt{noGmag}$ in Tab.\,\ref{TAB_COMP_PROPS}).

For the remaining 34 companions, we derived their masses and effective temperatures from their absolute G-band magnitudes via interpolation ($\texttt{inter}$ flag in Tab.\,\ref{TAB_COMP_PROPS}) using the $M_{\rm G}$-mass and $M_{\rm G}$-$T_{\rm eff}$ relations from \cite{pecaut2013}, adopting that these companions are main-sequence stars. In order to verify this hypothesis, we compared the obtained effective temperatures of the companions with either their Apsis-Priam temperature estimates if available, or with the effective temperatures of the companions, derived from their $(B_{\rm P}-R_{\rm P})$ colors and Apsis-Priam reddening estimates $E(B_{\rm P}-R_{\rm P})$ or if not available those of the associated (C)TOIs, using the $(B_{\rm P}-R_{\rm P})_0$-$T_{\rm eff}$ relation from \cite{pecaut2013}. For the temperature estimation of the companions, we have used preferably their Gaia DR2 colors if available, instead of those listed in the Gaia EDR3, because (1) the photometric passbands of both Gaia data releases are different \citep{lindegren2020}, (2) the reddening estimates are given only in Gaia DR2 colors, and (3) the used color-temperature relation is also based on Gaia DR2 photometry.

For all but five of these companions their effective temperatures, derived from their absolute magnitudes by assuming that they are main-sequence stars, agree well with either their Apsis-Priam temperature estimates or the temperatures, obtained from their colors. The typical deviation of the different temperature estimates is about 380\,K, well consistent with the average uncertainty of the derived effective temperatures. Hence, we conclude that these companions are all main-sequence stars.

In addition, also the Gaia EDR3 $(B_{\rm P}-R_{\rm P})$ colors of the (C)TOIs and their companions (if available) were compared with each other, indicated by the $\texttt{BPRP}$ flag in Tab.\,\ref{TAB_COMP_PROPS}. For main-sequence stars we expect that companions, which are fainter/brighter than the (C)TOIs, appear redder/bluer than the stars and this holds for the majority of all detected companions except for CTOI\,253040591\,B, CTOI\,341411516\,B, CTOI\,369376388\,C, TOI\,2092\,B, and TOI\,2127\,B. We summarize the photometric properties of these companions in Tab.\,\ref{TAB_WDS_PROPS}. The companions are all several magnitudes fainter than the associated (C)TOIs, but appear bluer than their primaries. Furthermore, the temperatures of these companions, derived from their absolute G-band magnitudes, adopting that they are main-sequence stars, is about 2800 to 7800\,K lower than the temperatures, obtained from their colors, which are listed in Tab.\,\ref{TAB_WDS_PROPS}.

\begin{figure}
\includegraphics[width=\linewidth]{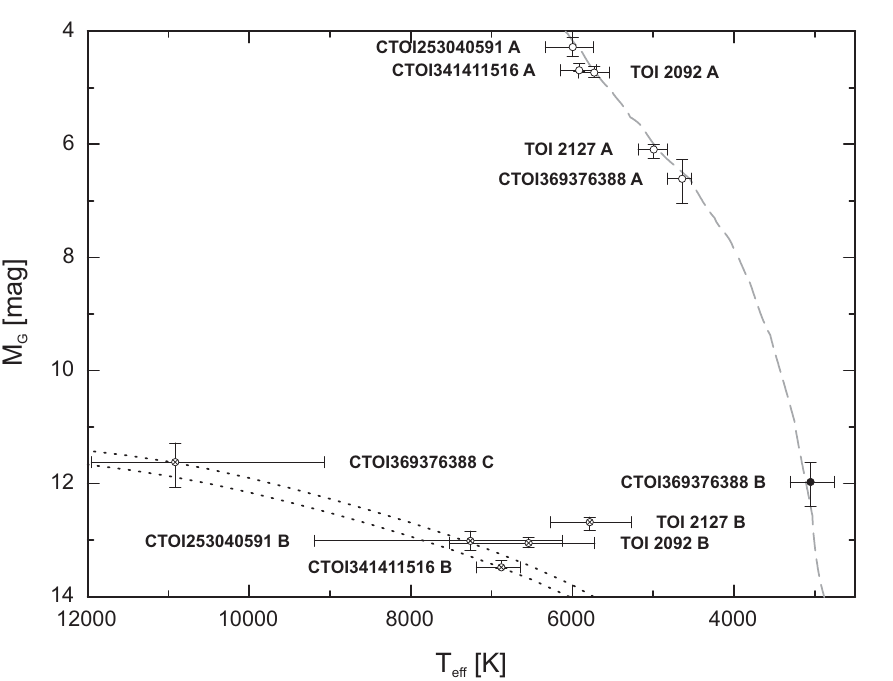}\caption{$T_{\rm eff}$-$M_{\rm G}$ diagram of the stellar systems with white dwarf components, detected in this survey. The main sequence is plotted as grey dashed line and the evolutionary mass tracks of DA white dwarfs with masses of 0.5 and 0.6\,$M_{\odot}$ as black dotted lines, respectively. The primaries of the systems are shown as white circles, main-sequence companions as black, and white dwarf companions as white crossed circles, respectively.}\label{HRD_WDS}
\end{figure}

\begin{figure*}
\includegraphics[width=\linewidth]{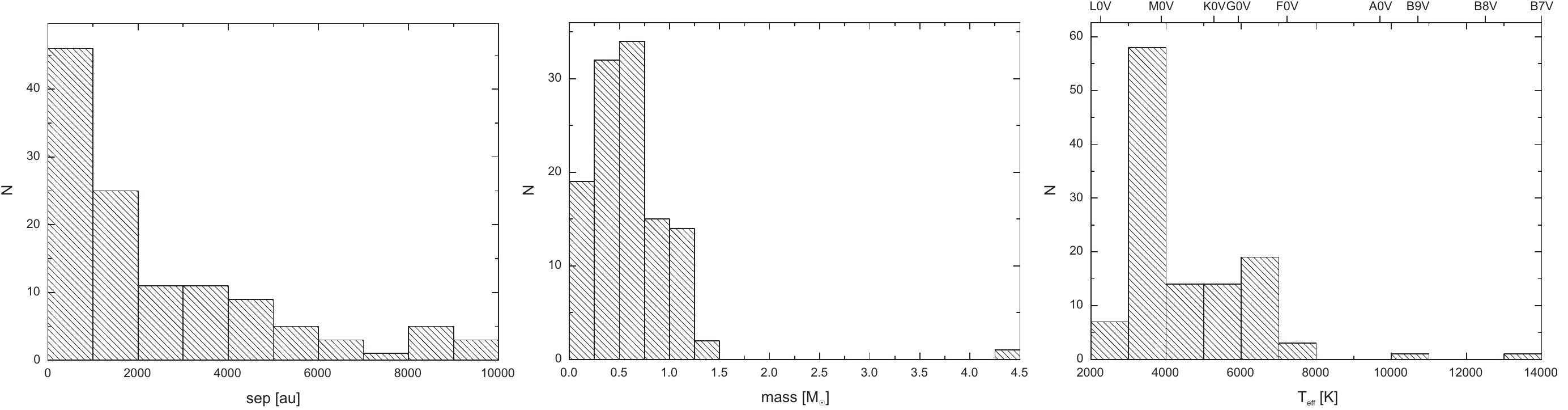}\caption{The histograms of the properties of the companions, detected in this survey.}\label{HIST_COMPS}
\end{figure*}

In Fig.\,\ref{HRD_WDS} we plot these companions together with the other components of their stellar systems, in a $T_{\rm eff}$-$M_{\rm G}$ diagram. For comparison we show in this diagram the main sequence from \cite{pecaut2013}, as well as evolutionary mass tracks of DA white dwarfs from the white dwarf models of \cite{holberg2006}, \cite{kowalski2006}, \cite{tremblay2011}, and \cite{bergeron2011}. While the brighter primary components of these systems as well as the secondary CTOI\,369376388\,B are all main-sequence stars the five faint companions are clearly located below the main sequence and their Gaia photometry is well consistent with that expected for white dwarfs, except for TOI\,2127\,B, which is discussed in more detail below.

\begin{figure*}
\begin{center}\includegraphics[height=20cm]{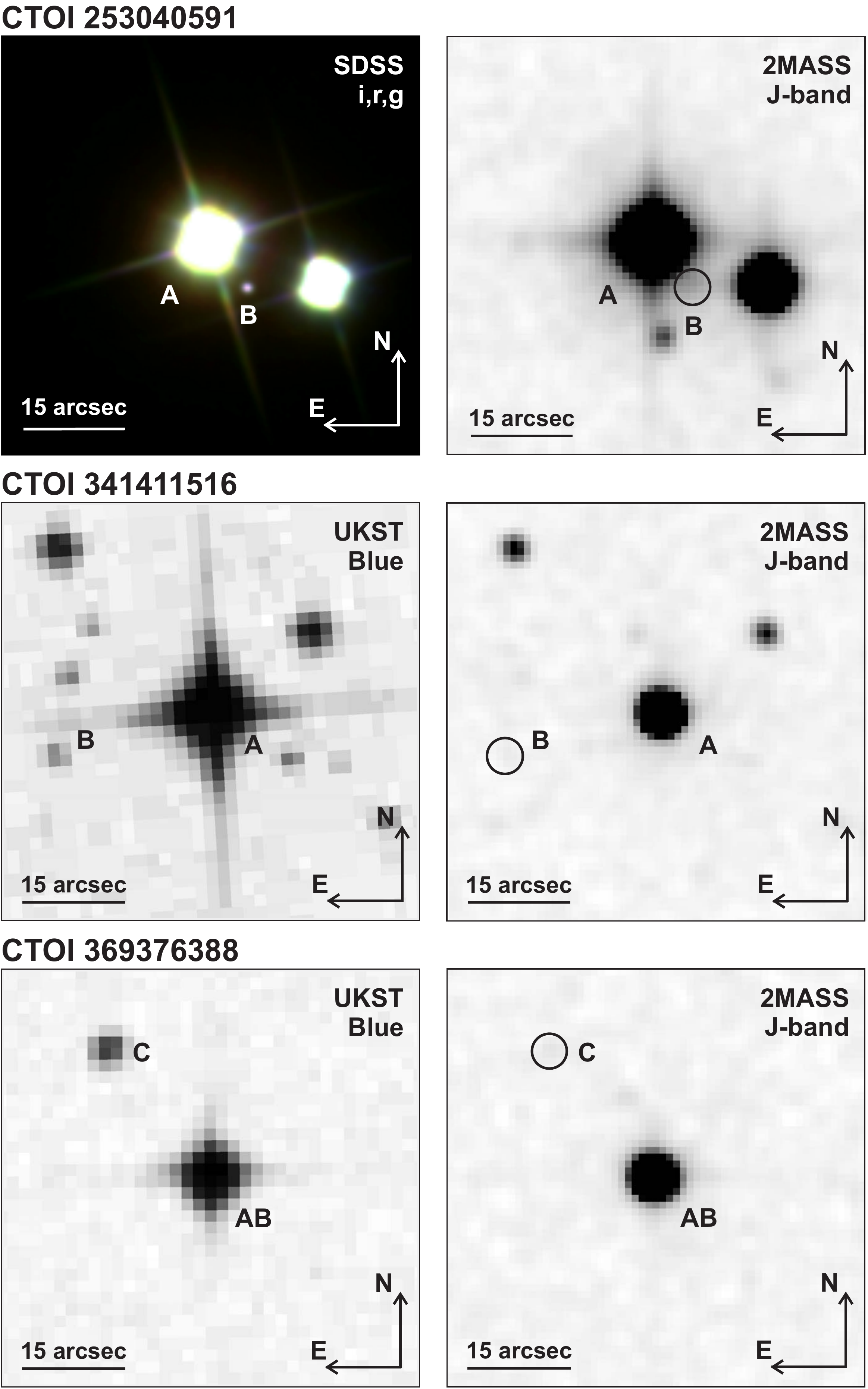}\end{center}\caption{A color(RGB)-composite image of the white dwarf CTOI\,253040591\,B, created from imaging data, taken in the course of the Sloan Digital Sky Survey (SDSS) in the i-, r-, and g-band together with images of the white dwarf companions CTOI\,341411516\,B, and CTOI\,369376388\,C, taken with the UK Schmidt Telescope (UKST) in the filter GG395. 2MASS J-band images of the CTOIs are shown for comparison with the expected position of the companions, indicated with black circles. While the companions are all well detected in the optical spectral range they remain invisible in the 2MASS images, consistent with the photometric properties expected for the companions in the case that they are white dwarfs.}\label{PICS}
\end{figure*}

As shown in Fig.\ref{PICS}, due to their wider angular separations ($\rho \gtrsim 8.8$\,arcsec) the faint companions CTOI\,253040591\,B, CTOI\,341411516\,B, and CTOI\,369376388\,C are all detected next to their primaries in all sky survey images, taken in the optical spectral range, while they remain invisible in the near infrared J-band images of the Two Micron All Sky Survey (2MASS). If these three companions would be low-mass main-sequence stars their absolute G-band magnitudes corresponds to masses in the range between 0.12 and 0.2\,$M_\odot$. Such low-mass companions would exhibit apparent J-band magnitudes between 14.0 and 16.2\,mag \citep[extinction is taken into account, adopting $A_{\rm J}/A_{\rm G}=0.34$, as derived by ][]{mugrauer2019}. As the 2MASS J-band images exhibit $SNR=10$ detection limits between 16.4 and 16.6\,mag, the companions should easily be detectable in these images, which is however not the case. Hence the companions exhibit an intrinsic faintness in both the optical and near infrared spectral range that clearly rule out that they are low-mass main-sequence stars. Therefore, we conclude that these stars are all white dwarf companions of the associated (C)TOIs, which is indicated with the $\texttt{WD}$ flag in Tab.\,\ref{TAB_COMP_PROPS}.

\begin{figure}
\includegraphics[width=\linewidth]{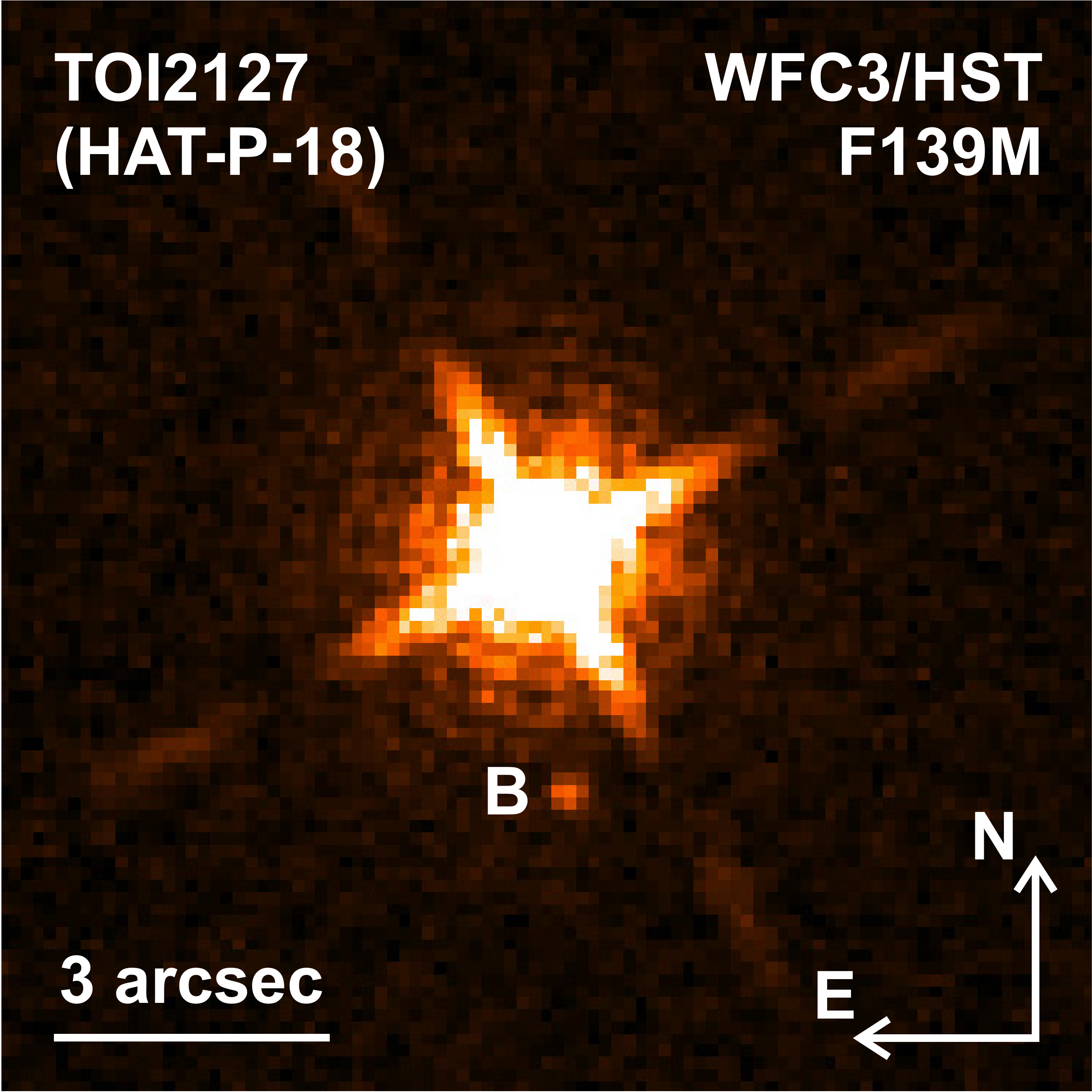}\caption{TOI\,2127 (alias HAT-P-18) observed on 11 February 2016 in the near infrared (F139M) with the WFC3 camera, aboard of the Hubble Space Telescope. The faint companion TOI\,2127\,B is well detected in the image south of the exoplanet host star.}\label{PIC_HATP18}
\end{figure}

As illustrated in the $T_{\rm eff}$-$M_{\rm G}$ diagram in Fig.\,\ref{HRD_WDS}, TOI\,2127\,B is located in the range between the main sequence and the white dwarf tracks slightly closer to these tracks than to the main sequence. Hence, from its Gaia photometry alone the classification of the nature of this companion remains uncertain. However, TOI\,2127 (alias HAT-P-18) was also observed in the near infrared (F139M) with the Wide Field Camera 3 (WFC3), aboard of the Hubble Space Telescope (HST). Two images of the star were taken on 11 February 2016, and on 12 January 2017, each with an integration time of 29.7\,s. Beside the exoplanet host star also its faint companion TOI\,2127\,B is detected in the HST images on average at $\rho=2.667\pm0.006$\,arcsec \& $PA=186.2\pm0.3\,^{\circ}$, very well consistent with the relative Gaia EDR3 astrometry of the companion, as listed in Tab.\,\ref{TAB_COMP_RELASTRO}. The first epoch WFC3 image of TOI\,2127 is shown in Fig.\ref{PIC_HATP18}.

We have used aperture photometry to determined the magnitude difference between the companion and the exoplanet host star in both WFC3 images and obtained $\Delta \text{F139M}=8.01\pm0.09$\,mag.
The exoplanet host star is listed in the SHC as a main-sequence star with a mass of $0.78_{-0.04}^{+0.02}\,M_{\odot}$. If its companion would be a low-mass main-sequence star as well, its absolute G-band magnitude, listed in Tab.\,\ref{TAB_COMP_PROPS}, corresponds to a mass of $0.15\pm0.01\,M_{\odot}$, following the $M_{\rm G}$-mass relation from \cite{pecaut2013}. According to the Dartmouth Stellar Evolution Database \citep{dotter2008}, for such a low-mass stellar companion we would expect a magnitude difference to HAT-P-18 of $\Delta \text{F139M}=4.7_{-0.3}^{+0.2}$\,mag at the age of the exoplanet host star of about 12\,Gyr \citep{hartmann2011}. However, the companion is significantly fainter (by about 3.3\,mag) in the near infrared than expected for a low-mass main-sequence star. Therefore, from its Gaia and HST photometry we conclude that TOI\,2127\,B is a white dwarf companion of the exoplanet host star. Follow-up spectroscopic observations are needed to further constrain its properties, as well as those of all the other degenerated companions, detected in this survey.

The histograms of the properties of all companions, presented here, are illustrated in Fig.\,\ref{HIST_COMPS}. The companions exhibit angular separations to the (C)TOIs, in the range between about 0.8 and 63\,arcsec, which corresponds to projected separations of 117 up to 9463\,au. According to the underlying cumulative distribution function, the frequency of the companions is the highest and constant up to about 500\,au while it continually decreases for larger projected separations. Half of all companions exhibit projected separations of less than 1300\,au. In total, 7 stellar systems (6 binaries and one hierarchical triple) are detected with projected separations below 200\,au, namely: TOI\,2009\,AB, TOI\,2072\,AB, TOI\,2183\,AB, TOI\,2299\,AB, TOI\,2384\,AB, TOI\,2422\,AB, and CTOI\,369376388\,AB+C(WD), i.e. these systems are the most challenging environments for planet formation, identified in this study.

The masses of the companions range between about 0.09\,$M_{\odot}$ and 4.5\,$M_{\odot}$ (average mass is $\sim$0.6\,$M_\odot$). The highest companion frequency is found in the cumulative distribution function in the mass range between 0.15 and 0.6\,$M_{\odot}$, which corresponds beside detected white dwarf companions mainly to mid M to late K dwarfs, according to the relation between mass and spectral type (SpT), described by \cite{pecaut2013}. For higher masses the companion frequency is lower but constant up to about 1.2\,$M_{\odot}$ from where it significantly decreases towards higher masses. This peak in the companion population is also detected in the distribution of their effective temperatures, which exhibits the highest frequency of companions in the temperature range between 3000 and 4000\,K. In this distribution also a second but fainter pile-up of companions is prominent, which is located between about 6000 and 6500\,K and corresponds to late to mid F type stars, according to the $T_{\rm eff}$-$\text{SpT}$ relation from \cite{pecaut2013}.

\begin{figure}[h]
\includegraphics[width=\linewidth]{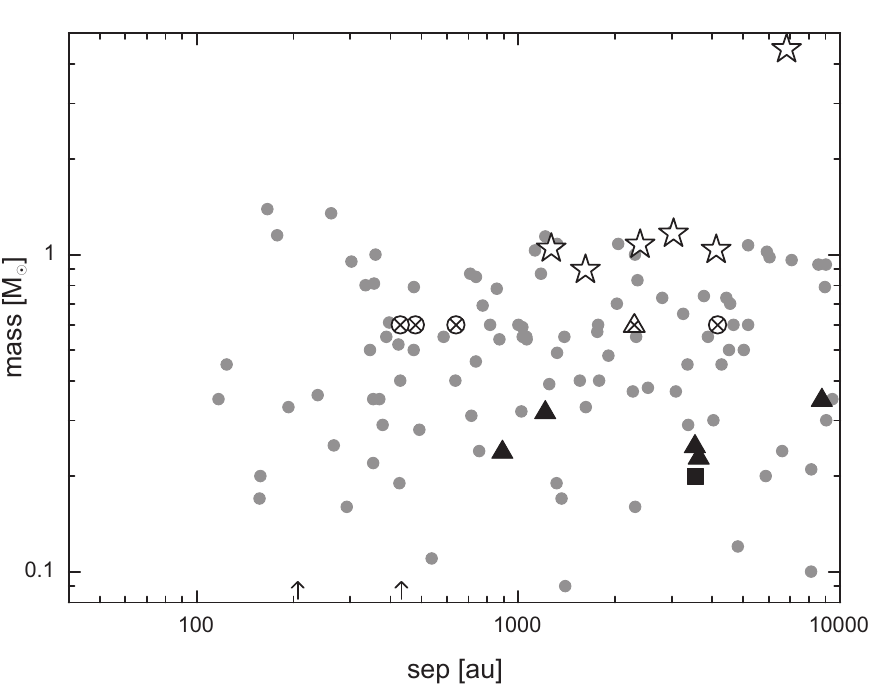}\caption{The separation-mass diagram of the companions, detected in this survey. Companions, which are the primary components of their stellar systems, are plotted as star symbols, those which are secondaries as circles, tertiary components as triangles, and quaternary components as squares, respectively. Detected white dwarf companions, for which a mass of 0.6\,$M_{\odot}$ is adopted, are plotted with white crossed symbols. The separations of the two companions TOI\,2001\,B, and TOI\,2115\,B, for which no masses could be determined, are indicated with black arrows.}\label{SEPMASS}
\end{figure}

As shown in the separation-mass diagram in Fig.\,\ref{SEPMASS}, among all 119 companions, presented here, 6 are the primary, 106 the secondary, 6 are the tertiary, and one is the quaternary component of their stellar systems.

\begin{figure}[h]
\includegraphics[width=\linewidth]{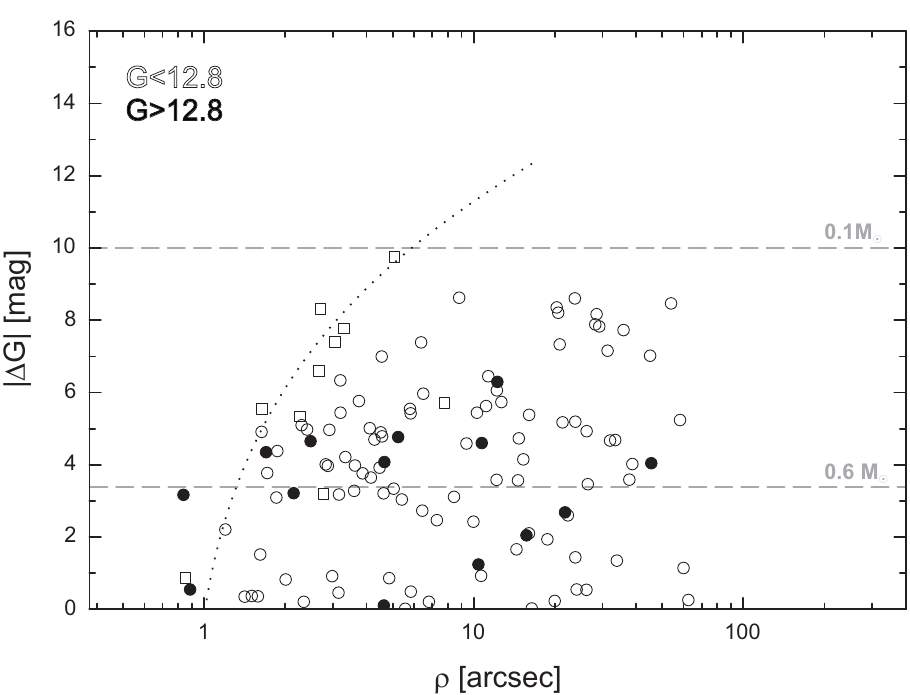}\caption{The magnitude differences of all detected companions plotted versus their angular separations to the associated (C)TOIs. The Gaia detection limit, found by \cite{mugrauer2020}, is shown as dotted line for comparison. The expected average magnitude difference for companions with 0.1 or 0.6\,$M_{\odot}$ is drawn as grey dashed horizontal lines. Companions of (C)TIOs brighter and fainter than G = 12.8 mag are plotted as open circles and filled black circles, respectively. Companions, whose astrometric solutions are only listed in the Gaia EDR3, are shown as open boxes.}\label{LIMIT}
\end{figure}

In order to characterize the detection limit, reached in this survey, we plot the magnitude differences of all detected companions over their angular separations to the associated (C)TOIs, as shown in Fig.\,\ref{LIMIT}.

For comparison we show the Gaia DR2 detection limit, determined by \cite{mugrauer2020} among (C)TOIs, which are brighter than $G=12.8$\,mag (about 88\% of the targets with detected companions of the survey, presented here). Companions with angular separations larger than about 1\,arcsec are detectable around bright (C)TOIs while closer companions slightly below this separation limit with magnitude differences up to about 3\,mag can be detected by Gaia around fainter targets. The companions, identified in this survey with astrometric solutions listed also in the Gaia DR2, all agree with the determined Gaia detection limit. In contrast, 4 companions detected in this survey around bright (C)TOIs exceed this specific limit. However, the astrometric solutions of these companions are listed only in the Gaia EDR3. In addition, all of the companions, whose astrometric solutions are listed only in this Gaia data release have angular separations from the associated (C)TOIs, which are smaller than about 8\,arcsec. Both indicate that the Gaia EDR3 has a higher sensitivity to close companions compared to its precursor.

The expected magnitude differences between the targets of this survey and low-mass main-sequence companions (indicated with grey dashed lines in Fig.\,\ref{LIMIT}) are estimated with the expected absolute G-band magnitudes of these stars, as listed by \cite{pecaut2013}, and the average absolute G-band magnitude of our targets ($M_{\rm G}=4.5$\,mag). As shown in Fig.\,\ref{LIMIT}, a magnitude difference of about 3.4\,mag is reached at an angular separation of about 1.4\,arcsec around the targets of this survey. This allows the detection of companions with masses down to about 0.6\,$M_\odot$ (average mass of all detected companions) which are separated from the (C)TOIs by more than 340\,au. Furthermore, companions with masses down to about 0.1\,$M_\odot$ are detectable beyond 6\,arcsec, which corresponds to a projected separation of 1440\,au at the average target distance of 240\,pc.

\section{Summary and Outlook}

The goal of the survey, whose latest results are presented here, is the detection and characterization of stellar companions of (C)TOIs, i.e. of potential exoplanet host stars. In this paper we have explored the multiplicity of 585 (C)TOIs, which were announced in the (C)TOI Release of the ExoFOP-TESS between the end of May and the beginning of December 2020.

In contrast to \cite{mugrauer2020}, who explored the Gaia DR2 to search for companions around (C)TOIs, the continuation of this survey, presented here, is based on Gaia EDR3 astro- and photometry. We have used the target sample as well as the detected companions to characterize the differences between both Gaia data releases regarding their sensitivity and the accuracy of their astrometric solutions. Within the applied search radius around the targets in total about 36100 sources with accurate astrometric solutions were identified in the Gaia EDR3 while only about 34000 such objects were found in the same regions on the sky in the Gaia DR2. In addition, among all 119 companions, detected in this survey, 12 have astrometric solutions only listed in the Gaia EDR3 (indicated with the $\texttt{EDR3}$ flag in Tab.\,\ref{TAB_COMP_PROPS}) and all of these companions are located within 8\,arcsec around the associated (C)TOIs. Therefore, we conclude that the Gaia EDR3 contains about 10\,\% more sources with accurate astrometric solutions and is more sensitive in particular for close companions at angular separations below about 10\,arcsec compared to its precursor. In addition, the astrometric solutions of the targets and their companions in the Gaia EDR3 exhibit uncertainties in their astrometric position, parallax, and proper motion, which are smaller by factors of about 0.6, 0.6, and 0.4, respectively, compared to the Gaia DR2. The larger number of detected sources, the higher sensitivity in particular for close companions, as well as the more accurate astrometry of the Gaia EDR3 compared to its precursor, found in this study, all agree well with the general characteristics of both Gaia data releases, as described by \cite{gaiadr2}, and \cite{gaiaedr3}.

In total we have detected companions around 113 of the 585 targets, whose multiplicity was studied here. Hence, the multiplicity rate of the investigated (C)TOIs is at least about $19\pm2$\,\%, which is consistent with the multiplicity rate of (C)TOIs, reported by \cite{mugrauer2020}.

Beside 107 binaries also 5 hierarchical triple star systems and one quadruple system were detected, in which either the (C)TOI exhibits a close and a wide companion or a close binary companion instead, which is located at a wider angular separation. As it is expected for the components of gravitationally bound stellar systems the (C)TOIs and the detected companions are equidistant and share a common proper motion, as proven with their accurate Gaia EDR3 parallaxes and proper motions. In particular, the direct proof of equidistance of the individual components of the stellar systems, as done in this survey by comparing their parallaxes, was not feasible in earlier multiplicity surveys before the release of the accurate Gaia data because in particular for the majority of the faint companions their parallaxes could not be measured by the ESA-Hipparcos mission \citep{perryman1997}.

However, 35 companions, identified in this survey, were already listed in the WDS, either as co-moving companions or as companion candidates of the (C)TOIs, which still needed confirmation of their companionship, eventually yielded by this survey. Although the WDS is currently the most complete available catalogue of multiple star systems, which contains relative astrometric measurements of multiple systems spanning a period of more than 300 years, in this study 84 (i.e. 70\,\% of all) companions were detected, which are not listed in the WDS, indicated with the $\bigstar$ flag in Tab.\,\ref{TAB_COMP_RELASTRO}. This demonstrates the great potential of the ESA-Gaia mission for multiplicity studies of stars, in particular for the detection of wide companions, as it is illustrated with the derived detection limit of this survey, shown in Fig.\,\ref{LIMIT}. On average, all stellar companions with masses down to about 0.1\,$M_\odot$ are detectable in this study around the targets beyond about 6\,arcsec (or 1440\,au of projected separation) and approximately half of all detected companions exhibit such separations. In total, companions are identified with projected separations between about 120 and 9500\,au and the frequency of companions is constant up to about 500\,au and continually decreases for larger projected separations. The companions, detected in this survey, exhibit masses in the range between about 0.09\,$M_\odot$ and 4.5\,$M_\odot$ and are most frequently found in the mass range between 0.15 and 0.6\,$M_\odot$. Beside low-mass main-sequence stars (mainly early to mid M dwarfs) also 5 white dwarfs could be identified as co-moving companions of the (C)TOIs, whose true nature was revealed in this survey, using their accurate astro- and photometric properties.

For 99 (i.e. about 83\,\% of all) companions, presented here, significant ($sig\text{-}\mu_{\rm rel} \geq 3$) differential proper motions $\mu_{\rm rel}$ relative to the associated (C)TOIs were detected. We derived the escape velocities $\mu_{\rm esc}$ of all these companions using the approximation, described in \cite{mugrauer2019}. The differential proper motion of most of these companions is consistent with orbital motion. In contrast for 19 companions, their differential proper motions significantly exceed the expected escape velocities, indicating an increased degree of multiplicity, as discussed in \cite{mugrauer2019}. Two of these companions are located in an already confirmed hierarchical triple star system. Follow-up high contrast imaging observations are needed to further explore the multiplicity status of all these particular systems and their companions, which are summarized in Tab.\,\ref{table_triples}.

\begin{table}[h!] \caption{List of all detected companions (sorted by their identifier), whose differential proper motions $\mu_{\rm rel}$ relative to the (C)TOIs significantly exceed the expected escape velocities $\mu_{\rm esc}$. Companions, which are already known to be members of hierarchical triple star systems, are indicated with $\bigstar\bigstar\bigstar$.}
\begin{center}
\begin{tabular}{lccc}
\hline
Companion          & $\mu_{\rm rel}$ [mas/yr] & $\mu_{\rm esc}$ [mas/yr] &\\
\hline
TOI\,1946\,B &                                       $5.65\pm0.22$ & $2.09 \pm 0.07$\\
TOI\,1966\,B &                                       $1.06\pm0.04$ & $0.52 \pm 0.02$\\
TOI\,2006\,B &                                       $0.68\pm0.02$ & $0.54 \pm 0.03$\\
TOI\,2050\,B &                                       $4.93\pm0.04$ & $2.51 \pm 0.06$\\
TOI\,2218\,A &                                       $2.42\pm0.05$ & $0.74 \pm 0.04$\\
TOI\,2239\,B &                                       $0.58\pm0.02$ & $0.30 \pm 0.02$\\
TOI\,2281\,B &                                       $1.26\pm0.02$ & $0.96 \pm 0.05$\\
TOI\,2325\,B &                                       $2.59\pm0.05$ & $1.73 \pm 0.01$\\
TOI\,2358\,B &                                       $1.75\pm0.16$ & $0.58 \pm 0.03$\\
TOI\,2409\,B &                                       $3.11\pm0.03$ & $0.87 \pm 0.02$\\
TOI\,2417\,B &                                       $3.23\pm0.03$ & $2.13 \pm 0.09$\\
{\fontsize{7}{0}\selectfont CTOI\,35703676\,B}  &    $3.71\pm0.10$ & $0.50 \pm 0.03$\\
{\fontsize{7}{0}\selectfont CTOI\,135145585\,B} &    $4.19\pm0.09$ & $0.42 \pm 0.02$\\
{\fontsize{7}{0}\selectfont CTOI\,151628217\,B} &    $1.35\pm0.02$ & $0.31 \pm 0.02$\\
{\fontsize{7}{0}\selectfont CTOI\,197760286\,B} &    $5.51\pm0.14$ & $1.92 \pm 0.13$\\
{\fontsize{7}{0}\selectfont CTOI\,230236827\,B} &    $6.56\pm0.05$ & $5.68 \pm 0.02$\\
{\fontsize{7}{0}\selectfont CTOI\,288240183\,B} &    $5.46\pm0.51$ & $0.98 \pm 0.10$ & $\bigstar\bigstar\bigstar$\\
{\fontsize{7}{0}\selectfont CTOI\,288240183\,C} &    $3.77\pm0.53$ & $0.87 \pm 0.05$ & $\bigstar\bigstar\bigstar$\\
{\fontsize{7}{0}\selectfont CTOI\,300116105\,B} &    $1.95\pm0.10$ & $1.24 \pm 0.05$\\
\hline
\end{tabular}
\end{center}
\label{table_triples}
\end{table}

The survey, whose latest results are presented here, is an ongoing project and its target list is steadily growing due to the continuing analysis of photometric data, collected by the TESS mission. The multiplicity of all these newly revealed (C)TOIs will be explored in the course of this survey and detected companions and their determined properties will be reported regularly in this journal and will also be made available online in the \verb"VizieR" database \citep{ochsenbein2000}\footnote{Online available at: \url{https://vizier.u-strasbg.fr/viz-bin/VizieR}}, and at the Webpage of the survey\footnote{Online available at: \url{https://www.astro.uni-jena.de/Users/markus/Multiplicity_of_(C)TOIs.html}}. The results of this survey combined with those of high-contrast imaging observations of the (C)TOIs, which can detect close companions with projected separations down to only a few au, will complete our knowledge of the multiplicity of all these potential exoplanet host stars.

\bibliography{paper}

\section*{Acknowledgments}

We made use of data from:

(1) the \verb"Simbad" and \verb"VizieR" databases operated at CDS in Strasbourg, France.

(2) the European Space Agency (ESA) mission Gaia (\url{https://www.cosmos.esa.int/gaia}), processed by the Gaia Data Processing and Analysis Consortium (DPAC, \url{https://www.cosmos.esa.int/web/gaia/dpac/consortium}). Funding for the DPAC has been provided by national institutions, in particular the institutions participating in the Gaia Multilateral Agreement.

(3) the \verb"Exoplanet Follow-up Observing Program" website, which is operated by the California Institute of Technology, under contract with the National Aeronautics and Space Administration under the Exoplanet Exploration Program.

(4) the Two Micron All Sky Survey, which is a joint project of the University of Massachusetts and the Infrared Processing and Analysis Center/California Institute of Technology, funded by the National Aeronautics and Space Administration and the National Science Foundation.

(5) the Digitized Sky Surveys, which were produced at the Space Telescope Science Institute (STScI) under U.S. Government grant NAG W-2166. The images of these surveys, used here, were taken with the UK Schmidt Telescope, which was operated by the Royal Observatory Edinburgh, with funding from the UK Science and Engineering Research Council (later the UK Particle Physics and Astronomy Research Council), until 1988 June and thereafter by the Anglo-Australian Observatory. Supplemental funding for sky-survey work at the STScI is provided by the European Southern Observatory.

(6) the Sloan Digital Sky Survey (SDSS) which has been provided by the Alfred P. Sloan Foundation, the Participating Institutions, the National Aeronautics and Space Administration, the National Science Foundation, the U.S. Department of Energy, the Japanese Monbukagakusho, and the Max Planck Society. The SDSS Web site is http://www.sdss.org/. The SDSS is managed by the Astrophysical Research Consortium (ARC) for the Participating Institutions. The Participating Institutions are The University of Chicago, Fermilab, the Institute for Advanced Study, the Japan Participation Group, The Johns Hopkins University, Los Alamos National Laboratory, the Max-Planck-Institute for Astronomy (MPIA), the Max-Planck-Institute for Astrophysics (MPA), New Mexico State University, University of Pittsburgh, Princeton University, the United States Naval Observatory, and the University of Washington.

(7) the Pan-STARRS1 surveys, which were made possible through contributions by the Institute for Astronomy, the University of Hawaii, the Pan-STARRS Project Office, the Max-Planck Society and its participating institutes, the Max Planck Institute for Astronomy, Heidelberg and the Max Planck Institute for Extraterrestrial Physics, Garching, The Johns Hopkins University, Durham University, the University of Edinburgh, the Queen's University Belfast, the Harvard-Smithsonian Center for Astrophysics, the Las Cumbres Observatory Global Telescope Network Incorporated, the National Central University of Taiwan, the Space Telescope Science Institute, and the National Aeronautics and Space Administration under Grant No. NNX08AR22G issued through the Planetary Science Division of the NASA Science Mission Directorate, the National Science Foundation Grant No. AST-1238877, the University of Maryland, Eotvos Lorand University (ELTE), and the Los Alamos National Laboratory. The Pan-STARRS1 Surveys are archived at the Space Telescope Science Institute (STScI) and can be accessed through MAST, the Mikulski Archive for Space Telescopes. Additional support for the Pan-STARRS1 public science archive is provided by the Gordon and Betty Moore Foundation.

\begin{center}
\begin{table*}[h]
\caption{This table summarizes for all (C)TOIs (listed at first) and their detected companions their Gaia EDR3 parallaxes $\pi$, proper motions $\mu$ in right ascension and declination, astrometric excess noises $epsi$, G-band magnitudes, as well as the used Apsis-Priam G-Band extinction estimates $A_{\rm G}$ or if not available the G-Band extinctions, as listed either in the SHC or in the SHC5, indicated with $\texttt{SHC}$ and $\texttt{SHC5}$, respectively.}\label{TAB_COMP_ASTROPHOTO}
\centering
\begin{tabular}{lccccccc}
\hline
TOI            & $\pi$              & $\mu_{\alpha}cos(\delta)$    & $\mu_{\delta}$ & $epsi$ & $G$     & $A_{\rm G}$\\
               & [mas]              & [mas/yr]                     & [mas/yr]       & [mas]  & [mag] & [mag]\\
\hline
1937\,A & $ 2.4113 \pm 0.0108 $ & $ -5.627 \pm 0.013 $ & $ 11.309 \pm 0.013 $ & 0.027 & $ 13.0048 \pm 0.0028 $ & $ 0.3990 _{ -0.1630 }^{ +0.2310 }$ &  \\
1937\,B & $ 2.3514 \pm 0.0891 $ & $ -5.387 \pm 0.104 $ & $ 11.349 \pm 0.096 $ & 0.373 & $ 17.6530 \pm 0.0034 $ & $  _{  }^{  }$ &  \\
\hline
1940\,A & $ 3.7497 \pm 0.0177 $ & $ 4.967 \pm 0.021 $ & $ -2.265 \pm 0.016 $ & 0.097 & $ 9.8591 \pm 0.0028 $ & $ 0.3259 _{ -0.2092 }^{ +0.2092 }$ & $\texttt{SHC}$ \\
1940\,B & $ 2.9986 \pm 0.1211 $ & $ 4.181 \pm 0.152 $ & $ -0.663 \pm 0.161 $ & 0.712 & $ 13.0511 \pm 0.0045 $ & $  _{  }^{  }$ &  \\
\hline
1943\,A & $ 7.6634 \pm 0.0125 $ & $ -91.364 \pm 0.012 $ & $ -24.450 \pm 0.014 $ & 0.025 & $ 10.6662 \pm 0.0028 $ & $ 0.2485 _{ -0.1046 }^{ +0.0889 }$ &  \\
1943\,B & $ 7.5589 \pm 0.0458 $ & $ -92.307 \pm 0.045 $ & $ -24.881 \pm 0.057 $ & 0.397 & $ 14.6400 \pm 0.0031 $ & $  _{  }^{  }$ &  \\
\hline
1946\,A & $ 3.9857 \pm 0.1243 $ & $ -32.551 \pm 0.126 $ & $ -20.338 \pm 0.124 $ & 0.918 & $ 8.7606 \pm 0.0028 $ & $ 0.1566 _{ -0.1566 }^{ +0.2006 }$ & $\texttt{SHC}$ \\
1946\,B & $ 3.3840 \pm 0.1615 $ & $ -26.909 \pm 0.176 $ & $ -20.078 \pm 0.132 $ & 0.952 & $ 12.7681 \pm 0.0034 $ & $  _{  }^{  }$ &  \\
\hline
1953\,A & $ 3.7995 \pm 0.0314 $ & $ -6.971 \pm 0.036 $ & $ -3.961 \pm 0.031 $ & 0.223 & $ 9.8517 \pm 0.0028 $ & $ 1.2685 _{ -0.1765 }^{ +0.0835 }$ &  \\
1953\,B & $ 3.8352 \pm 0.1079 $ & $ -7.352 \pm 0.133 $ & $ -3.312 \pm 0.107 $ & 0.311 & $ 17.5745 \pm 0.0033 $ & $  _{  }^{  }$ &  \\
\hline
1964\,A & $ 2.5086 \pm 0.0198 $ & $ -19.280 \pm 0.022 $ & $ 3.241 \pm 0.016 $ & 0.136 & $ 11.1381 \pm 0.0028 $ & $ 0.0105 _{ -0.0105 }^{ +0.1119 }$ & $\texttt{SHC}$ \\
1964\,B & $ 2.4861 \pm 0.0628 $ & $ -19.105 \pm 0.066 $ & $ 3.314 \pm 0.057 $ & 0.237 & $ 16.5546 \pm 0.0036 $ & $  _{  }^{  }$ &  \\
\hline
1966\,A & $ 3.7310 \pm 0.0380 $ & $ -39.460 \pm 0.037 $ & $ 1.372 \pm 0.046 $ & 0.285 & $ 9.3827 \pm 0.0028 $ & $  _{  }^{  }$ &  \\
1966\,B & $ 3.3062 \pm 0.0196 $ & $ -38.402 \pm 0.020 $ & $ 1.287 \pm 0.025 $ & 0.000 & $ 14.0628 \pm 0.0028 $ & $ 0.7567 _{ -0.2802 }^{ +0.1699 }$ &  \\
\hline
1970\,A & $ 2.4254 \pm 0.0365 $ & $ -19.008 \pm 0.029 $ & $ -0.431 \pm 0.028 $ & 0.317 & $ 11.3380 \pm 0.0028 $ & $ 0.8230 _{ -0.2274 }^{ +0.2910 }$ &  \\
1970\,B & $ 2.4923 \pm 0.0828 $ & $ -19.215 \pm 0.066 $ & $ -0.414 \pm 0.070 $ & 0.000 & $ 17.3905 \pm 0.0029 $ & $  _{  }^{  }$ &  \\
\hline
1972\,A & $ 5.0597 \pm 0.0123 $ & $ -9.982 \pm 0.013 $ & $ 34.161 \pm 0.014 $ & 0.056 & $ 10.1464 \pm 0.0028 $ & $  _{  }^{  }$ &  \\
1972\,B & $ 5.0479 \pm 0.0115 $ & $ -9.781 \pm 0.013 $ & $ 33.776 \pm 0.013 $ & 0.050 & $ 10.6799 \pm 0.0028 $ & $ 0.0780 _{ -0.0610 }^{ +0.2566 }$ &  \\
\hline
1984\,A & $ 4.0450 \pm 0.0135 $ & $ -12.779 \pm 0.010 $ & $ -24.344 \pm 0.011 $ & 0.083 & $ 10.7127 \pm 0.0028 $ & $ 0.0260 _{ -0.0260 }^{ +0.1245 }$ &  \\
1984\,B & $ 4.0759 \pm 0.2280 $ & $ -12.242 \pm 0.172 $ & $ -24.758 \pm 0.172 $ & 0.905 & $ 18.1146 \pm 0.0081 $ & $  _{  }^{  }$ &  \\
\hline
1992\,A & $ 2.6006 \pm 0.0153 $ & $ 0.481 \pm 0.012 $ & $ -7.645 \pm 0.014 $ & 0.099 & $ 10.8569 \pm 0.0028 $ & $  _{  }^{  }$ &  \\
1992\,B & $ 2.6098 \pm 0.0234 $ & $ 0.092 \pm 0.020 $ & $ -7.758 \pm 0.022 $ & 0.204 & $ 11.3123 \pm 0.0028 $ & $ 0.4967 _{ -0.3633 }^{ +0.2533 }$ &  \\
\hline
2001\,A & $ 2.2236 \pm 0.0196 $ & $ -2.694 \pm 0.024 $ & $ 2.659 \pm 0.024 $ & 0.185 & $ 9.1756 \pm 0.0028 $ & $ 0.3032 _{ -0.3032 }^{ +0.3928 }$ & $\texttt{SHC}$ \\
2001\,B & $ 2.4396 \pm 0.0925 $ & $ -3.651 \pm 0.108 $ & $ 3.820 \pm 0.227 $ & 0.649 & $    $ & $  _{  }^{  }$ &  \\
\hline
2006\,A & $ 2.0109 \pm 0.0117 $ & $ 8.281 \pm 0.015 $ & $ 22.092 \pm 0.013 $ & 0.045 & $ 9.9077 \pm 0.0028 $ & $  _{  }^{  }$ &  \\
2006\,B & $ 1.9621 \pm 0.0159 $ & $ 7.662 \pm 0.017 $ & $ 22.363 \pm 0.019 $ & 0.079 & $ 13.1132 \pm 0.0028 $ & $ 0.4653 _{ -0.1830 }^{ +0.1268 }$ &  \\
\hline
2009\,A & $ 48.6807 \pm 0.0323 $ & $ 103.168 \pm 0.036 $ & $ -490.281 \pm 0.025 $ & 0.161 & $ 8.0339 \pm 0.0028 $ & $  _{  }^{  }$ &  \\
2009\,B & $ 48.7040 \pm 0.0478 $ & $ 96.607 \pm 0.049 $ & $ -495.495 \pm 0.035 $ & 0.315 & $ 12.6158 \pm 0.0028 $ & $ 0.3915 _{ -0.1796 }^{ +0.1630 }$ &  \\
\hline
2033\,A & $ 3.5400 \pm 0.0118 $ & $ 24.900 \pm 0.010 $ & $ -2.831 \pm 0.012 $ & 0.054 & $ 9.7429 \pm 0.0028 $ & $  _{  }^{  }$ &  \\
2033\,B & $ 3.5679 \pm 0.0118 $ & $ 24.779 \pm 0.010 $ & $ -2.912 \pm 0.012 $ & 0.062 & $ 10.6656 \pm 0.0028 $ & $ 0.6020 _{ -0.3314 }^{ +0.3616 }$ &  \\
\hline
2036\,A & $ 4.4161 \pm 0.0173 $ & $ -17.432 \pm 0.014 $ & $ 15.230 \pm 0.013 $ & 0.121 & $ 9.3132 \pm 0.0028 $ & $ 1.1860 _{ -0.4734 }^{ +0.2960 }$ &  \\
2036\,B & $ 4.3064 \pm 0.0662 $ & $ -16.975 \pm 0.063 $ & $ 14.628 \pm 0.060 $ & 0.336 & $ 16.3031 \pm 0.0042 $ & $  _{  }^{  }$ &  \\
\hline
2050\,A & $ 8.7474 \pm 0.0114 $ & $ 20.878 \pm 0.013 $ & $ 2.096 \pm 0.013 $ & 0.086 & $ 10.1196 \pm 0.0028 $ & $  _{  }^{  }$ &  \\
2050\,B & $ 8.6148 \pm 0.0366 $ & $ 23.547 \pm 0.042 $ & $ 6.240 \pm 0.039 $ & 0.403 & $ 13.6954 \pm 0.0028 $ & $ 0.3630 _{ -0.1361 }^{ +0.2810 }$ &  \\
\hline
2056\,A & $ 10.7827 \pm 0.0191 $ & $ -98.730 \pm 0.017 $ & $ -9.845 \pm 0.019 $ & 0.139 & $ 7.6570 \pm 0.0028 $ & $ 0.1313 _{ -0.1313 }^{ +0.1649 }$ & $\texttt{SHC}$ \\
2056\,B & $ 10.7935 \pm 0.0175 $ & $ -96.775 \pm 0.016 $ & $ -10.659 \pm 0.018 $ & 0.147 & $ 12.3521 \pm 0.0030 $ & $  _{  }^{  }$ &  \\
\hline
2068\,A & $ 18.8696 \pm 0.0131 $ & $ -197.943 \pm 0.013 $ & $ -6.062 \pm 0.013 $ & 0.097 & $ 12.2109 \pm 0.0028 $ & $  _{  }^{  }$ &  \\
2068\,B & $ 18.8558 \pm 0.0144 $ & $ -200.442 \pm 0.015 $ & $ -7.292 \pm 0.015 $ & 0.116 & $ 12.4366 \pm 0.0028 $ & $ 0.3745 _{ -0.1616 }^{ +0.3516 }$ &  \\
\hline
2072\,A & $ 25.6289 \pm 0.0192 $ & $ -184.967 \pm 0.022 $ & $ -87.241 \pm 0.023 $ & 0.169 & $ 12.6971 \pm 0.0028 $ & $  _{  }^{  }$ &  \\
2072\,B & $ 25.5547 \pm 0.0192 $ & $ -184.594 \pm 0.021 $ & $ -92.030 \pm 0.023 $ & 0.143 & $ 13.6121 \pm 0.0028 $ & $ 0.1350 _{ -0.1014 }^{ +0.1140 }$ &  \\
\hline
2084\,A & $ 8.7499 \pm 0.0166 $ & $ 47.731 \pm 0.018 $ & $ 36.756 \pm 0.019 $ & 0.000 & $ 14.3890 \pm 0.0028 $ & $ 0.0210 _{ -0.0140 }^{ +0.1170 }$ &  \\
2084\,B & $ 8.7935 \pm 0.7475 $ & $ 49.097 \pm 0.825 $ & $ 38.893 \pm 0.967 $ & 1.419 & $ 20.6793 \pm 0.0093 $ & $  _{  }^{  }$ &  \\
\hline
2092\,A & $ 5.6461 \pm 0.0133 $ & $ 96.066 \pm 0.011 $ & $ -29.136 \pm 0.014 $ & 0.110 & $ 11.0975 \pm 0.0028 $ & $ 0.1137 _{ -0.0718 }^{ +0.0977 }$ &  \\
2092\,B & $ 4.1900 \pm 0.7059 $ & $ 96.657 \pm 0.693 $ & $ -30.618 \pm 0.776 $ & 1.273 & $ 19.4080 \pm 0.0092 $ & $  _{  }^{  }$ &  \\
\hline
\end{tabular}
\end{table*}
\end{center}

\setcounter{table}{2}

\begin{center}
\begin{table*}[h]
\caption{continued}
\centering
\begin{tabular}{lccccccc}
\hline
TOI            & $\pi$              & $\mu_{\alpha}cos(\delta)$    & $\mu_{\delta}$ & $epsi$ & $G$     & $A_{\rm G}$\\
               & [mas]              & [mas/yr]                     & [mas/yr]       & [mas]  & [mag] & [mag]\\
\hline
2094\,A & $ 19.9111 \pm 0.0129 $ & $ -55.300 \pm 0.016 $ & $ -0.125 \pm 0.017 $ & 0.072 & $ 13.4332 \pm 0.0028 $ & $ 0.3315 _{ -0.0416 }^{ +0.1336 }$ &  \\
2094\,B & $ 20.0622 \pm 0.1183 $ & $ -54.380 \pm 0.154 $ & $ 1.269 \pm 0.150 $ & 0.896 & $ 18.0302 \pm 0.0033 $ & $  _{  }^{  }$ &  \\
\hline
2106\,A & $ 8.3310 \pm 0.0136 $ & $ -28.444 \pm 0.011 $ & $ -63.918 \pm 0.013 $ & 0.103 & $ 10.3233 \pm 0.0028 $ & $ 0.1190 _{ -0.0930 }^{ +0.0940 }$ &  \\
2106\,B & $ 8.4985 \pm 0.4626 $ & $ -25.845 \pm 0.384 $ & $ -60.689 \pm 0.403 $ & 4.903 & $ 14.2890 \pm 0.0047 $ & $  _{  }^{  }$ &  \\
\hline
2108\,A & $ 3.9652 \pm 0.0149 $ & $ -0.348 \pm 0.013 $ & $ 7.484 \pm 0.014 $ & 0.106 & $ 10.5998 \pm 0.0028 $ & $ 0.4700 _{ -0.2686 }^{ +0.4091 }$ &  \\
2108\,B & $ 3.9693 \pm 0.0541 $ & $ -0.694 \pm 0.110 $ & $ 8.224 \pm 0.046 $ & 0.330 & $ 12.8030 \pm 0.0034 $ & $  _{  }^{  }$ &  \\
\hline
2113\,A & $ 3.9296 \pm 0.0172 $ & $ -2.477 \pm 0.020 $ & $ -63.678 \pm 0.025 $ & 0.182 & $ 10.9145 \pm 0.0028 $ & $  _{  }^{  }$ &  \\
2113\,B & $ 3.9522 \pm 0.0240 $ & $ -2.283 \pm 0.027 $ & $ -62.231 \pm 0.036 $ & 0.255 & $ 11.2621 \pm 0.0029 $ & $ 0.0445 _{ -0.0445 }^{ +0.1313 }$ & $\texttt{SHC}$ \\
\hline
2115\,A & $ 4.6375 \pm 0.0252 $ & $ -5.432 \pm 0.019 $ & $ 9.584 \pm 0.027 $ & 0.216 & $ 8.4507 \pm 0.0028 $ & $ 0.7949 _{ -0.3851 }^{ +0.3851 }$ & $\texttt{SHC}$ \\
2115\,B & $ 5.0970 \pm 0.1099 $ & $ -4.066 \pm 0.079 $ & $ 9.996 \pm 0.204 $ & 0.698 & $    $ & $  _{  }^{  }$ &  \\
\hline
2127\,A & $ 6.1863 \pm 0.0093 $ & $ -14.002 \pm 0.009 $ & $ -36.751 \pm 0.011 $ & 0.052 & $ 12.3582 \pm 0.0028 $ & $ 0.2303 _{ -0.1490 }^{ +0.0858 }$ &  \\
2127\,B & $ 6.4227 \pm 0.2978 $ & $ -14.782 \pm 0.261 $ & $ -35.939 \pm 0.356 $ & 1.378 & $ 18.9491 \pm 0.0058 $ & $  _{  }^{  }$ &  \\
\hline
2128\,A & $ 27.2686 \pm 0.0151 $ & $ -161.884 \pm 0.016 $ & $ -42.106 \pm 0.017 $ & 0.140 & $ 7.0810 \pm 0.0028 $ & $ 0.0685 _{ -0.0605 }^{ +0.1156 }$ &  \\
2128\,B & $ 27.2456 \pm 0.0397 $ & $ -165.242 \pm 0.046 $ & $ -44.975 \pm 0.043 $ & 0.462 & $ 13.0419 \pm 0.0028 $ & $  _{  }^{  }$ &  \\
\hline
2144\,A & $ 9.4696 \pm 0.0112 $ & $ 74.073 \pm 0.014 $ & $ 77.519 \pm 0.013 $ & 0.084 & $ 10.3998 \pm 0.0028 $ & $ 0.0813 _{ -0.0590 }^{ +0.1628 }$ &  \\
2144\,B & $ 9.4699 \pm 0.0208 $ & $ 75.337 \pm 0.026 $ & $ 76.393 \pm 0.023 $ & 0.061 & $ 15.1281 \pm 0.0028 $ & $  _{  }^{  }$ &  \\
\hline
2149\,A & $ 4.5418 \pm 0.0145 $ & $ 13.482 \pm 0.016 $ & $ 15.150 \pm 0.018 $ & 0.146 & $ 10.5168 \pm 0.0028 $ & $ 0.6150 _{ -0.4568 }^{ +0.3468 }$ &  \\
2149\,B & $ 4.5903 \pm 0.0264 $ & $ 15.393 \pm 0.027 $ & $ 12.735 \pm 0.036 $ & 0.247 & $ 12.0274 \pm 0.0028 $ & $  _{  }^{  }$ &  \\
\hline
2152\,A & $ 3.1166 \pm 0.0169 $ & $ 27.534 \pm 0.017 $ & $ -11.833 \pm 0.019 $ & 0.151 & $ 11.2435 \pm 0.0028 $ & $ 0.7547 _{ -0.1955 }^{ +0.1955 }$ & $\texttt{SHC}$ \\
2152\,B & $ 3.2220 \pm 0.3165 $ & $ 26.996 \pm 0.291 $ & $ -12.013 \pm 0.326 $ & 1.280 & $ 19.4442 \pm 0.0040 $ & $  _{  }^{  }$ &  \\
\hline
2169\,A & $ 2.7494 \pm 0.0169 $ & $ 5.505 \pm 0.012 $ & $ -31.042 \pm 0.016 $ & 0.156 & $ 10.9589 \pm 0.0028 $ & $  _{  }^{  }$ &  \\
2169\,B & $ 2.7704 \pm 0.0140 $ & $ 5.273 \pm 0.010 $ & $ -30.728 \pm 0.016 $ & 0.000 & $ 13.6850 \pm 0.0028 $ & $ 0.1838 _{ -0.1013 }^{ +0.1992 }$ &  \\
\hline
2183\,A & $ 9.5169 \pm 0.0190 $ & $ -15.514 \pm 0.020 $ & $ 43.704 \pm 0.020 $ & 0.134 & $ 8.4875 \pm 0.0028 $ & $ 0.6354 _{ -0.3262 }^{ +0.3262 }$ & $\texttt{SHC}$ \\
2183\,B & $ 9.5902 \pm 0.0204 $ & $ -21.330 \pm 0.027 $ & $ 40.148 \pm 0.021 $ & 0.156 & $ 8.8416 \pm 0.0028 $ & $  _{  }^{  }$ &  \\
\hline
2193\,A & $ 2.9294 \pm 0.0094 $ & $ -2.465 \pm 0.009 $ & $ -0.911 \pm 0.011 $ & 0.050 & $ 11.8060 \pm 0.0028 $ & $ 0.1520 _{ -0.1170 }^{ +0.3031 }$ &  \\
2193\,B & $ 2.9032 \pm 0.0601 $ & $ -2.530 \pm 0.052 $ & $ -0.987 \pm 0.064 $ & 0.403 & $ 16.1816 \pm 0.0029 $ & $  _{  }^{  }$ &  \\
\hline
2195\,A & $ 5.7048 \pm 0.0100 $ & $ -16.224 \pm 0.012 $ & $ -55.807 \pm 0.011 $ & 0.080 & $ 11.5123 \pm 0.0028 $ & $ 0.1810 _{ -0.1180 }^{ +0.2771 }$ &  \\
2195\,B & $ 5.6714 \pm 0.0382 $ & $ -15.625 \pm 0.048 $ & $ -54.966 \pm 0.044 $ & 0.339 & $ 15.7231 \pm 0.0029 $ & $  _{  }^{  }$ &  \\
\hline
2205\,A & $ 2.4099 \pm 0.0295 $ & $ -2.328 \pm 0.038 $ & $ 27.633 \pm 0.042 $ & 0.027 & $ 16.1296 \pm 0.0029 $ & $ 0.1342 _{ -0.0601 }^{ +0.0601 }$ & $\texttt{SHC}$ \\
2205\,B & $ 2.3861 \pm 0.1390 $ & $ -2.123 \pm 0.165 $ & $ 27.466 \pm 0.203 $ & 0.601 & $ 18.8104 \pm 0.0035 $ & $  _{  }^{  }$ &  \\
2205\,C & $ 2.5016 \pm 0.1934 $ & $ -2.725 \pm 0.223 $ & $ 28.340 \pm 0.287 $ & 0.000 & $ 19.3401 \pm 0.0041 $ & $  _{  }^{  }$ &  \\
\hline
2215\,A & $ 14.1234 \pm 0.0160 $ & $ -8.274 \pm 0.016 $ & $ -6.212 \pm 0.013 $ & 0.069 & $ 10.5895 \pm 0.0028 $ & $  _{  }^{  }$ &  \\
2215\,B & $ 14.0973 \pm 0.0171 $ & $ -8.401 \pm 0.016 $ & $ -6.279 \pm 0.014 $ & 0.071 & $ 10.8414 \pm 0.0028 $ & $ 0.0320 _{ -0.0107 }^{ +0.1923 }$ &  \\
\hline
2218\,B & $ 2.8549 \pm 0.0356 $ & $ -17.669 \pm 0.046 $ & $ 33.504 \pm 0.040 $ & 0.388 & $ 11.9424 \pm 0.0028 $ & $  _{  }^{  }$ &  \\
2218\,A & $ 2.5423 \pm 0.0110 $ & $ -15.254 \pm 0.014 $ & $ 33.429 \pm 0.012 $ & 0.082 & $ 11.7408 \pm 0.0028 $ & $ 0.1377 _{ -0.1310 }^{ +0.2053 }$ &  \\
\hline
2233\,A & $ 5.1165 \pm 0.0291 $ & $ 1.165 \pm 0.035 $ & $ -24.974 \pm 0.023 $ & 0.201 & $ 11.5776 \pm 0.0030 $ & $ 0.5892 _{ -0.3520 }^{ +0.2088 }$ &  \\
2233\,B & $ 5.2052 \pm 0.0379 $ & $ 1.061 \pm 0.048 $ & $ -25.646 \pm 0.030 $ & 0.000 & $ 15.4905 \pm 0.0029 $ & $  _{  }^{  }$ &  \\
\hline
2239\,A & $ 2.1906 \pm 0.0118 $ & $ 2.987 \pm 0.017 $ & $ -4.658 \pm 0.013 $ & 0.096 & $ 11.3658 \pm 0.0028 $ & $  _{  }^{  }$ &  \\
2239\,B & $ 2.0299 \pm 0.0106 $ & $ 2.982 \pm 0.017 $ & $ -4.083 \pm 0.012 $ & 0.000 & $ 13.2956 \pm 0.0028 $ & $ 0.1557 _{ -0.0681 }^{ +0.1704 }$ &  \\
\hline
2244\,A & $ 5.7439 \pm 0.0188 $ & $ 15.007 \pm 0.017 $ & $ -11.535 \pm 0.018 $ & 0.104 & $ 9.5576 \pm 0.0029 $ & $ 0.1198 _{ -0.1198 }^{ +0.1538 }$ & $\texttt{SHC}$ \\
2244\,B & $ 5.7633 \pm 0.0222 $ & $ 16.129 \pm 0.019 $ & $ -11.662 \pm 0.018 $ & 0.119 & $ 9.9215 \pm 0.0029 $ & $  _{  }^{  }$ &  \\
2244\,C & $ 5.7981 \pm 0.1555 $ & $ 15.022 \pm 0.144 $ & $ -10.886 \pm 0.148 $ & 0.438 & $ 17.9098 \pm 0.0033 $ & $  _{  }^{  }$ &  \\
\hline
2246\,A & $ 4.4280 \pm 0.0121 $ & $ 16.025 \pm 0.015 $ & $ 51.047 \pm 0.016 $ & 0.107 & $ 10.3992 \pm 0.0028 $ & $  _{  }^{  }$ &  \\
2246\,B & $ 4.4457 \pm 0.0100 $ & $ 16.939 \pm 0.013 $ & $ 50.845 \pm 0.013 $ & 0.062 & $ 10.8843 \pm 0.0028 $ & $ 0.4560 _{ -0.2243 }^{ +0.2521 }$ &  \\
\hline
2248\,A & $ 2.6398 \pm 0.0110 $ & $ 3.956 \pm 0.013 $ & $ 15.787 \pm 0.014 $ & 0.000 & $ 10.7376 \pm 0.0028 $ & $ 0.8535 _{ -0.1875 }^{ +0.2826 }$ &  \\
2248\,B & $ 2.8055 \pm 0.2212 $ & $ 3.562 \pm 0.260 $ & $ 15.319 \pm 0.296 $ & 0.831 & $ 18.5134 \pm 0.0119 $ & $  _{  }^{  }$ &  \\
\hline
2253\,A & $ 5.8108 \pm 0.0532 $ & $ 9.455 \pm 0.056 $ & $ 17.639 \pm 0.066 $ & 0.567 & $ 10.4163 \pm 0.0028 $ & $  _{  }^{  }$ &  \\
2253\,B & $ 6.5645 \pm 0.0428 $ & $ 8.891 \pm 0.045 $ & $ 17.588 \pm 0.056 $ & 0.309 & $ 15.3387 \pm 0.0028 $ & $ 0.3200 _{ -0.1481 }^{ +0.4791 }$ &  \\
\hline
\end{tabular}
\end{table*}
\end{center}

\setcounter{table}{2}

\begin{center}
\begin{table*}[h]
\caption{continued}
\centering
\begin{tabular}{lccccccc}
\hline
TOI            & $\pi$              & $\mu_{\alpha}cos(\delta)$    & $\mu_{\delta}$ & $epsi$ & $G$     & $A_{\rm G}$\\
               & [mas]              & [mas/yr]                     & [mas/yr]       & [mas]  & [mag] & [mag]\\
\hline
2279\,A & $ 10.6363 \pm 0.0097 $ & $ 22.635 \pm 0.011 $ & $ 16.147 \pm 0.013 $ & 0.054 & $ 10.8215 \pm 0.0028 $ & $ 0.1605 _{ -0.0848 }^{ +0.2243 }$ &  \\
2279\,B & $ 10.6551 \pm 0.0296 $ & $ 23.264 \pm 0.036 $ & $ 15.201 \pm 0.044 $ & 0.172 & $ 15.6025 \pm 0.0029 $ & $  _{  }^{  }$ &  \\
\hline
2281\,A & $ 5.6262 \pm 0.0116 $ & $ 28.656 \pm 0.013 $ & $ -14.733 \pm 0.014 $ & 0.090 & $ 9.4515 \pm 0.0028 $ & $  _{  }^{  }$ &  \\
2281\,B & $ 5.6049 \pm 0.0108 $ & $ 28.760 \pm 0.013 $ & $ -13.475 \pm 0.013 $ & 0.072 & $ 10.7952 \pm 0.0028 $ & $ 0.0600 _{ -0.0430 }^{ +0.1118 }$ &  \\
\hline
2283\,A & $ 20.1663 \pm 0.0123 $ & $ -56.281 \pm 0.016 $ & $ 109.172 \pm 0.016 $ & 0.104 & $ 12.2841 \pm 0.0028 $ & $  _{  }^{  }$ &  \\
2283\,B & $ 20.1454 \pm 0.0280 $ & $ -57.264 \pm 0.037 $ & $ 107.463 \pm 0.035 $ & 0.000 & $ 15.7433 \pm 0.0028 $ & $ 0.1090 _{ -0.0321 }^{ +0.0897 }$ &  \\
\hline
2289\,A & $ 7.1011 \pm 0.0144 $ & $ 4.525 \pm 0.016 $ & $ 75.018 \pm 0.017 $ & 0.111 & $ 10.4503 \pm 0.0028 $ & $  _{  }^{  }$ &  \\
2289\,B & $ 7.1855 \pm 0.0410 $ & $ 4.324 \pm 0.045 $ & $ 75.200 \pm 0.051 $ & 0.193 & $ 16.1822 \pm 0.0029 $ & $ 0.7560 _{ -0.3173 }^{ +0.2162 }$ &  \\
\hline
2293\,A & $ 15.9360 \pm 0.0156 $ & $ 45.215 \pm 0.011 $ & $ -120.703 \pm 0.015 $ & 0.111 & $ 12.8578 \pm 0.0028 $ & $  _{  }^{  }$ &  \\
2293\,B & $ 16.0752 \pm 0.0617 $ & $ 45.112 \pm 0.046 $ & $ -123.739 \pm 0.059 $ & 0.206 & $ 16.9297 \pm 0.0029 $ & $ 0.3100 _{ -0.2540 }^{ +0.3101 }$ &  \\
\hline
2299\,A & $ 29.1809 \pm 0.0120 $ & $ -14.452 \pm 0.015 $ & $ 195.077 \pm 0.015 $ & 0.085 & $ 9.2915 \pm 0.0028 $ & $ 0.0347 _{ -0.0248 }^{ +0.2433 }$ &  \\
2299\,B & $ 29.2165 \pm 0.0328 $ & $ -13.038 \pm 0.042 $ & $ 197.713 \pm 0.050 $ & 0.338 & $ 12.5649 \pm 0.0029 $ & $  _{  }^{  }$ &  \\
\hline
2307\,A & $ 2.2094 \pm 0.0158 $ & $ 21.744 \pm 0.016 $ & $ -20.084 \pm 0.013 $ & 0.068 & $ 12.6917 \pm 0.0028 $ & $ 0.5075 _{ -0.0861 }^{ +0.1503 }$ &  \\
2307\,B & $ 2.0919 \pm 0.1416 $ & $ 22.161 \pm 0.179 $ & $ -19.523 \pm 0.110 $ & 0.408 & $ 17.6546 \pm 0.0054 $ & $  _{  }^{  }$ &  \\
\hline
2321\,A & $ 8.4506 \pm 0.0281 $ & $ -3.922 \pm 0.032 $ & $ -24.932 \pm 0.042 $ & 0.154 & $ 10.2853 \pm 0.0028 $ & $  _{  }^{  }$ &  \\
2321\,B & $ 8.4306 \pm 0.0421 $ & $ -3.760 \pm 0.051 $ & $ -25.597 \pm 0.063 $ & 0.143 & $ 15.4528 \pm 0.0028 $ & $ 0.0370 _{ -0.0300 }^{ +0.1211 }$ &  \\
\hline
2325\,A & $ 4.5153 \pm 0.0233 $ & $ -7.670 \pm 0.026 $ & $ -18.459 \pm 0.024 $ & 0.171 & $ 9.8309 \pm 0.0028 $ & $ 0.1113 _{ -0.1113 }^{ +0.1734 }$ & $\texttt{SHC}$ \\
2325\,B & $ 4.5677 \pm 0.0401 $ & $ -5.116 \pm 0.041 $ & $ -18.875 \pm 0.036 $ & 0.208 & $ 14.7230 \pm 0.0030 $ & $  _{  }^{  }$ &  \\
\hline
2327\,A & $ 9.9809 \pm 0.0105 $ & $ 6.518 \pm 0.013 $ & $ -46.525 \pm 0.014 $ & 0.087 & $ 9.8614 \pm 0.0028 $ & $  _{  }^{  }$ &  \\
2327\,B & $ 9.9334 \pm 0.0112 $ & $ 6.023 \pm 0.014 $ & $ -46.330 \pm 0.015 $ & 0.000 & $ 13.8791 \pm 0.0028 $ & $ 0.2367 _{ -0.0824 }^{ +0.1853 }$ &  \\
\hline
2328\,A & $ 4.3202 \pm 0.0090 $ & $ 31.870 \pm 0.011 $ & $ -1.815 \pm 0.011 $ & 0.040 & $ 12.1029 \pm 0.0028 $ & $ 0.2085 _{ -0.1730 }^{ +0.1521 }$ &  \\
2328\,B & $ 4.1584 \pm 0.2225 $ & $ 31.636 \pm 0.273 $ & $ -0.560 \pm 0.276 $ & 0.780 & $ 17.6401 \pm 0.0055 $ & $  _{  }^{  }$ &  \\
\hline
2335\,A & $ 2.1997 \pm 0.0138 $ & $ -4.027 \pm 0.012 $ & $ 9.270 \pm 0.016 $ & 0.112 & $ 12.5012 \pm 0.0028 $ & $ 0.0433 _{ -0.0368 }^{ +0.1303 }$ &  \\
2335\,B & $ 2.3676 \pm 0.0587 $ & $ -3.618 \pm 0.053 $ & $ 9.002 \pm 0.073 $ & 0.472 & $ 16.2577 \pm 0.0029 $ & $  _{  }^{  }$ &  \\
\hline
2340\,A & $ 5.6344 \pm 0.0152 $ & $ 59.660 \pm 0.017 $ & $ 84.574 \pm 0.020 $ & 0.079 & $ 12.3635 \pm 0.0028 $ & $ 0.1287 _{ -0.1127 }^{ +0.1263 }$ &  \\
2340\,B & $ 5.1835 \pm 0.0570 $ & $ 60.650 \pm 0.058 $ & $ 83.952 \pm 0.063 $ & 0.416 & $ 14.7856 \pm 0.0028 $ & $  _{  }^{  }$ &  \\
\hline
2350\,A & $ 5.3462 \pm 0.0200 $ & $ -5.301 \pm 0.013 $ & $ -8.965 \pm 0.019 $ & 0.180 & $ 11.1384 \pm 0.0028 $ & $ 0.2785 _{ -0.1049 }^{ +0.2486 }$ &  \\
2350\,B & $ 5.1138 \pm 0.1399 $ & $ -5.123 \pm 0.087 $ & $ -9.298 \pm 0.129 $ & 0.619 & $ 18.2885 \pm 0.0030 $ & $  _{  }^{  }$ &  \\
\hline
2358\,A & $ 2.6084 \pm 0.0180 $ & $ -6.499 \pm 0.019 $ & $ -14.015 \pm 0.019 $ & 0.094 & $ 11.8802 \pm 0.0028 $ & $  _{  }^{  }$ &  \\
2358\,B & $ 2.4728 \pm 0.1570 $ & $ -8.215 \pm 0.162 $ & $ -13.653 \pm 0.137 $ & 1.138 & $ 14.3423 \pm 0.0034 $ & $ 0.7293 _{ -0.2026 }^{ +0.1268 }$ &  \\
\hline
2374\,B & $ 7.3592 \pm 0.0180 $ & $ -11.470 \pm 0.018 $ & $ -31.421 \pm 0.016 $ & 0.085 & $ 11.8186 \pm 0.0028 $ & $  _{  }^{  }$ &  \\
2374\,A & $ 7.3387 \pm 0.0186 $ & $ -11.949 \pm 0.018 $ & $ -31.746 \pm 0.018 $ & 0.071 & $ 9.2231 \pm 0.0028 $ & $ 0.4100 _{ -0.2125 }^{ +0.2141 }$ &  \\
\hline
2380\,A & $ 3.7225 \pm 0.0252 $ & $ 2.432 \pm 0.023 $ & $ 9.437 \pm 0.019 $ & 0.105 & $ 9.8239 \pm 0.0028 $ & $ 0.1857 _{ -0.1703 }^{ +0.1703 }$ & $\texttt{SHC}$ \\
2380\,B & $ 3.4875 \pm 0.5810 $ & $ 1.827 \pm 0.718 $ & $ 9.278 \pm 0.437 $ & 0.000 & $ 19.5744 \pm 0.0062 $ & $  _{  }^{  }$ &  \\
\hline
2383\,B & $ 5.8490 \pm 0.0182 $ & $ -6.728 \pm 0.017 $ & $ -49.801 \pm 0.013 $ & 0.093 & $ 10.9057 \pm 0.0031 $ & $  _{  }^{  }$ &  \\
2383\,A & $ 5.8524 \pm 0.0169 $ & $ -6.653 \pm 0.016 $ & $ -49.965 \pm 0.012 $ & 0.082 & $ 10.3605 \pm 0.0028 $ & $ 0.0637 _{ -0.0510 }^{ +0.2353 }$ &  \\
\hline
2384\,A & $ 5.3219 \pm 0.0440 $ & $ 9.257 \pm 0.061 $ & $ -52.662 \pm 0.053 $ & 0.455 & $ 14.3880 \pm 0.0029 $ & $ 1.4283 _{ -0.1100 }^{ +0.1100 }$ & $\texttt{SHC}$ \\
2384\,B & $ 5.0517 \pm 0.1666 $ & $ 8.013 \pm 0.623 $ & $ -51.041 \pm 0.255 $ & 0.995 & $ 17.5517 \pm 0.0051 $ & $  _{  }^{  }$ &  \\
\hline
2409\,A & $ 5.2313 \pm 0.0283 $ & $ -19.893 \pm 0.025 $ & $ -18.523 \pm 0.033 $ & 0.299 & $ 11.8352 \pm 0.0028 $ & $  _{  }^{  }$ &  \\
2409\,B & $ 5.2737 \pm 0.0105 $ & $ -17.476 \pm 0.010 $ & $ -20.485 \pm 0.012 $ & 0.000 & $ 13.2702 \pm 0.0028 $ & $ 0.1167 _{ -0.0831 }^{ +0.1299 }$ &  \\
\hline
2417\,A & $ 3.9122 \pm 0.0119 $ & $ 33.622 \pm 0.010 $ & $ 11.780 \pm 0.013 $ & 0.094 & $ 10.7059 \pm 0.0028 $ & $ 0.9090 _{ -0.2897 }^{ +0.3721 }$ &  \\
2417\,B & $ 3.9841 \pm 0.0324 $ & $ 36.848 \pm 0.025 $ & $ 11.989 \pm 0.035 $ & 0.300 & $ 13.7951 \pm 0.0089 $ & $  _{  }^{  }$ &  \\
\hline
2419\,A & $ 4.0433 \pm 0.0097 $ & $ -12.828 \pm 0.012 $ & $ -19.732 \pm 0.011 $ & 0.078 & $ 10.8435 \pm 0.0028 $ & $ 0.3953 _{ -0.2484 }^{ +0.2457 }$ &  \\
2419\,B & $ 4.0702 \pm 0.0334 $ & $ -13.367 \pm 0.048 $ & $ -20.741 \pm 0.037 $ & 0.293 & $ 14.6100 \pm 0.0029 $ & $  _{  }^{  }$ &  \\
\hline
2422\,A & $ 4.7712 \pm 0.0251 $ & $ 21.109 \pm 0.027 $ & $ -27.759 \pm 0.022 $ & 0.148 & $ 9.9060 \pm 0.0028 $ & $ 0.1762 _{ -0.0208 }^{ +0.0208 }$ & $\texttt{SHC5}$ \\
2422\,B & $ 4.9353 \pm 0.0593 $ & $ 21.982 \pm 0.085 $ & $ -29.180 \pm 0.082 $ & 0.290 & $ 10.7782 \pm 0.0042 $ & $  _{  }^{  }$ &  \\
\hline
2425\,A & $ 6.2150 \pm 0.0192 $ & $ 4.997 \pm 0.022 $ & $ -12.557 \pm 0.020 $ & 0.094 & $ 11.7689 \pm 0.0028 $ & $ 0.0620 _{ -0.0457 }^{ +0.1270 }$ &  \\
2425\,B & $ 6.3902 \pm 0.1064 $ & $ 5.512 \pm 0.135 $ & $ -11.979 \pm 0.169 $ & 0.510 & $ 16.8583 \pm 0.0044 $ & $  _{  }^{  }$ &  \\
\hline
\end{tabular}
\end{table*}
\end{center}

\setcounter{table}{2}

\begin{center}
\begin{table*}[h]
\caption{continued}
\centering
\begin{tabular}{lccccccc}
\hline
CTOI            & $\pi$              & $\mu_{\alpha}cos(\delta)$    & $\mu_{\delta}$ & $epsi$ & $G$     & $A_{\rm G}$\\
               & [mas]              & [mas/yr]                     & [mas/yr]       & [mas]  & [mag] & [mag]\\
\hline
{\fontsize{7}{0}\selectfont 35703676\,A} & $ 2.5970 \pm 0.0923 $ & $ 2.843 \pm 0.119 $ & $ -30.120 \pm 0.074 $ & 0.727 & $ 12.7286 \pm 0.0028 $ & $ 0.0630 _{ -0.0407 }^{ +0.2071 }$ &  \\
{\fontsize{7}{0}\selectfont 35703676\,B} & $ 2.8329 \pm 0.0428 $ & $ 5.202 \pm 0.052 $ & $ -27.253 \pm 0.031 $ & 0.000 & $ 15.8373 \pm 0.0029 $ & $  _{  }^{  }$ &  \\
\hline
{\fontsize{7}{0}\selectfont 83839341\,A} & $ 2.7223 \pm 0.0181 $ & $ -33.140 \pm 0.017 $ & $ -26.355 \pm 0.017 $ & 0.068 & $ 12.2518 \pm 0.0028 $ & $ 0.6090 _{ -0.1481 }^{ +0.2281 }$ &  \\
{\fontsize{7}{0}\selectfont 83839341\,B} & $ 2.6648 \pm 0.0207 $ & $ -33.144 \pm 0.020 $ & $ -25.801 \pm 0.019 $ & 0.089 & $ 12.2623 \pm 0.0028 $ & $  _{  }^{  }$ &  \\
\hline
{\fontsize{7}{0}\selectfont 98957720\,A} & $ 4.7871 \pm 0.0316 $ & $ -42.613 \pm 0.038 $ & $ -3.408 \pm 0.019 $ & 0.247 & $ 11.1592 \pm 0.0028 $ & $  _{  }^{  }$ &  \\
{\fontsize{7}{0}\selectfont 98957720\,B} & $ 4.7377 \pm 0.0559 $ & $ -42.468 \pm 0.064 $ & $ -3.353 \pm 0.034 $ & 0.000 & $ 16.5348 \pm 0.0028 $ & $ 0.3110 _{ -0.0211 }^{ +0.1541 }$ &  \\
\hline
{\fontsize{7}{0}\selectfont 105850602\,AB} & $ 8.0429 \pm 0.2139 $ & $ -23.862 \pm 0.185 $ & $ -29.428 \pm 0.168 $ & 1.796 & $ 8.9251 \pm 0.0033 $ & $ 0.0175 _{ -0.0122 }^{ +0.0726 }$ &  \\
{\fontsize{7}{0}\selectfont 105850602\,C} & $ 6.0423 \pm 0.2164 $ & $ -23.125 \pm 0.176 $ & $ -30.583 \pm 0.171 $ & 1.733 & $ 16.7498 \pm 0.0033 $ & $  _{  }^{  }$ &  \\
{\fontsize{7}{0}\selectfont 105850602\,D} & $ 7.1762 \pm 0.1470 $ & $ -22.662 \pm 0.091 $ & $ -29.601 \pm 0.112 $ & 0.571 & $ 17.0850 \pm 0.0034 $ & $  _{  }^{  }$ &  \\
\hline
{\fontsize{7}{0}\selectfont 117644481\,A} & $ 2.9517 \pm 0.0113 $ & $ 33.782 \pm 0.009 $ & $ 43.201 \pm 0.009 $ & 0.046 & $ 12.4883 \pm 0.0028 $ & $  _{  }^{  }$ &  \\
{\fontsize{7}{0}\selectfont 117644481\,B} & $ 2.9461 \pm 0.0555 $ & $ 33.909 \pm 0.048 $ & $ 42.964 \pm 0.043 $ & 0.000 & $ 16.6315 \pm 0.0030 $ & $ 1.9848 _{ -0.4761 }^{ +0.4293 }$ &  \\
\hline
{\fontsize{7}{0}\selectfont 135145585\,A} & $ 2.5969 \pm 0.0517 $ & $ -34.591 \pm 0.042 $ & $ 2.421 \pm 0.027 $ & 0.440 & $ 11.9949 \pm 0.0028 $ & $ 0.6600 _{ -0.1301 }^{ +0.2310 }$ &  \\
{\fontsize{7}{0}\selectfont 135145585\,B} & $ 2.7543 \pm 0.0986 $ & $ -30.442 \pm 0.083 $ & $ 1.842 \pm 0.060 $ & 0.000 & $ 17.6175 \pm 0.0029 $ & $  _{  }^{  }$ &  \\
\hline
{\fontsize{7}{0}\selectfont 139444326\,A} & $ 2.0667 \pm 0.0322 $ & $ 1.731 \pm 0.022 $ & $ 4.759 \pm 0.028 $ & 0.327 & $ 12.0059 \pm 0.0028 $ & $ 1.0452 _{ -0.4019 }^{ +0.3279 }$ &  \\
{\fontsize{7}{0}\selectfont 139444326\,B} & $ 2.4099 \pm 0.2909 $ & $ 2.201 \pm 0.191 $ & $ 4.395 \pm 0.244 $ & 1.182 & $ 19.3892 \pm 0.0050 $ & $  _{  }^{  }$ &  \\
\hline
{\fontsize{7}{0}\selectfont 142443425\,A} & $ 8.1451 \pm 0.0159 $ & $ -77.368 \pm 0.013 $ & $ 34.110 \pm 0.019 $ & 0.129 & $ 11.5055 \pm 0.0028 $ & $ 0.1930 _{ -0.0941 }^{ +0.0880 }$ &  \\
{\fontsize{7}{0}\selectfont 142443425\,B} & $ 8.1393 \pm 0.0462 $ & $ -75.775 \pm 0.035 $ & $ 33.880 \pm 0.047 $ & 0.383 & $ 14.6761 \pm 0.0030 $ & $  _{  }^{  }$ &  \\
\hline
{\fontsize{7}{0}\selectfont 144164538\,A} & $ 4.7965 \pm 0.0181 $ & $ 16.116 \pm 0.019 $ & $ 33.129 \pm 0.022 $ & 0.058 & $ 14.6250 \pm 0.0028 $ & $ 1.0210 _{ -0.1641 }^{ +0.4031 }$ &  \\
{\fontsize{7}{0}\selectfont 144164538\,B} & $ 4.8422 \pm 0.2773 $ & $ 15.012 \pm 0.322 $ & $ 33.089 \pm 0.341 $ & 1.580 & $ 18.9670 \pm 0.0053 $ & $  _{  }^{  }$ &  \\
\hline
{\fontsize{7}{0}\selectfont 151628217\,A} & $ 2.0385 \pm 0.0136 $ & $ 3.714 \pm 0.016 $ & $ -5.446 \pm 0.020 $ & 0.127 & $ 11.4970 \pm 0.0028 $ & $  _{  }^{  }$ &  \\
{\fontsize{7}{0}\selectfont 151628217\,B} & $ 2.0334 \pm 0.0095 $ & $ 4.262 \pm 0.011 $ & $ -6.683 \pm 0.013 $ & 0.000 & $ 13.1539 \pm 0.0028 $ & $ 0.1683 _{ -0.0596 }^{ +0.0944 }$ &  \\
\hline
{\fontsize{7}{0}\selectfont 152226055\,A} & $ 4.2468 \pm 0.0190 $ & $ 11.080 \pm 0.016 $ & $ -12.086 \pm 0.014 $ & 0.103 & $ 11.7271 \pm 0.0028 $ & $  _{  }^{  }$ &  \\
{\fontsize{7}{0}\selectfont 152226055\,B} & $ 4.2203 \pm 0.0189 $ & $ 10.447 \pm 0.016 $ & $ -12.161 \pm 0.014 $ & 0.000 & $ 13.8270 \pm 0.0028 $ & $ 0.5720 _{ -0.3110 }^{ +0.2630 }$ &  \\
\hline
{\fontsize{7}{0}\selectfont 164781040\,A} & $ 2.6525 \pm 0.4136 $ & $ 10.367 \pm 0.457 $ & $ 0.886 \pm 0.499 $ & 4.212 & $ 14.5887 \pm 0.0030 $ & $  _{  }^{  }$ &  \\
{\fontsize{7}{0}\selectfont 164781040\,B} & $ 1.0362 \pm 0.0417 $ & $ 8.515 \pm 0.062 $ & $ 3.140 \pm 0.083 $ & 0.155 & $ 15.1346 \pm 0.0029 $ & $ 0.0000 _{ -0.0000 }^{ +0.4742 }$ & $\texttt{SHC}$ \\
\hline
{\fontsize{7}{0}\selectfont 178367145\,B} & $ 3.8421 \pm 0.0302 $ & $ -13.234 \pm 0.030 $ & $ -2.789 \pm 0.022 $ & 0.203 & $ 11.7726 \pm 0.0029 $ & $  _{  }^{  }$ &  \\
{\fontsize{7}{0}\selectfont 178367145\,A} & $ 3.9752 \pm 0.0190 $ & $ -13.887 \pm 0.020 $ & $ -2.818 \pm 0.015 $ & 0.117 & $ 10.9126 \pm 0.0028 $ & $ 0.1957 _{ -0.1437 }^{ +0.1133 }$ &  \\
\hline
{\fontsize{7}{0}\selectfont 197760286\,A} & $ 5.4658 \pm 0.1691 $ & $ 34.814 \pm 0.178 $ & $ -18.655 \pm 0.131 $ & 1.392 & $ 9.9572 \pm 0.0028 $ & $  _{  }^{  }$ &  \\
{\fontsize{7}{0}\selectfont 197760286\,B} & $ 5.1856 \pm 0.0372 $ & $ 36.545 \pm 0.043 $ & $ -13.420 \pm 0.029 $ & 0.081 & $ 15.5006 \pm 0.0028 $ & $ 0.0280 _{ -0.0187 }^{ +0.4341 }$ &  \\
\hline
{\fontsize{7}{0}\selectfont 202712304\,A} & $ 15.3450 \pm 0.0168 $ & $ 107.379 \pm 0.016 $ & $ -84.082 \pm 0.013 $ & 0.071 & $ 13.0651 \pm 0.0028 $ & $ 0.4505 _{ -0.3870 }^{ +0.4125 }$ &  \\
{\fontsize{7}{0}\selectfont 202712304\,B} & $ 14.9765 \pm 0.0652 $ & $ 107.234 \pm 0.063 $ & $ -86.297 \pm 0.049 $ & 0.446 & $ 15.1065 \pm 0.0028 $ & $  _{  }^{  }$ &  \\
\hline
{\fontsize{7}{0}\selectfont 224327878\,A} & $ 8.4853 \pm 0.0145 $ & $ -4.061 \pm 0.014 $ & $ 35.502 \pm 0.014 $ & 0.107 & $ 11.0964 \pm 0.0028 $ & $ 0.0620 _{ -0.0540 }^{ +0.2797 }$ &  \\
{\fontsize{7}{0}\selectfont 224327878\,B} & $ 8.5937 \pm 0.0854 $ & $ -4.003 \pm 0.077 $ & $ 37.640 \pm 0.104 $ & 0.556 & $ 16.4315 \pm 0.0056 $ & $  _{  }^{  }$ &  \\
{\fontsize{7}{0}\selectfont 224327878\,C} & $ 8.4661 \pm 0.0467 $ & $ -3.193 \pm 0.043 $ & $ 35.728 \pm 0.045 $ & 0.000 & $ 16.5314 \pm 0.0029 $ & $  _{  }^{  }$ &  \\
\hline
{\fontsize{7}{0}\selectfont 230236827\,A} & $ 11.7407 \pm 0.0122 $ & $ 5.556 \pm 0.015 $ & $ 37.566 \pm 0.017 $ & 0.119 & $ 11.6854 \pm 0.0028 $ & $  _{  }^{  }$ &  \\
{\fontsize{7}{0}\selectfont 230236827\,B} & $ 11.7974 \pm 0.0336 $ & $ 0.372 \pm 0.045 $ & $ 33.544 \pm 0.040 $ & 0.289 & $ 15.3282 \pm 0.0028 $ & $ 0.0900 _{ -0.0630 }^{ +0.2905 }$ &  \\
\hline
{\fontsize{7}{0}\selectfont 238235254\,B} & $ 2.4073 \pm 0.0403 $ & $ -5.316 \pm 0.049 $ & $ 8.021 \pm 0.046 $ & 0.360 & $ 6.3244 \pm 0.0028 $ & $  _{  }^{  }$ &  \\
{\fontsize{7}{0}\selectfont 238235254\,A} & $ 2.5463 \pm 0.0433 $ & $ -5.435 \pm 0.055 $ & $ 8.186 \pm 0.059 $ & 0.429 & $ 6.3063 \pm 0.0028 $ & $ 0.0588 _{ -0.0393 }^{ +0.1099 }$ &  \\
\hline
{\fontsize{7}{0}\selectfont 238920872\,A} & $ 9.4737 \pm 0.0106 $ & $ 10.410 \pm 0.013 $ & $ 19.547 \pm 0.013 $ & 0.053 & $ 13.1190 \pm 0.0032 $ & $ 0.2471 _{ -0.1271 }^{ +0.1271 }$ & $\texttt{SHC}$ \\
{\fontsize{7}{0}\selectfont 238920872\,B} & $ 9.4336 \pm 0.0559 $ & $ 10.074 \pm 0.073 $ & $ 19.829 \pm 0.070 $ & 0.117 & $ 17.1604 \pm 0.0031 $ & $  _{  }^{  }$ &  \\
\hline
{\fontsize{7}{0}\selectfont 253040591\,A} & $ 13.8011 \pm 0.3445 $ & $ 25.446 \pm 0.374 $ & $ -35.205 \pm 0.282 $ & 2.826 & $ 9.1333 \pm 0.0028 $ & $ 0.4387 _{ -0.1509 }^{ +0.1509 }$ & $\texttt{SHC}$ \\
{\fontsize{7}{0}\selectfont 253040591\,B} & $ 14.5199 \pm 0.1306 $ & $ 25.997 \pm 0.120 $ & $ -29.287 \pm 0.124 $ & 0.103 & $ 17.7513 \pm 0.0031 $ & $  _{  }^{  }$ &  \\
\hline
{\fontsize{7}{0}\selectfont 257605131\,A} & $ 8.0993 \pm 0.0108 $ & $ -11.061 \pm 0.010 $ & $ 12.347 \pm 0.014 $ & 0.065 & $ 10.7561 \pm 0.0028 $ & $ 0.0732 _{ -0.0607 }^{ +0.1586 }$ &  \\
{\fontsize{7}{0}\selectfont 257605131\,B} & $ 8.1199 \pm 0.0172 $ & $ -11.481 \pm 0.016 $ & $ 12.229 \pm 0.023 $ & 0.095 & $ 14.3404 \pm 0.0028 $ & $  _{  }^{  }$ &  \\
\hline
{\fontsize{7}{0}\selectfont 259376845\,A} & $ 3.9444 \pm 0.0112 $ & $ 11.525 \pm 0.015 $ & $ -23.759 \pm 0.016 $ & 0.072 & $ 12.8546 \pm 0.0028 $ & $  _{  }^{  }$ &  \\
{\fontsize{7}{0}\selectfont 259376845\,B} & $ 3.9732 \pm 0.0133 $ & $ 10.182 \pm 0.017 $ & $ -23.315 \pm 0.019 $ & 0.110 & $ 12.9574 \pm 0.0031 $ & $ 0.5850 _{ -0.1631 }^{ +0.1238 }$ &  \\
\hline
{\fontsize{7}{0}\selectfont 282502866\,A} & $ 13.3826 \pm 0.0258 $ & $ -25.115 \pm 0.035 $ & $ -26.723 \pm 0.030 $ & 0.145 & $ 7.8098 \pm 0.0028 $ & $  _{  }^{  }$ &  \\
{\fontsize{7}{0}\selectfont 282502866\,B} & $ 13.1925 \pm 0.0320 $ & $ -24.914 \pm 0.035 $ & $ -26.448 \pm 0.031 $ & 0.186 & $ 14.8252 \pm 0.0028 $ & $ 0.8100 _{ -0.1288 }^{ +0.0760 }$ &  \\
\hline
\end{tabular}
\end{table*}
\end{center}

\setcounter{table}{2}

\begin{center}
\begin{table*}[h]
\caption{continued}
\centering
\begin{tabular}{lccccccc}
\hline
CTOI            & $\pi$              & $\mu_{\alpha}cos(\delta)$    & $\mu_{\delta}$ & $epsi$ & $G$     & $A_{\rm G}$\\
               & [mas]              & [mas/yr]                     & [mas/yr]       & [mas]  & [mag] & [mag]\\
\hline
{\fontsize{7}{0}\selectfont 288240183\,A} & $ 6.6408 \pm 0.3602 $ & $ -51.336 \pm 0.520 $ & $ 24.806 \pm 0.444 $ & 3.859 & $ 9.4662 \pm 0.0041 $ & $ 0.3020 _{ -0.2134 }^{ +0.2134 }$ & $\texttt{SHC}$ \\
{\fontsize{7}{0}\selectfont 288240183\,B} & $ 7.0498 \pm 0.0153 $ & $ -46.377 \pm 0.020 $ & $ 22.520 \pm 0.020 $ & 0.123 & $ 10.6051 \pm 0.0028 $ & $ 0.1530 _{ -0.0978 }^{ +0.1120 }$ &  \\
{\fontsize{7}{0}\selectfont 288240183\,C} & $ 7.0320 \pm 0.0574 $ & $ -47.646 \pm 0.108 $ & $ 24.037 \pm 0.074 $ & 0.477 & $ 14.7007 \pm 0.0032 $ & $  _{  }^{  }$ &  \\
\hline
{\fontsize{7}{0}\selectfont 290596728\,A} & $ 6.6127 \pm 0.0186 $ & $ -8.669 \pm 0.015 $ & $ -3.979 \pm 0.013 $ & 0.087 & $ 10.7333 \pm 0.0028 $ & $  _{  }^{  }$ &  \\
{\fontsize{7}{0}\selectfont 290596728\,B} & $ 6.6471 \pm 0.0195 $ & $ -8.455 \pm 0.016 $ & $ -3.969 \pm 0.014 $ & 0.006 & $ 13.7674 \pm 0.0028 $ & $ 0.6563 _{ -0.2434 }^{ +0.1948 }$ &  \\
\hline
{\fontsize{7}{0}\selectfont 300116105\,A} & $ 3.2646 \pm 0.0113 $ & $ -1.698 \pm 0.013 $ & $ 16.463 \pm 0.013 $ & 0.079 & $ 12.1874 \pm 0.0028 $ & $ 0.5350 _{ -0.1068 }^{ +0.0971 }$ &  \\
{\fontsize{7}{0}\selectfont 300116105\,B} & $ 3.1741 \pm 0.0817 $ & $ 0.148 \pm 0.101 $ & $ 17.099 \pm 0.085 $ & 0.449 & $ 17.1562 \pm 0.0034 $ & $  _{  }^{  }$ &  \\
\hline
{\fontsize{7}{0}\selectfont 308301091\,A} & $ 4.4783 \pm 0.0138 $ & $ -7.295 \pm 0.011 $ & $ -6.361 \pm 0.015 $ & 0.109 & $ 10.1343 \pm 0.0028 $ & $ 0.1750 _{ -0.1710 }^{ +0.1561 }$ &  \\
{\fontsize{7}{0}\selectfont 308301091\,B} & $ 4.7173 \pm 0.1050 $ & $ -6.105 \pm 0.110 $ & $ -6.001 \pm 0.161 $ & 0.612 & $ 16.4662 \pm 0.0079 $ & $  _{  }^{  }$ &  \\
\hline
{\fontsize{7}{0}\selectfont 312091232\,A} & $ 3.4329 \pm 0.0169 $ & $ -30.516 \pm 0.021 $ & $ -24.666 \pm 0.015 $ & 0.084 & $ 11.7888 \pm 0.0028 $ & $ 0.1763 _{ -0.1268 }^{ +0.1394 }$ &  \\
{\fontsize{7}{0}\selectfont 312091232\,B} & $ 3.2928 \pm 0.1112 $ & $ -30.899 \pm 0.130 $ & $ -24.713 \pm 0.098 $ & 0.129 & $ 17.4970 \pm 0.0030 $ & $  _{  }^{  }$ &  \\
\hline
{\fontsize{7}{0}\selectfont 326092637\,A} & $ 2.5628 \pm 0.0201 $ & $ -9.956 \pm 0.025 $ & $ -16.293 \pm 0.026 $ & 0.093 & $ 12.3868 \pm 0.0028 $ & $ 0.1720 _{ -0.0951 }^{ +0.2470 }$ &  \\
{\fontsize{7}{0}\selectfont 326092637\,B} & $ 2.5714 \pm 0.4362 $ & $ -9.452 \pm 0.597 $ & $ -15.607 \pm 0.632 $ & 0.521 & $ 19.7135 \pm 0.0041 $ & $  _{  }^{  }$ &  \\
\hline
{\fontsize{7}{0}\selectfont 341411516\,A} & $ 5.7036 \pm 0.0219 $ & $ 30.074 \pm 0.031 $ & $ 7.168 \pm 0.029 $ & 0.260 & $ 11.1204 \pm 0.0028 $ & $ 0.0170 _{ -0.0117 }^{ +0.1161 }$ &  \\
{\fontsize{7}{0}\selectfont 341411516\,B} & $ 5.8132 \pm 0.2458 $ & $ 31.061 \pm 0.335 $ & $ 7.253 \pm 0.294 $ & 0.000 & $ 19.7210 \pm 0.0045 $ & $  _{  }^{  }$ &  \\
\hline
{\fontsize{7}{0}\selectfont 345324572\,B} & $ 19.8554 \pm 0.0166 $ & $ -40.646 \pm 0.016 $ & $ 23.672 \pm 0.016 $ & 0.109 & $ 13.6824 \pm 0.0028 $ & $  _{  }^{  }$ &  \\
{\fontsize{7}{0}\selectfont 345324572\,A} & $ 19.8231 \pm 0.0108 $ & $ -41.421 \pm 0.010 $ & $ 22.536 \pm 0.011 $ & 0.052 & $ 9.0158 \pm 0.0028 $ & $ 0.2843 _{ -0.1894 }^{ +0.1402 }$ &  \\
\hline
{\fontsize{7}{0}\selectfont 349793830\,A} & $ 12.2347 \pm 0.0211 $ & $ -166.404 \pm 0.022 $ & $ -46.521 \pm 0.014 $ & 0.109 & $ 9.5710 \pm 0.0028 $ & $ 0.0680 _{ -0.0261 }^{ +0.0783 }$ &  \\
{\fontsize{7}{0}\selectfont 349793830\,B} & $ 12.2912 \pm 0.1075 $ & $ -167.068 \pm 0.121 $ & $ -44.129 \pm 0.087 $ & 0.000 & $ 17.4495 \pm 0.0029 $ & $  _{  }^{  }$ &  \\
\hline
{\fontsize{7}{0}\selectfont 352915304\,A} & $ 8.7944 \pm 0.0102 $ & $ 64.929 \pm 0.016 $ & $ 151.830 \pm 0.014 $ & 0.072 & $ 11.5368 \pm 0.0028 $ & $ 0.0980 _{ -0.0183 }^{ +0.1220 }$ &  \\
{\fontsize{7}{0}\selectfont 352915304\,B} & $ 8.7595 \pm 0.0839 $ & $ 65.107 \pm 0.138 $ & $ 152.160 \pm 0.134 $ & 0.409 & $ 17.2980 \pm 0.0041 $ & $  _{  }^{  }$ &  \\
\hline
{\fontsize{7}{0}\selectfont 369376388\,A} & $ 10.3996 \pm 0.0104 $ & $ 153.458 \pm 0.009 $ & $ 24.005 \pm 0.012 $ & 0.074 & $ 11.9448 \pm 0.0028 $ & $ 0.6005 _{ -0.4346 }^{ +0.3379 }$ &  \\
{\fontsize{7}{0}\selectfont 369376388\,B} & $ 10.6997 \pm 0.1053 $ & $ 150.662 \pm 0.098 $ & $ 21.621 \pm 0.117 $ & 0.766 & $ 16.8507 \pm 0.0037 $ & $  _{  }^{  }$ &  \\
{\fontsize{7}{0}\selectfont 369376388\,C} & $ 10.3986 \pm 0.0553 $ & $ 153.085 \pm 0.053 $ & $ 22.296 \pm 0.063 $ & 0.129 & $ 17.1337 \pm 0.0029 $ & $  _{  }^{  }$ &  \\
\hline
{\fontsize{7}{0}\selectfont 372913337\,A} & $ 2.4642 \pm 0.0158 $ & $ -4.695 \pm 0.020 $ & $ 11.038 \pm 0.019 $ & 0.137 & $ 8.9919 \pm 0.0028 $ & $  _{  }^{  }$ &  \\
{\fontsize{7}{0}\selectfont 372913337\,B} & $ 2.4294 \pm 0.0114 $ & $ -5.120 \pm 0.014 $ & $ 10.868 \pm 0.014 $ & 0.072 & $ 12.5578 \pm 0.0028 $ & $ 0.2790 _{ -0.1070 }^{ +0.2363 }$ &  \\
\hline
{\fontsize{7}{0}\selectfont 374352402\,A} & $ 2.0320 \pm 0.0129 $ & $ 12.702 \pm 0.009 $ & $ -4.057 \pm 0.013 $ & 0.068 & $ 12.3260 \pm 0.0028 $ & $ 0.3583 _{ -0.2818 }^{ +0.3858 }$ &  \\
{\fontsize{7}{0}\selectfont 374352402\,B} & $ 1.8700 \pm 0.0916 $ & $ 12.937 \pm 0.068 $ & $ -4.224 \pm 0.095 $ & 0.401 & $ 17.3357 \pm 0.0030 $ & $  _{  }^{  }$ &  \\
\hline
{\fontsize{7}{0}\selectfont 374732772\,A} & $ 6.5180 \pm 0.0305 $ & $ -9.891 \pm 0.038 $ & $ -25.283 \pm 0.027 $ & 0.213 & $ 10.6796 \pm 0.0030 $ & $ 1.2515 _{ -0.4849 }^{ +0.4256 }$ &  \\
{\fontsize{7}{0}\selectfont 374732772\,B} & $ 6.3821 \pm 0.0307 $ & $ -9.741 \pm 0.037 $ & $ -23.746 \pm 0.023 $ & 0.148 & $ 14.0143 \pm 0.0028 $ & $  _{  }^{  }$ &  \\
\hline
{\fontsize{7}{0}\selectfont 394721720\,A} & $ 2.8051 \pm 0.0114 $ & $ -15.344 \pm 0.013 $ & $ -26.763 \pm 0.013 $ & 0.098 & $ 12.2568 \pm 0.0028 $ & $ 0.4284 _{ -0.1410 }^{ +0.1410 }$ & $\texttt{SHC}$ \\
{\fontsize{7}{0}\selectfont 394721720\,B} & $ 2.4257 \pm 0.1402 $ & $ -15.359 \pm 0.168 $ & $ -26.140 \pm 0.165 $ & 0.366 & $ 18.7052 \pm 0.0034 $ & $  _{  }^{  }$ &  \\
\hline
{\fontsize{7}{0}\selectfont 399913539\,A} & $ 3.2337 \pm 0.0099 $ & $ 1.850 \pm 0.013 $ & $ 28.329 \pm 0.013 $ & 0.000 & $ 13.2471 \pm 0.0031 $ & $ 0.2455 _{ -0.2291 }^{ +0.3096 }$ &  \\
{\fontsize{7}{0}\selectfont 399913539\,B} & $ 3.3647 \pm 0.0954 $ & $ 1.427 \pm 0.141 $ & $ 28.147 \pm 0.132 $ & 0.329 & $ 18.0096 \pm 0.0031 $ & $  _{  }^{  }$ &  \\
\hline
{\fontsize{7}{0}\selectfont 453455638\,A} & $ 5.4637 \pm 0.0172 $ & $ -113.184 \pm 0.023 $ & $ 101.703 \pm 0.020 $ & 0.000 & $ 14.8091 \pm 0.0028 $ & $  _{  }^{  }$ &  \\
{\fontsize{7}{0}\selectfont 453455638\,B} & $ 6.1329 \pm 0.1091 $ & $ -114.144 \pm 0.149 $ & $ 102.925 \pm 0.131 $ & 1.237 & $ 16.0432 \pm 0.0037 $ & $ 0.5827 _{ -0.2471 }^{ +0.1686 }$ &  \\
\hline
{\fontsize{7}{0}\selectfont 460950389\,A} & $ 6.6459 \pm 0.0093 $ & $ -17.586 \pm 0.010 $ & $ 10.490 \pm 0.010 $ & 0.024 & $ 12.3172 \pm 0.0029 $ & $ 0.5819 _{ -0.1431 }^{ +0.1431 }$ & $\texttt{SHC}$ \\
{\fontsize{7}{0}\selectfont 460950389\,B} & $ 5.3938 \pm 1.3984 $ & $ -17.024 \pm 1.756 $ & $ 7.556 \pm 1.400 $ & 9.219 & $ 20.7782 \pm 0.0205 $ & $  _{  }^{  }$ &  \\
\hline
{\fontsize{7}{0}\selectfont 467785319\,A} & $ 6.4966 \pm 0.0268 $ & $ -1.972 \pm 0.030 $ & $ -30.034 \pm 0.021 $ & 0.208 & $ 11.4273 \pm 0.0028 $ & $ 0.1590 _{ -0.1340 }^{ +0.1350 }$ &  \\
{\fontsize{7}{0}\selectfont 467785319\,B} & $ 6.2976 \pm 0.1264 $ & $ -0.804 \pm 0.161 $ & $ -28.542 \pm 0.144 $ & 0.349 & $ 16.8640 \pm 0.0059 $ & $  _{  }^{  }$ &  \\
\hline
{\fontsize{7}{0}\selectfont 738065944\,A} & $ 2.3423 \pm 0.0097 $ & $ 35.434 \pm 0.011 $ & $ 26.888 \pm 0.011 $ & 0.053 & $ 12.7926 \pm 0.0028 $ & $ 0.3543 _{ -0.3159 }^{ +0.2020 }$ &  \\
{\fontsize{7}{0}\selectfont 738065944\,B} & $ 2.2919 \pm 0.0120 $ & $ 35.904 \pm 0.013 $ & $ 27.026 \pm 0.017 $ & 0.041 & $ 13.6147 \pm 0.0028 $ & $  _{  }^{  }$ &  \\
\hline
{\fontsize{7}{0}\selectfont 901674675\,A} & $ 2.0794 \pm 0.0133 $ & $ -0.017 \pm 0.012 $ & $ -27.116 \pm 0.012 $ & 0.053 & $ 12.5042 \pm 0.0028 $ & $  _{  }^{  }$ &  \\
{\fontsize{7}{0}\selectfont 901674675\,B} & $ 2.0663 \pm 0.0162 $ & $ -0.002 \pm 0.014 $ & $ -27.103 \pm 0.014 $ & 0.088 & $ 12.7075 \pm 0.0028 $ & $ 0.4010 _{ -0.2281 }^{+ 0.0981 }$ &  \\
\hline
\end{tabular}
\end{table*}
\end{center}

\begin{table*} \caption{This table lists for each detected companion (sorted by its identifier) the angular separation $\rho$ and position angle $PA$ to the associated (C)TOI, the difference between its parallax and that of the (C)TOI $\Delta\pi$ with its significance (in brackets calculated by taking into account also the Gaia astrometric excess noise), the differential proper motion of the companion relative to the (C)TOI $\mu_{\rm rel}$ with its significance, as well as its $cpm\text{-}index$. The last column indicates ($\bigstar$) if the detected companion is not listed in the WDS as companion(-candidate) of the (C)TOI.}\label{TAB_COMP_RELASTRO}
\begin{tabular}{lcccccccc}
\hline
TOI   & $\rho$    & $PA$        & $\Delta\pi$ & $sig$-      & $\mu_{\rm rel}$ & $sig$-             & $cpm$-   & not in\\
             & [arcsec]  & [$^\circ$]  & [mas]       & $\Delta\pi$ & [mas/yr]    & $\mu_{\rm rel}$ & $index$  & WDS\\
\hline
1937\,B& $2.48355\pm0.00008$ & $355.88654\pm0.00228$ & $0.06\pm0.09$ & 0.7 (0.2)& $0.24\pm0.10$ & 2.3&104&$\bigstar$\\
1940\,B& $2.77116\pm0.00012$ & $242.00318\pm0.00294$ & $0.75\pm0.12$ & 6.1 (1.0)& $1.78\pm0.16$ & 11.1&5.4\\
1943\,B& $3.62638\pm0.00003$ & $114.43687\pm0.00060$ & $0.10\pm0.05$ & 2.2 (0.3)& $1.04\pm0.05$ & 21.2&183&$\bigstar$\\
1946\,B& $2.82716\pm0.00014$ & $343.39102\pm0.00338$ & $0.60\pm0.20$ & 3.0 (0.4)& $5.65\pm0.22$ & 26.1&13\\
1953\,B& $35.95409\pm0.00009$ & $200.19546\pm0.00017$ & $0.04\pm0.11$ & 0.3 (0.1)& $0.75\pm0.12$ & 6.3&21&$\bigstar$\\
1964\,B& $5.83383\pm0.00006$ & $53.32376\pm0.00053$ & $0.02\pm0.07$ & 0.3 (0.1)& $0.19\pm0.07$ & 2.8&205&$\bigstar$\\
1966\,B& $33.48745\pm0.00004$ & $359.63816\pm0.00007$ & $0.42\pm0.04$ & 9.9 (1.5)& $1.06\pm0.04$ & 25.2&73&$\bigstar$\\
1970\,B& $12.18765\pm0.00005$ & $321.04307\pm0.00025$ & $0.07\pm0.09$ & 0.7 (0.2)& $0.21\pm0.07$ & 2.9&184&$\bigstar$\\
1972\,B& $26.20433\pm0.00001$ & $305.73275\pm0.00003$ & $0.01\pm0.02$ & 0.7 (0.2)& $0.43\pm0.02$ & 22.9&163\\
1984\,B& $3.05695\pm0.00014$ & $259.78789\pm0.00245$ & $0.03\pm0.23$ & 0.1 (0.0)& $0.68\pm0.17$ & 3.9&81&$\bigstar$\\
1992\,B& $3.15370\pm0.00002$ & $202.28009\pm0.00033$ & $0.01\pm0.03$ & 0.3 (0.0)& $0.41\pm0.02$ & 17.2&38\\
2001\,B& $0.96202\pm0.00007$ & $280.20210\pm0.00766$ & $0.22\pm0.09$ & 2.3 (0.3)& $1.50\pm0.19$ & 7.9&6.0\\
2006\,B& $4.63500\pm0.00002$ & $246.83942\pm0.00026$ & $0.05\pm0.02$ & 2.5 (0.5)& $0.68\pm0.02$ & 29.7&70&$\bigstar$\\
2009\,B& $9.41072\pm0.00005$ & $73.21041\pm0.00023$ & $0.02\pm0.06$ & 0.4 (0.1)& $8.38\pm0.05$ & 153.5&120\\
2033\,B& $10.65305\pm0.00001$ & $21.98648\pm0.00006$ & $0.03\pm0.02$ & 1.7 (0.3)& $0.15\pm0.02$ & 9.7&343\\
2036\,B& $4.55841\pm0.00005$ & $196.21520\pm0.00055$ & $0.11\pm0.07$ & 1.6 (0.3)& $0.76\pm0.06$ & 12.1&60&$\bigstar$\\
2050\,B& $12.16939\pm0.00004$ & $108.63467\pm0.00016$ & $0.13\pm0.04$ & 3.5 (0.3)& $4.93\pm0.04$ & 117.5&9.2&$\bigstar$\\
2056\,B& $4.28319\pm0.00002$ & $92.44090\pm0.00027$ & $0.01\pm0.03$ & 0.4 (0.1)& $2.12\pm0.02$ & 89.0&93\\
2068\,B& $19.98712\pm0.00002$ & $65.48869\pm0.00004$ & $0.01\pm0.02$ & 0.7 (0.1)& $2.79\pm0.02$ & 140.3&143\\
2072\,B& $2.98886\pm0.00002$ & $130.50157\pm0.00043$ & $0.07\pm0.03$ & 2.7 (0.3)& $4.80\pm0.03$ & 147.7&86\\
2084\,B& $12.24533\pm0.00057$ & $191.19264\pm0.00314$ & $0.04\pm0.75$ & 0.1 (0.0)& $2.54\pm0.93$ & 2.7&48&$\bigstar$\\
2092\,B& $2.70349\pm0.00062$ & $272.26872\pm0.01306$ & $1.46\pm0.71$ & 2.1 (1.0)& $1.60\pm0.77$ & 2.1&126&$\bigstar$\\
2094\,B& $10.71371\pm0.00011$ & $215.71577\pm0.00062$ & $0.15\pm0.12$ & 1.3 (0.2)& $1.67\pm0.15$ & 11.0&66&$\bigstar$\\
2106\,B& $2.87903\pm0.00035$ & $1.74172\pm0.00667$ & $0.17\pm0.46$ & 0.4 (0.0)& $4.15\pm0.40$ & 10.5&33&$\bigstar$\\
2108\,B& $1.19997\pm0.00003$ & $185.80397\pm0.00356$ & $0.00\pm0.06$ & 0.1 (0.0)& $0.82\pm0.06$ & 12.8&19&$\bigstar$\\
2113\,B& $1.41338\pm0.00003$ & $274.78287\pm0.00130$ & $0.02\pm0.03$ & 0.8 (0.1)& $1.46\pm0.04$ & 33.4&86\\
2115\,B& $0.95546\pm0.00007$ & $88.20173\pm0.00946$ & $0.46\pm0.11$ & 4.1 (0.6)& $1.43\pm0.10$ & 14.6&15\\
2127\,B& $2.66269\pm0.00028$ & $185.95004\pm0.00475$ & $0.24\pm0.30$ & 0.8 (0.2)& $1.13\pm0.31$ & 3.6&69\\
2128\,B& $6.49976\pm0.00004$ & $196.73203\pm0.00030$ & $0.02\pm0.04$ & 0.5 (0.0)& $4.42\pm0.05$ & 92.6&77\\
2144\,B& $14.70306\pm0.00002$ & $227.52790\pm0.00008$ & $0.00\pm0.02$ & 0.0 (0.0)& $1.69\pm0.03$ & 60.0&127\\
2149\,B& $1.61679\pm0.00003$ & $110.20239\pm0.00108$ & $0.05\pm0.03$ & 1.6 (0.2)& $3.08\pm0.04$ & 83.0&13\\
2152\,B& $20.59743\pm0.00025$ & $302.41777\pm0.00070$ & $0.11\pm0.32$ & 0.3 (0.1)& $0.57\pm0.30$ & 1.9&105&$\bigstar$\\
2169\,B& $6.45111\pm0.00001$ & $233.68918\pm0.00015$ & $0.02\pm0.02$ & 1.0 (0.1)& $0.39\pm0.02$ & 19.1&161&$\bigstar$\\
2183\,B& $1.58405\pm0.00002$ & $164.18066\pm0.00086$ & $0.07\pm0.03$ & 2.6 (0.4)& $6.82\pm0.03$ & 210.3&13\\
2193\,B& $1.86909\pm0.00005$ & $124.24580\pm0.00160$ & $0.03\pm0.06$ & 0.4 (0.1)& $0.10\pm0.06$ & 1.7&53&$\bigstar$\\
2195\,B& $3.34320\pm0.00004$ & $210.34890\pm0.00064$ & $0.03\pm0.04$ & 0.8 (0.1)& $1.03\pm0.05$ & 22.1&112&$\bigstar$\\
2205\,B& $21.83293\pm0.00014$ & $255.40300\pm0.00039$ & $0.02\pm0.14$ & 0.2 (0.0)& $0.26\pm0.19$ & 1.4&209&$\bigstar$\\
2205\,C& $2.14975\pm0.00018$ & $83.39429\pm0.00570$ & $0.09\pm0.20$ & 0.5 (0.5)& $0.81\pm0.28$ & 2.9&69&$\bigstar$\\
2215\,B& $62.61249\pm0.00002$ & $223.79323\pm0.00002$ & $0.03\pm0.02$ & 1.1 (0.3)& $0.14\pm0.02$ & 6.6&145&$\bigstar$\\
2218\,A& $6.82084\pm0.00004$ & $269.12176\pm0.00030$ & $0.31\pm0.04$ & 8.4 (0.8)& $2.42\pm0.05$ & 50.3&31\\
2233\,B& $4.47377\pm0.00003$ & $188.08418\pm0.00054$ & $0.09\pm0.05$ & 1.9 (0.4)& $0.68\pm0.04$ & 17.7&75&$\bigstar$\\
2239\,B& $18.77303\pm0.00002$ & $233.49052\pm0.00005$ & $0.16\pm0.02$ & 10.1 (1.7)& $0.58\pm0.02$ & 32.5&18&$\bigstar$\\
2244\,B& $1.50266\pm0.00002$ & $49.25479\pm0.00077$ & $0.02\pm0.03$ & 0.7 (0.1)& $1.13\pm0.03$ & 44.3&34\\
2244\,C& $20.32665\pm0.00012$ & $307.25633\pm0.00033$ & $0.05\pm0.16$ & 0.3 (0.1)& $0.65\pm0.15$ & 4.4&58&$\bigstar$\\
2246\,B& $5.84629\pm0.00002$ & $339.01233\pm0.00015$ & $0.02\pm0.02$ & 1.1 (0.1)& $0.94\pm0.02$ & 47.1&114\\
\hline
\end{tabular}
\end{table*}

\setcounter{table}{3}

\begin{table*} \caption{continued}
\begin{tabular}{lccccccccc}
\hline
TOI   & $\rho$    & $PA$        & $\Delta\pi$ & $sig$-      & $\mu_{\rm rel}$ & $sig$-             & $cpm$-   & not in\\
             & [arcsec]  & [$^\circ$]  & [mas]       & $\Delta\pi$ & [mas/yr]    & $\mu_{\rm rel}$ & $index$  & WDS\\
\hline
2248\,B& $3.29667\pm0.00021$ & $49.71924\pm0.00369$ & $0.17\pm0.22$ & 0.7 (0.2)& $0.61\pm0.28$ & 2.2&52&$\bigstar$\\
2253\,B& $26.26311\pm0.00006$ & $291.05107\pm0.00015$ & $0.75\pm0.07$ & 11.0 (1.2)& $0.57\pm0.07$ & 7.9&70&$\bigstar$\\
2279\,B& $4.57139\pm0.00003$ & $319.06660\pm0.00037$ & $0.02\pm0.03$ & 0.6 (0.1)& $1.14\pm0.04$ & 26.1&49&$\bigstar$\\
2281\,B& $33.97687\pm0.00002$ & $206.74198\pm0.00002$ & $0.02\pm0.02$ & 1.3 (0.2)& $1.26\pm0.02$ & 66.1&51\\
2283\,B& $26.52255\pm0.00003$ & $336.13432\pm0.00007$ & $0.02\pm0.03$ & 0.7 (0.2)& $1.97\pm0.04$ & 50.6&124\\
2289\,B& $12.66428\pm0.00004$ & $131.12439\pm0.00018$ & $0.08\pm0.04$ & 1.9 (0.4)& $0.27\pm0.05$ & 5.4&555&$\bigstar$\\
2293\,B& $4.66336\pm0.00005$ & $1.24380\pm0.00045$ & $0.14\pm0.06$ & 2.2 (0.6)& $3.04\pm0.06$ & 49.9&86&$\bigstar$\\
2299\,B& $3.60503\pm0.00003$ & $280.93394\pm0.00056$ & $0.04\pm0.03$ & 1.0 (0.1)& $2.99\pm0.05$ & 59.1&132\\
2307\,B& $2.91691\pm0.00009$ & $205.82182\pm0.00202$ & $0.12\pm0.14$ & 0.8 (0.3)& $0.70\pm0.14$ & 5.0&85&$\bigstar$\\
2321\,B& $21.37467\pm0.00003$ & $150.86054\pm0.00009$ & $0.02\pm0.05$ & 0.4 (0.1)& $0.68\pm0.07$ & 9.1&75&$\bigstar$\\
2325\,B& $4.52007\pm0.00003$ & $300.86111\pm0.00045$ & $0.05\pm0.05$ & 1.1 (0.2)& $2.59\pm0.05$ & 53.4&15&$\bigstar$\\
2327\,B& $38.79962\pm0.00002$ & $156.28735\pm0.00002$ & $0.05\pm0.02$ & 3.1 (0.5)& $0.53\pm0.02$ & 27.6&176\\
2328\,B& $1.63846\pm0.00023$ & $258.49775\pm0.00812$ & $0.16\pm0.22$ & 0.7 (0.2)& $1.28\pm0.28$ & 4.6&50&$\bigstar$\\
2335\,B& $3.87308\pm0.00005$ & $304.83135\pm0.00073$ & $0.17\pm0.06$ & 2.8 (0.3)& $0.49\pm0.06$ & 8.0&41&$\bigstar$\\
2340\,B& $9.97872\pm0.00005$ & $125.88685\pm0.00029$ & $0.45\pm0.06$ & 7.6 (1.1)& $1.17\pm0.06$ & 18.8&177&$\bigstar$\\
2350\,B& $31.40562\pm0.00005$ & $269.01582\pm0.00019$ & $0.23\pm0.14$ & 1.6 (0.4)& $0.38\pm0.12$ & 3.1&56&$\bigstar$\\
2358\,B& $7.30773\pm0.00012$ & $139.11250\pm0.00091$ & $0.14\pm0.16$ & 0.9 (0.1)& $1.75\pm0.16$ & 10.8&18&$\bigstar$\\
2374\,A& $22.34477\pm0.00002$ & $46.57715\pm0.00005$ & $0.02\pm0.03$ & 0.8 (0.2)& $0.58\pm0.03$ & 23.1&116\\
2380\,B& $5.07723\pm0.00047$ & $256.80573\pm0.00470$ & $0.24\pm0.58$ & 0.4 (0.4)& $0.63\pm0.70$ & 0.9&31&$\bigstar$\\
2383\,A& $24.09397\pm0.00002$ & $39.17459\pm0.00005$ & $0.00\pm0.02$ & 0.1 (0.0)& $0.18\pm0.02$ & 9.6&558\\
2384\,B& $0.83848\pm0.00020$ & $350.29022\pm0.02994$ & $0.27\pm0.17$ & 1.6 (0.2)& $2.04\pm0.43$ & 4.7&51&$\bigstar$\\
2409\,B& $23.83756\pm0.00002$ & $270.74528\pm0.00006$ & $0.04\pm0.03$ & 1.4 (0.1)& $3.11\pm0.03$ & 102.3&17&$\bigstar$\\
2417\,B& $1.85419\pm0.00002$ & $285.16794\pm0.00090$ & $0.07\pm0.03$ & 2.1 (0.2)& $3.23\pm0.03$ & 119.8&23&$\bigstar$\\
2419\,B& $1.71429\pm0.00003$ & $331.75352\pm0.00133$ & $0.03\pm0.03$ & 0.8 (0.1)& $1.14\pm0.04$ & 27.7&42&$\bigstar$\\
2422\,B& $0.85018\pm0.00007$ & $140.97694\pm0.00437$ & $0.16\pm0.06$ & 2.5 (0.5)& $1.67\pm0.09$ & 19.4&43\\
2425\,B& $2.30113\pm0.00009$ & $72.60363\pm0.00229$ & $0.18\pm0.11$ & 1.6 (0.3)& $0.77\pm0.16$ & 5.0&34&$\bigstar$\\
\hline
\end{tabular}
\end{table*}

\setcounter{table}{3}

\begin{table*} \caption{continued}
\begin{tabular}{lccccccccc}
\hline
CTOI   & $\rho$    & $PA$        & $\Delta\pi$ & $sig$-      & $\mu_{\rm rel}$ & $sig$-             & $cpm$-   & not in\\
             & [arcsec]  & [$^\circ$]  & [mas]       & $\Delta\pi$ & [mas/yr]    & $\mu_{\rm rel}$ & $index$  & WDS\\
\hline
{\fontsize{7}{0}\selectfont  35703676\,B} & $8.45241\pm0.00008$ & $71.11569\pm0.00045$ & $0.24\pm0.10$ & 2.3 (0.3)& $3.71\pm0.10$ & 36.0&16&$\bigstar$\\
{\fontsize{7}{0}\selectfont  83839341\,B} & $5.56932\pm0.00002$ & $207.66328\pm0.00026$ & $0.06\pm0.03$ & 2.1 (0.5)& $0.55\pm0.03$ & 21.7&152&$\bigstar$\\
{\fontsize{7}{0}\selectfont  98957720\,B} & $16.06523\pm0.00004$ & $231.40252\pm0.00014$ & $0.05\pm0.06$ & 0.8 (0.2)& $0.16\pm0.07$ & 2.2&550&$\bigstar$\\
{\fontsize{7}{0}\selectfont  105850602\,C} & $29.25593\pm0.00023$ & $106.60682\pm0.00041$ & $2.00\pm0.30$ & 6.6 (0.8)& $1.37\pm0.24$ & 5.6&56&$\bigstar$\\
{\fontsize{7}{0}\selectfont  105850602\,D} & $28.57790\pm0.00018$ & $106.83131\pm0.00036$ & $0.87\pm0.26$ & 3.3 (0.5)& $1.21\pm0.21$ & 5.9&62&$\bigstar$\\
{\fontsize{7}{0}\selectfont  117644481\,B} & $15.28674\pm0.00004$ & $224.70015\pm0.00014$ & $0.01\pm0.06$ & 0.1 (0.1)& $0.27\pm0.05$ & 6.0&408&$\bigstar$\\
{\fontsize{7}{0}\selectfont  135145585\,B} & $11.11883\pm0.00006$ & $29.94000\pm0.00040$ & $0.16\pm0.11$ & 1.4 (0.3)& $4.19\pm0.09$ & 45.3&16&$\bigstar$\\
{\fontsize{7}{0}\selectfont  139444326\,B} & $6.38457\pm0.00020$ & $174.59532\pm0.00130$ & $0.34\pm0.29$ & 1.2 (0.3)& $0.59\pm0.21$ & 2.8&17&$\bigstar$\\
{\fontsize{7}{0}\selectfont  142443425\,B} & $3.16439\pm0.00003$ & $115.19061\pm0.00072$ & $0.01\pm0.05$ & 0.1 (0.0)& $1.61\pm0.04$ & 42.7&104&$\bigstar$\\
{\fontsize{7}{0}\selectfont  144164538\,B} & $1.69843\pm0.00028$ & $48.92419\pm0.00936$ & $0.05\pm0.28$ & 0.2 (0.0)& $1.10\pm0.32$ & 3.4&66&$\bigstar$\\
{\fontsize{7}{0}\selectfont  151628217\,B} & $14.42224\pm0.00002$ & $213.85689\pm0.00006$ & $0.01\pm0.02$ & 0.3 (0.0)& $1.35\pm0.02$ & 58.4&11&$\bigstar$\\
{\fontsize{7}{0}\selectfont  152226055\,B} & $16.03023\pm0.00002$ & $319.68424\pm0.00006$ & $0.03\pm0.03$ & 1.0 (0.2)& $0.64\pm0.02$ & 28.2&51&$\bigstar$\\
{\fontsize{7}{0}\selectfont  164781040\,B} & $0.88788\pm0.00038$ & $47.48872\pm0.02454$ & $1.62\pm0.42$ & 3.9 (0.4)& $2.92\pm0.49$ & 6.0&6.6\\
{\fontsize{7}{0}\selectfont  178367145\,A} & $4.86196\pm0.00002$ & $318.91640\pm0.00029$ & $0.13\pm0.04$ & 3.7 (0.6)& $0.65\pm0.04$ & 18.1&42\\
{\fontsize{7}{0}\selectfont  197760286\,B} & $5.80368\pm0.00013$ & $243.46265\pm0.00105$ & $0.28\pm0.17$ & 1.6 (0.2)& $5.51\pm0.14$ & 39.5&14&$\bigstar$\\
{\fontsize{7}{0}\selectfont  202712304\,B} & $15.70443\pm0.00004$ & $221.78081\pm0.00014$ & $0.37\pm0.07$ & 5.5 (0.8)& $2.22\pm0.05$ & 43.7&123\\
{\fontsize{7}{0}\selectfont  224327878\,B} & $2.26527\pm0.00006$ & $98.18354\pm0.00250$ & $0.11\pm0.09$ & 1.3 (0.2)& $2.14\pm0.10$ & 20.4&34&$\bigstar$\\
{\fontsize{7}{0}\selectfont  224327878\,C} & $10.28639\pm0.00004$ & $35.10678\pm0.00021$ & $0.02\pm0.05$ & 0.4 (0.2)& $0.90\pm0.05$ & 19.8&80&$\bigstar$\\
{\fontsize{7}{0}\selectfont  230236827\,B} & $4.15576\pm0.00003$ & $343.54497\pm0.00047$ & $0.06\pm0.04$ & 1.6 (0.2)& $6.56\pm0.05$ & 142.7&11&$\bigstar$\\
{\fontsize{7}{0}\selectfont  238235254\,A} & $16.42331\pm0.00006$ & $227.56205\pm0.00020$ & $0.14\pm0.06$ & 2.3 (0.2)& $0.20\pm0.07$ & 2.7&96\\
{\fontsize{7}{0}\selectfont  238920872\,B} & $45.59533\pm0.00006$ & $121.75295\pm0.00007$ & $0.04\pm0.06$ & 0.7 (0.3)& $0.44\pm0.07$ & 6.0&101&$\bigstar$\\
{\fontsize{7}{0}\selectfont  253040591\,B} & $8.83229\pm0.00027$ & $221.32306\pm0.00178$ & $0.72\pm0.37$ & 2.0 (0.3)& $5.94\pm0.31$ & 19.2&14&$\bigstar$\\
{\fontsize{7}{0}\selectfont  257605131\,B} & $37.81874\pm0.00002$ & $239.83724\pm0.00003$ & $0.02\pm0.02$ & 1.0 (0.2)& $0.44\pm0.02$ & 22.3&76&$\bigstar$\\
{\fontsize{7}{0}\selectfont  259376845\,B} & $4.64337\pm0.00002$ & $242.40365\pm0.00021$ & $0.03\pm0.02$ & 1.7 (0.2)& $1.41\pm0.02$ & 61.8&37&$\bigstar$\\
{\fontsize{7}{0}\selectfont  282502866\,B} & $45.17057\pm0.00004$ & $41.42217\pm0.00005$ & $0.19\pm0.04$ & 4.6 (0.8)& $0.34\pm0.05$ & 7.5&214&$\bigstar$\\
{\fontsize{7}{0}\selectfont  288240183\,B} & $60.05636\pm0.00036$ & $332.60249\pm0.00036$ & $0.41\pm0.36$ & 1.1 (0.1)& $5.46\pm0.51$ & 10.8&20\\
{\fontsize{7}{0}\selectfont  288240183\,C} & $58.24718\pm0.00036$ & $332.54920\pm0.00037$ & $0.39\pm0.36$ & 1.1 (0.1)& $3.77\pm0.53$ & 7.1&29&$\bigstar$\\
{\fontsize{7}{0}\selectfont  290596728\,B} & $5.41246\pm0.00002$ & $135.75015\pm0.00020$ & $0.03\pm0.03$ & 1.3 (0.4)& $0.21\pm0.02$ & 9.8&88&$\bigstar$\\
{\fontsize{7}{0}\selectfont  300116105\,B} & $2.41041\pm0.00008$ & $277.14941\pm0.00162$ & $0.09\pm0.08$ & 1.1 (0.2)& $1.95\pm0.10$ & 19.5&17&$\bigstar$\\
{\fontsize{7}{0}\selectfont  308301091\,B} & $3.20160\pm0.00009$ & $31.36135\pm0.00133$ & $0.24\pm0.11$ & 2.3 (0.4)& $1.24\pm0.12$ & 10.7&15&$\bigstar$\\
{\fontsize{7}{0}\selectfont  312091232\,B} & $7.79773\pm0.00010$ & $56.36780\pm0.00069$ & $0.14\pm0.11$ & 1.2 (0.7)& $0.39\pm0.13$ & 2.9&204&$\bigstar$\\
{\fontsize{7}{0}\selectfont  326092637\,B} & $20.85385\pm0.00027$ & $16.93096\pm0.00093$ & $0.01\pm0.44$ & 0.0 (0.0)& $0.85\pm0.62$ & 1.4&44&$\bigstar$\\
{\fontsize{7}{0}\selectfont  341411516\,B} & $23.73899\pm0.00024$ & $105.66263\pm0.00061$ & $0.11\pm0.25$ & 0.4 (0.3)& $0.99\pm0.34$ & 2.9&63&$\bigstar$\\
{\fontsize{7}{0}\selectfont  345324572\,A} & $32.07658\pm0.00002$ & $19.00568\pm0.00003$ & $0.03\pm0.02$ & 1.6 (0.3)& $1.38\pm0.02$ & 71.5&68&$\bigstar$\\
{\fontsize{7}{0}\selectfont  349793830\,B} & $28.23873\pm0.00009$ & $264.01259\pm0.00016$ & $0.06\pm0.11$ & 0.5 (0.4)& $2.48\pm0.09$ & 27.3&139\\
{\fontsize{7}{0}\selectfont  352915304\,B} & $3.75547\pm0.00009$ & $206.52765\pm0.00136$ & $0.03\pm0.08$ & 0.4 (0.1)& $0.37\pm0.14$ & 2.8&882\\
{\fontsize{7}{0}\selectfont  369376388\,B} & $1.63306\pm0.00008$ & $242.28326\pm0.00304$ & $0.30\pm0.11$ & 2.8 (0.4)& $3.67\pm0.11$ & 34.4&84&$\bigstar$\\
{\fontsize{7}{0}\selectfont  369376388\,C} & $23.83109\pm0.00005$ & $39.23375\pm0.00011$ & $0.00\pm0.06$ & 0.0 (0.0)& $1.75\pm0.06$ & 27.5&177\\
{\fontsize{7}{0}\selectfont  372913337\,B} & $14.59903\pm0.00002$ & $322.97056\pm0.00007$ & $0.03\pm0.02$ & 1.8 (0.2)& $0.46\pm0.02$ & 18.8&52&$\bigstar$\\
{\fontsize{7}{0}\selectfont  374352402\,B} & $4.11475\pm0.00005$ & $252.26653\pm0.00097$ & $0.16\pm0.09$ & 1.8 (0.4)& $0.29\pm0.08$ & 3.7&93&$\bigstar$\\
{\fontsize{7}{0}\selectfont  374732772\,B} & $5.05320\pm0.00003$ & $328.64658\pm0.00041$ & $0.14\pm0.04$ & 3.1 (0.5)& $1.54\pm0.04$ & 43.3&34&$\bigstar$\\
{\fontsize{7}{0}\selectfont  394721720\,B} & $11.33222\pm0.00013$ & $151.28537\pm0.00068$ & $0.38\pm0.14$ & 2.7 (0.9)& $0.62\pm0.17$ & 3.8&98&$\bigstar$\\
{\fontsize{7}{0}\selectfont  399913539\,B} & $5.24516\pm0.00009$ & $136.57882\pm0.00100$ & $0.13\pm0.10$ & 1.4 (0.4)& $0.46\pm0.14$ & 3.3&123&$\bigstar$\\
{\fontsize{7}{0}\selectfont  453455638\,B} & $10.42144\pm0.00011$ & $83.96279\pm0.00057$ & $0.67\pm0.11$ & 6.1 (0.5)& $1.55\pm0.14$ & 11.1&197&$\bigstar$\\
{\fontsize{7}{0}\selectfont  460950389\,B} & $54.01510\pm0.00121$ & $202.76271\pm0.00130$ & $1.25\pm1.40$ & 0.9 (0.1)& $2.99\pm1.41$ & 2.1&13&$\bigstar$\\
{\fontsize{7}{0}\selectfont  467785319\,B} & $3.20547\pm0.00012$ & $24.51624\pm0.00263$ & $0.20\pm0.13$ & 1.5 (0.5)& $1.89\pm0.15$ & 12.4&31&$\bigstar$\\
{\fontsize{7}{0}\selectfont  738065944\,B} & $2.00627\pm0.00001$ & $276.64459\pm0.00043$ & $0.05\pm0.02$ & 3.3 (0.7)& $0.49\pm0.02$ & 28.3&183&$\bigstar$\\
{\fontsize{7}{0}\selectfont  901674675\,B} & $2.34422\pm0.00002$ & $217.51342\pm0.00037$ & $0.01\pm0.02$ & 0.6 (0.1)& $0.02\pm0.02$ & 1.1&2732&$\bigstar$\\
\hline
\end{tabular}
\end{table*}

\begin{table*} \caption{This table lists the equatorial coordinates ($\alpha$, $\delta$ for epoch 2016.0) of all detected co-moving companions (sorted by their identifiers) together with their derived absolute G-band magnitudes $M_{\rm G}$, masses, and effective temperatures $T_{\rm eff}$. The flags for all companions, as defined in the text, are listed in the last column of this table.}\label{TAB_COMP_PROPS}
\begin{tabular}{lcccccccc}
\hline
TOI     & $\alpha$   & $\delta$   & $M_{\rm G}$ & $sep$ & $mass$        &  $T_{\rm eff}$ & Flags \\
& [$^\circ$] & [$^\circ$] & [mag]   & [au]  & [$M_{\odot}$] &  [K]       & \\
\hline
1937\,B & 116.37063898358 & -52.38256904657 & $ 7.62 _{ -0.24 }^{ +0.17 } $ & 1030 & $ 0.59 _{ -0.05 }^{ +0.02 }$ & $ 4177 _{ -250 }^{ +276 }$ & $\texttt{BPRP}$ \\
1940\,B & 212.64705044663 & -27.99343384615 & $ 5.60 _{ -0.21 }^{ +0.21 } $ & 739 & $ 0.85 _{ -0.06 }^{ +0.04 }$ & $ 5200 _{ -120 }^{ +124 }$ & $\texttt{inter}$ $\texttt{EDR3}$  \\
1943\,B & 183.70611541557 & -63.08713763185 & $ 7.81 _{ -0.09 }^{ +0.11 } $ & 473 & $ 0.50 _{ -0.01 }^{ +0.01 }$ & $ 4411 _{ -157 }^{ +168 }$ & $\texttt{BPRP}$ \\
1946\,B & 217.27930985483 & -43.36122919815 & $ 5.09 _{ -0.22 }^{ +0.18 } $ & 709 & $ 0.87 _{ -0.07 }^{ +0.09 }$ & $ 5714 _{ -380 }^{ +386   }$ \\
1953\,B & 215.14092879184 & -41.99269221369 & $ 10.26 _{ -0.09 }^{ +0.18 } $ & 9463 & $ 0.35 _{ -0.01 }^{ +0.05 }$ & $ 3318 _{ -48 }^{ +59 }$ & $\texttt{BPRP}$ \\
1964\,B & 201.64389030315 & -39.62891769667 & $ 7.83 _{ -0.12 }^{ +0.03 } $ & 2326 & $ 0.55 _{ -0.02 }^{ +0.01 }$ & $ 4193 _{ -214 }^{ +307 }$ & $\texttt{BPRP}$ \\
1966\,B & 209.98243641458 & -43.31801392589 & $ 6.11 _{ -0.18 }^{ +0.29 } $ & 8975 & $ 0.79 _{ -0.03 }^{ +0.04 }$ & $ 4906 _{ -84 }^{ +192 }$ & $\texttt{BPRP}$ $\texttt{PRI}$ \\
1970\,B & 190.71676597317 & -53.68567116786 & $ 9.08 _{ -0.30 }^{ +0.23 } $ & 5025 & $ 0.50 _{ -0.05 }^{ +0.01 }$ & $ 3550 _{ -57 }^{ +64 }$ & $\texttt{BPRP}$ \\
1972\,B & 176.05846644906 & -59.90759743827 & $ 3.89 _{ -0.26 }^{ +0.07 } $ & 5179 & $ 1.07 _{ -0.12 }^{ +0.12 }$ & $ 6277 _{ -315 }^{ +362 }$ & $\texttt{BPRP}$ $\texttt{PRI}$  \\
1984\,B & 168.87036013723 & -44.42914110070 & $ 11.12 _{ -0.13 }^{ +0.03 } $ & 756 & $ 0.24 _{ -0.01 }^{ +0.01 }$ & $ 3228 _{ -3 }^{ +13 }$ & $\texttt{inter}$ $\texttt{EDR3}$  $\texttt{BPRP}$ \\
1992\,B & 162.91641069316 & -44.98555236333 & $ 3.42 _{ -0.26 }^{ +0.37 } $ & 1213 & $ 1.14 _{ -0.17 }^{ +0.19 }$ & $ 6381 _{ -490 }^{ +560 }$ & $\texttt{BPRP}$ $\texttt{PRI}$  \\
2001\,B & 128.18085020302 & -58.57546310498 & $   _{   }^{   } $ & 433 & $   _{   }^{   }$ & $   _{   }^{   }$ & $\texttt{EDR3}$ $\texttt{noGmag}$ \\
2006\,B & 91.22129476743 & -69.36082554536 & $ 4.21 _{ -0.13 }^{ +0.19 } $ & 2305 & $ 1.00 _{ -0.10 }^{ +0.10 }$ & $ 6135 _{ -325 }^{ +371 }$ & $\texttt{BPRP}$ $\texttt{PRI}$  \\
2009\,B & 16.91102529334 & 22.95355541328 & $ 10.88 _{ -0.17 }^{ +0.18 } $ & 193 & $ 0.33 _{ -0.01 }^{ +0.01 }$ & $ 3072 _{ -31 }^{ +29 }$ & $\texttt{BPRP}$ $\texttt{PRI}$ \\
2033\,B & 17.68275285338 & 65.25239187480 & $ 3.36 _{ -0.37 }^{ +0.34 } $ & 3009 & $ 1.18 _{ -0.16 }^{ +0.17 }$ & $ 6372 _{ -423 }^{ +547 }$ & $\texttt{BPRP}$ $\texttt{PRI}$  \\
2036\,B & 233.01453280974 & 24.08380188116 & $ 7.33 _{ -0.30 }^{ +0.48 } $ & 1032 & $ 0.55 _{ -0.01 }^{ +0.01 }$ & $ 5038 _{ -188 }^{ +69 }$ & $\texttt{BPRP}$ \\
2050\,B & 323.91600940223 & 65.43645611145 & $ 8.24 _{ -0.29 }^{ +0.14 } $ & 1391 & $ 0.55 _{ -0.01 }^{ +0.01 }$ & $ 3881 _{ -32 }^{ +52 }$ & $\texttt{BPRP}$ $\texttt{PRI}$ \\
2056\,B & 2.60181476661 & 58.48938732348 & $ 7.14 _{ -0.17 }^{ +0.14 } $ & 397 & $ 0.61 _{ -0.01 }^{ +0.04 }$ & $ 4600 _{ -199 }^{ +357 }$ & $\texttt{BPRP}$ \\
2068\,B & 186.28373776065 & 60.42064810535 & $ 8.80 _{ -0.36 }^{ +0.17 } $ & 1059 & $ 0.55 _{ -0.01 }^{ +0.01 }$ & $ 3572 _{ -31 }^{ +45 }$ & $\texttt{BPRP}$ $\texttt{PRI}$ \\
2072\,B & 173.98897259356 & 75.54607768850 & $ 10.44 _{ -0.12 }^{ +0.11 } $ & 117 & $ 0.35 _{ -0.01 }^{ +0.04 }$ & $ 3247 _{ -51 }^{ +246 }$ & $\texttt{BPRP}$ $\texttt{PRI}$ \\
2084\,B & 259.25233532656 & 72.74369142724 & $ 15.37 _{ -0.12 }^{ +0.02 } $ & 1399 & $ 0.09 _{ -0.01 }^{ +0.01 }$ & $ 2590 _{ -5 }^{ +18 }$ & $\texttt{inter}$ $\texttt{BPRP}$ \\
2092\,B & 212.54643389257 & 45.56389415042 & $ 13.05 _{ -0.10 }^{ +0.08 } $ & 479 & $   _{  }^{  }$ & $   _{   }^{  }$ & $\texttt{WD}$ $\texttt{EDR3}$ $\texttt{BPRP}$ \\
2094\,B & 254.13742491435 & 70.02489938048 & $ 14.19 _{ -0.14 }^{ +0.05 } $ & 538 & $ 0.11 _{ -0.01 }^{ +0.01 }$ & $ 2894 _{ -8 }^{ +21 }$ & $\texttt{inter}$ $\texttt{BPRP}$ \\
2106\,B & 207.17848145517 & 44.91254228578 & $ 8.77 _{ -0.10 }^{ +0.10 } $ & 346 & $ 0.50 _{ -0.01 }^{ +0.01 }$ & $ 3724 _{ 24 }^{ +24 }$ & $\texttt{inter}$ $\texttt{BPRP}$ \\
2108\,B & 247.48156446857 & 21.49483143158 & $ 3.11 _{ -0.41 }^{ +0.27 } $ & 303 & $ 0.95 _{ -0.19 }^{ +0.30 }$ & $ 5954 _{ 740 }^{ +800   }$ \\
2113\,B & 267.35407815488 & 40.45712910786 & $ 4.21 _{ -0.14 }^{ +0.05 } $ & 360 & $ 1.00 _{ -0.11 }^{ +0.10 }$ & $ 6384 _{ 299 }^{ +350 }$ & $\texttt{BPRP}$ \\
2115\,B & 19.59598675578 & 63.54509607880 & $   _{   }^{   } $ & 206 & $   _{   }^{   }$ & $   _{   }^{   }$ & $\texttt{EDR3}$ $\texttt{noGmag}$ \\
2127\,B & 256.34628231343 & 33.01158434380 & $ 12.68 _{ -0.09 }^{ +0.15 } $ & 430 & $   _{  }^{  }$ & $   _{   }^{  }$ & $\texttt{WD}$ $\texttt{EDR3}$ $\texttt{BPRP}$ \\
2128\,B & 256.98115969078 & 32.10356773055 & $ 9.41 _{ -0.12 }^{ +0.07 } $ & 238 & $ 0.36 _{ -0.01 }^{ +0.04 }$ & $ 3748 _{ -161 }^{ +229 }$ & $\texttt{BPRP}$ \\
2144\,B & 350.23675703392 & 79.43903525527 & $ 9.90 _{ -0.17 }^{ +0.06 } $ & 1553 & $ 0.40 _{ -0.01 }^{ +0.01 }$ & $ 3360 _{ -37 }^{ +35 }$ & $\texttt{BPRP}$ $\texttt{PRI}$ \\
2149\,B & 285.59708263433 & 42.50726078542 & $ 5.51 _{ -0.35 }^{ +0.46 } $ & 356 & $ 0.81 _{ -0.19 }^{ +0.21 }$ & $ 5425 _{ -1052 }^{ +776   }$ \\
2152\,B & 26.31614962469 & 77.79318868748 & $ 11.16 _{ -0.20 }^{ +0.20 } $ & 6609 & $ 0.24 _{ -0.02 }^{ +0.02 }$ & $ 3223 _{ -20 }^{ +20 }$ &  $\texttt{inter}$ $\texttt{BPRP}$ \\
2169\,B & 279.16854084831 & 23.25755382074 & $ 5.52 _{ -0.20 }^{ +0.11 } $ & 2346 & $ 0.83 _{ -0.05 }^{ +0.05 }$ & $ 5374 _{ -206 }^{ +262 }$ &     $\texttt{BPRP}$ $\texttt{PRI}$ \\
2183\,B & 283.48984559533 & 37.38040766442 & $ 2.77 _{ -0.33 }^{ +0.33 } $ & 166 & $ 1.39 _{ -0.25 }^{ +0.25 }$ & $ 7066 _{ -812 }^{ +1196 }$ &     $\texttt{BPRP}$ \\
2193\,B & 313.69268269724 & -72.80493275516 & $ 9.65 _{ -0.31 }^{ +0.12 } $ & 638 & $ 0.40 _{ -0.14 }^{ +0.16 }$ & $ 3541 _{ -312 }^{ +496   }$ \\
2195\,B & 34.83851060185 & -72.70970300127 & $ 7.83 _{ -0.28 }^{ +0.12 } $ & 586 & $ 0.55 _{ -0.04 }^{ +0.04 }$ & $ 4119 _{ -250 }^{ +150 }$ &     $\texttt{BPRP}$ \\
2205\,B & 94.30504481936 & -56.51138346415 & $ 10.59 _{ -0.07 }^{ +0.07 } $ & 9060 & $ 0.30 _{ -0.01 }^{ +0.01 }$ & $ 3308 _{ -13 }^{ +13 }$ &  $\texttt{inter}$ $\texttt{BPRP}$ \\
2205\,C & 94.31675618323 & -56.50978635189 & $ 11.12 _{ -0.07 }^{ +0.07 } $ & 892 & $ 0.24 _{ -0.01 }^{ +0.01 }$ & $ 3228 _{ -7 }^{ +7 }$ &  $\texttt{inter}$ $\texttt{BPRP}$ \\
2215\,B & 286.05887028356 & -32.36683422030 & $ 6.49 _{ -0.20 }^{ +0.02 } $ & 4433 & $ 0.73 _{ -0.03 }^{ +0.02 }$ & $ 4759 _{ -187 }^{ +182 }$ &     $\texttt{BPRP}$ $\texttt{PRI}$ \\
2218\,A & 107.15134082021 & -64.23282491835 & $ 3.64 _{ -0.21 }^{ +0.14 } $ & 2389 & $ 1.08 _{ -0.13 }^{ +0.16 }$ & $ 6299 _{ -395 }^{ +477 }$ &     $\texttt{BPRP}$ $\texttt{PRI}$ \\
2233\,B & 296.91945139989 & -32.57720015101 & $ 8.22 _{ -0.21 }^{ +0.36 } $ & 874 & $ 0.54 _{ -0.02 }^{ +0.01 }$ & $ 3950 _{ -29 }^{ +41 }$ &     $\texttt{BPRP}$ \\
2239\,B & 65.39925931554 & -67.47014069910 & $ 4.49 _{ -0.18 }^{ +0.07 } $ & 8570 & $ 0.93 _{ -0.09 }^{ +0.08 }$ & $ 6125 _{ -230 }^{ +266 }$ &     $\texttt{BPRP}$ $\texttt{PRI}$ \\
2244\,B & 308.56398032092 & -48.27456998081 & $ 3.00 _{ -0.16 }^{ +0.13 } $ & 262 & $ 1.35 _{ -0.24 }^{ +0.20 }$ & $ 7145 _{ -777 }^{ +1156 }$ &     $\texttt{BPRP}$ \\
2244\,C & 308.55675305431 & -48.27142425630 & $ 11.52 _{ -0.16 }^{ +0.12 } $ & 3539 & $ 0.25 _{ -0.05 }^{ +0.01 }$ & $ 3041 _{ -40 }^{ +13 }$ &     $\texttt{BPRP}$ \\
2246\,B & 113.04208107725 & -65.80987155300 & $ 3.76 _{ -0.26 }^{ +0.23 } $ & 1320 & $ 1.08 _{ -0.13 }^{ +0.13 }$ & $ 6346 _{ -334 }^{ +386 }$ &     $\texttt{BPRP}$ $\texttt{PRI}$ \\
\hline
\end{tabular}
\end{table*}

\setcounter{table}{4}

\begin{table*} \caption{continued}
\begin{tabular}{lcccccccc}
\hline
TOI     & $\alpha$   & $\delta$   & $M_{\rm G}$ & $sep$ & $mass$        &  $T_{\rm eff}$ & Flags \\
& [$^\circ$] & [$^\circ$] & [mag]   & [au]  & [$M_{\odot}$] &  [K]       & \\
\hline
2248\,B & 80.62608655037 & -70.25373433972 & $ 9.77 _{ -0.29 }^{ +0.19 } $ & 1249 & $ 0.39 _{ -0.02 }^{ +0.04 }$ & $ 3478 _{ -46 }^{ +55 }$ &  $\texttt{inter}$ $\texttt{EDR3}$ $\texttt{BPRP}$ \\
2253\,B & 276.84450187297 & 73.26719741598 & $ 9.28 _{ -0.48 }^{ +0.15 } $ & 4520 & $ 0.50 _{ -0.01 }^{ +0.01 }$ & $ 3428 _{ -9 }^{ +47 }$ &     $\texttt{BPRP}$ $\texttt{PRI}$ \\
2279\,B & 255.24461493115 & 45.38724425492 & $ 9.35 _{ -0.23 }^{ +0.09 } $ & 430 & $ 0.40 _{ -0.02 }^{ +0.01 }$ & $ 3687 _{ -194 }^{ +151 }$ &     $\texttt{BPRP}$ \\
2281\,B & 294.37931940875 & 66.96300161303 & $ 4.48 _{ -0.12 }^{ +0.05 } $ & 6039 & $ 0.98 _{ -0.10 }^{ +0.07 }$ & $ 6027 _{ -280 }^{ +275 }$ &     $\texttt{BPRP}$ $\texttt{PRI}$ \\
2283\,B & 248.32656847020 & 63.53124834111 & $ 12.24 _{ -0.09 }^{ +0.04 } $ & 1315 & $ 0.19 _{ -0.07 }^{ +0.01 }$ & $ 2909 _{ -14 }^{ +13 }$ &     $\texttt{BPRP}$ $\texttt{PRI}$ \\
2289\,B & 255.10100397713 & 39.72794649361 & $ 10.35 _{ -0.22 }^{ +0.32 } $ & 1783 & $ 0.40 _{ -0.01 }^{ +0.01 }$ & $ 3139 _{ -11 }^{ +31 }$ &     $\texttt{BPRP}$ $\texttt{PRI}$ \\
2293\,B & 115.59347921340 & 70.40535489770 & $ 12.43 _{ -0.32 }^{ +0.26 } $ & 293 & $ 0.16 _{ -0.01 }^{ +0.01 }$ & $ 2983 _{ -56 }^{ +8 }$ &     $\texttt{BPRP}$ $\texttt{PRI}$ \\
2299\,B & 286.22240974084 & 79.75619277781 & $ 8.88 _{ -0.25 }^{ +0.03 } $ & 124 & $ 0.45 _{ -0.01 }^{ +0.04 }$ & $ 3830 _{ -110 }^{ +193 }$ &     $\texttt{BPRP}$ \\
2307\,B & 330.13554334398 & -29.14932158011 & $ 8.87 _{ -0.16 }^{ +0.09 } $ & 1320 & $ 0.49 _{ -0.01 }^{ +0.02 }$ & $ 3700 _{ -18 }^{ +39 }$ &  $\texttt{inter}$ $\texttt{BPRP}$ \\
2321\,B & 196.73942411562 & 16.54415038318 & $ 10.13 _{ -0.13 }^{ +0.04 } $ & 2529 & $ 0.38 _{ -0.08 }^{ +0.01 }$ & $ 3305 _{ -37 }^{ +12 }$ &     $\texttt{BPRP}$ $\texttt{PRI}$ \\
2325\,B & 343.54329331845 & -36.28713515058 & $ 6.94 _{ -0.18 }^{ +0.12 } $ & 1001 & $ 0.60 _{ -0.01 }^{ +0.04 }$ & $ 4829 _{ -256 }^{ +206 }$ &     $\texttt{BPRP}$ \\
2327\,B & 103.43704505805 & -73.08558755467 & $ 8.79 _{ -0.19 }^{ +0.09 } $ & 3887 & $ 0.55 _{ -0.01 }^{ +0.01 }$ & $ 3592 _{ -37 }^{ +68 }$ &     $\texttt{BPRP}$ $\texttt{PRI}$ \\
2328\,B & 31.68784113786 & -81.24746699057 & $ 10.61 _{ -0.16 }^{ +0.18 } $ & 379 & $ 0.29 _{ -0.02 }^{ +0.02 }$ & $ 3305 _{ -34 }^{ +30 }$ &  $\texttt{inter}$ $\texttt{EDR3}$ $\texttt{BPRP}$ \\
2335\,B & 50.82546214680 & -40.45290191998 & $ 7.93 _{ -0.14 }^{ +0.04 } $ & 1761 & $ 0.57 _{ -0.01 }^{ +0.02 }$ & $ 3966 _{ -15 }^{ +47 }$ &  $\texttt{inter}$ $\texttt{BPRP}$ \\
2340\,B & 42.26379287408 & -16.55926854837 & $ 8.40 _{ -0.13 }^{ +0.12 } $ & 1771 & $ 0.60 _{ -0.01 }^{ +0.03 }$ & $ 3661 _{ -153 }^{ +90 }$ &     $\texttt{BPRP}$ $\texttt{PRI}$  \\
2350\,B & 97.39871390344 & -21.99453651654 & $ 11.65 _{ -0.25 }^{ +0.11 } $ & 5874 & $ 0.20 _{ -0.01 }^{ +0.02 }$ & $ 3159 _{ -17 }^{ +40 }$ &  $\texttt{inter}$ $\texttt{BPRP}$ \\
2358\,B & 196.28739985402 & -31.98691483902 & $ 5.32 _{ -0.13 }^{ +0.21 } $ & 2802 & $ 0.73 _{ -0.04 }^{ +0.08 }$ & $ 5716 _{ -727 }^{ +255 }$ &     $\texttt{BPRP}$ $\texttt{PRI}$ \\
2374\,A & 319.50310270344 & -22.04556079240 & $ 3.50 _{ -0.22 }^{ +0.22 } $ & 3036 & $ 1.17 _{ -0.16 }^{ +0.17 }$ & $ 6343 _{ -478 }^{ +585 }$ &     $\texttt{BPRP}$ $\texttt{PRI}$ \\
2380\,B & 0.32437138931 & -8.92657181230 & $ 12.24 _{ -0.18 }^{ +0.18 } $ & 1364 & $ 0.17 _{ -0.01 }^{ +0.01 }$ & $ 3071 _{ -23 }^{ +23 }$ &  $\texttt{inter}$ $\texttt{EDR3}$ \\
2383\,A & 1.48935329659 & -20.64877382679 & $ 4.10 _{ -0.24 }^{ +0.06 } $ & 4119 & $ 1.04 _{ -0.11 }^{ +0.11 }$ & $ 6118 _{ -275 }^{ +345 }$ &     $\texttt{BPRP}$ $\texttt{PRI}$ \\
2384\,B & 36.15605640406 & -64.99997342715 & $ 11.39 _{ -0.12 }^{ +0.12 } $ & 158 & $ 0.20 _{ -0.07 }^{ +0.13 }$ & $ 3170 _{ -293 }^{ +292   }$ \\
2409\,B & 44.32959674872 & -41.19168974979 & $ 6.75 _{ -0.14 }^{ +0.09 } $ & 4557 & $ 0.70 _{ -0.02 }^{ +0.01 }$ & $ 4609 _{ -175 }^{ +210 }$ &     $\texttt{BPRP}$ $\texttt{PRI}$ \\
2417\,B & 37.88458406443 & -40.01606774547 & $ 5.85 _{ -0.38 }^{ +0.29 } $ & 474 & $ 0.79 _{ -0.03 }^{ +0.09 }$ & $ 5060 _{ -143 }^{ +239 }$ &  $\texttt{inter}$ \\
2419\,B & 53.26996227934 & -56.58801950869 & $ 8.52 _{ -0.25 }^{ +0.25 } $ & 424 & $ 0.52 _{ -0.18 }^{ +0.16 }$ & $ 3853 _{ -428 }^{ +867   }$ \\
2422\,B & 24.88466502593 & -7.39491592998 & $ 4.00 _{ -0.03 }^{ +0.03 } $ & 178 & $ 1.15 _{ -0.01 }^{ +0.01 }$ & $ 6091 _{ -24 }^{ +24 }$ &  $\texttt{inter}$ $\texttt{EDR3}$ $\texttt{BPRP}$ \\
2425\,B & 19.13130751686 & -0.28991369508 & $ 9.60 _{ -0.13 }^{ +0.05 } $ & 370 & $ 0.35 _{ -0.05 }^{ +0.05 }$ & $ 3612 _{ -177 }^{ +204 }$ &     $\texttt{BPRP}$ \\                                                                                      \hline
\end{tabular}
\end{table*}

\setcounter{table}{4}

\begin{table*} \caption{continued}
\begin{tabular}{lccccccc}
\hline
CTOI     & $\alpha$   & $\delta$   & $M_{\rm G}$ & $sep$ & $mass$        &  $T_{\rm eff}$ & Flags \\
& [$^\circ$] & [$^\circ$] & [mag]   & [au]  & [$M_{\odot}$] &  [K]       & \\
\hline
{\fontsize{7}{0}\selectfont 35703676\,B} & 23.79838607967 & -0.88660913129 & $ 7.60 _{ -0.23 }^{ +0.09 } $ & 3255 & $ 0.65 _{ -0.03 }^{ +0.04 }$ & $ 3823 _{ -197 }^{ +50 }$ & $\texttt{BPRP}$ $\texttt{PRI}$ \\
{\fontsize{7}{0}\selectfont 83839341\,B} & 217.53558345727 & -25.55056842979 & $ 3.94 _{ -0.23 }^{ +0.15 } $ & 2046 & $ 1.08 _{ -0.12 }^{ +0.11 }$ & $ 6298 _{ -322 }^{ +385 }$ & $\texttt{BPRP}$ $\texttt{PRI}$ \\
{\fontsize{7}{0}\selectfont 98957720\,B} & 180.99683654878 & -28.32813097238 & $ 9.80 _{ -0.16 }^{ +0.03 } $ & 3356 & $ 0.45 _{ -0.05 }^{ +0.01 }$ & $ 3317 _{ -46 }^{ +19 }$ & $\texttt{BPRP}$ $\texttt{PRI}$ \\
{\fontsize{7}{0}\selectfont 105850602\,C} & 244.33420561134 & 16.42671038186 & $ 11.26 _{ -0.10 }^{ +0.06 } $ & 3637 & $ 0.23 _{ -0.01 }^{ +0.01 }$ & $ 3213 _{ -6 }^{ +10 }$ & $\texttt{inter}$ $\texttt{BPRP}$ \\
{\fontsize{7}{0}\selectfont 105850602\,D} & 244.33400811636 & 16.42673442250 & $ 11.72 _{ -0.10 }^{ +0.06 } $ & 3553 & $ 0.20 _{ -0.07 }^{ +0.11 }$ & $ 3106 _{ -295 }^{ +270   }$ \\
{\fontsize{7}{0}\selectfont 117644481\,B} & 67.74861870851 & -14.36826105677 & $ 8.32 _{ -0.43 }^{ +0.48 } $ & 5179 & $ 0.60 _{ -0.05 }^{ +0.01 }$ & $ 3731 _{ -77 }^{ +75 }$ & $\texttt{BPRP}$ $\texttt{PRI}$ \\
{\fontsize{7}{0}\selectfont 135145585\,B} & 185.41006070851 & -41.19261307186 & $ 9.81 _{ -0.24 }^{ +0.14 } $ & 4282 & $ 0.45 _{ -0.01 }^{ +0.01 }$ & $ 3280 _{ -25 }^{ +83 }$ & $\texttt{BPRP}$ \\
{\fontsize{7}{0}\selectfont 139444326\,B} & 80.73136663361 & -21.95745238328 & $ 9.92 _{ -0.33 }^{ +0.41 } $ & 3089 & $ 0.37 _{ -0.05 }^{ +0.04 }$ & $ 3442 _{ -84 }^{ +73 }$ & $\texttt{inter}$ $\texttt{BPRP}$ \\
{\fontsize{7}{0}\selectfont 142443425\,B} & 146.55289564051 & 67.09725083384 & $ 7.24 _{ -0.09 }^{ +0.10 } $ & 389 & $ 0.55 _{ -0.01 }^{ +0.03 }$ & $ 4617 _{ -508 }^{ +229 }$ & $\texttt{BPRP}$ \\
{\fontsize{7}{0}\selectfont 144164538\,B} & 61.18284430216 & -47.16584470949 & $ 11.35 _{ -0.41 }^{ +0.17 } $ & 354 & $ 0.22 _{ -0.01 }^{ +0.03 }$ & $ 3204 _{ -25 }^{ +42 }$ & $\texttt{inter}$ \\
{\fontsize{7}{0}\selectfont 151628217\,B} & 88.60524481432 & -46.42448223827 & $ 4.30 _{ -0.10 }^{ +0.07 } $ & 7075 & $ 0.96 _{ -0.10 }^{ +0.10 }$ & $ 6236 _{ -248 }^{ +276 }$ & $\texttt{BPRP}$ $\texttt{PRI}$ \\
{\fontsize{7}{0}\selectfont 152226055\,B} & 312.48730424663 & -33.27779249937 & $ 6.70 _{ -0.27 }^{ +0.32 } $ & 3775 & $ 0.74 _{ -0.04 }^{ +0.01 }$ & $ 4562 _{ -71 }^{ +124 }$ & $\texttt{BPRP}$ $\texttt{PRI}$ \\
{\fontsize{7}{0}\selectfont 164781040\,B} & 284.69989267282 & 46.37521416536 & $ 5.49 _{ -0.59 }^{ +0.34 } $ & 335 & $ 0.80 _{ -0.17 }^{ +0.22 }$ & $ 5434 _{ -1004 }^{ +782   }$ \\
{\fontsize{7}{0}\selectfont 178367145\,A} & 123.39224456462 & -1.98278782520 & $ 3.92 _{ -0.12 }^{ +0.15 } $ & 1265 & $ 1.05 _{ -0.11 }^{ +0.14 }$ & $ 6207 _{ -323 }^{ +418 }$ & $\texttt{BPRP}$ $\texttt{PRI}$ \\
{\fontsize{7}{0}\selectfont 197760286\,B} & 331.50570859756 & -37.28596009795 & $ 8.69 _{ -0.44 }^{ +0.07 } $ & 1062 & $ 0.54 _{ -0.04 }^{ +0.02 }$ & $ 3711 _{ -98 }^{ +28 }$ & $\texttt{BPRP}$ $\texttt{PRI}$ \\
{\fontsize{7}{0}\selectfont 202712304\,B} & 7.89840211300 & 50.95370147188 & $ 10.94 _{ -0.42 }^{ +0.39 } $ & 1023 & $ 0.32 _{ -0.16 }^{ +0.03 }$ & $ 3017 _{ -20 }^{ +40 }$ & $\texttt{BPRP}$ $\texttt{PRI}$ \\
{\fontsize{7}{0}\selectfont 224327878\,B} & 251.02883064726 & 30.00805300731 & $ 11.01 _{ -0.28 }^{ +0.06 } $ & 267 & $ 0.25 _{ -0.01 }^{ +0.03 }$ & $ 3239 _{ -6 }^{ +43 }$ & $\texttt{inter}$ $\texttt{EDR3}$ $\texttt{BPRP}$\\
{\fontsize{7}{0}\selectfont 224327878\,C} & 251.03000905102 & 30.01048010687 & $ 10.99 _{ -0.28 }^{ +0.06 } $ & 1212 & $ 0.32 _{ -0.01 }^{ +0.01 }$ & $ 3055 _{ -32 }^{ +51 }$ & $\texttt{BPRP}$ $\texttt{PRI}$ \\
{\fontsize{7}{0}\selectfont 230236827\,B} & 261.53342682463 & 42.59471357813 & $ 10.43 _{ -0.30 }^{ +0.07 } $ & 354 & $ 0.35 _{ -0.01 }^{ +0.01 }$ & $ 3248 _{ -6 }^{ +6 }$ & $\texttt{BPRP}$ $\texttt{PRI}$ \\
{\fontsize{7}{0}\selectfont 238235254\,A} & 119.80123285388 & -49.97681348175 & $ -1.69 _{ -0.12 }^{ +0.06 } $ & 6822 & $ 4.47 _{ -0.36 }^{ +0.69 }$ & $ 13518 _{ -1222 }^{ +1839 }$ & $\texttt{BPRP}$ $\texttt{PRI}$ \\
{\fontsize{7}{0}\selectfont 238920872\,B} & 94.29032954304 & -52.34601641705 & $ 11.82 _{ -0.13 }^{ +0.13 } $ & 4813 & $ 0.12 _{ -0.01 }^{ +0.13 }$ & $ 2823 _{ -16 }^{ +58 }$ & $\texttt{BPRP}$ \\
{\fontsize{7}{0}\selectfont 253040591\,B} & 159.25714983803 & 42.12868405630 & $ 13.01 _{ -0.17 }^{ +0.17 } $ & 640 & $   _{  }^{  }$ & $   _{   }^{  }$ & $\texttt{WD}$ $\texttt{BPRP}$ \\
{\fontsize{7}{0}\selectfont 257605131\,B} & 62.95486583450 & -37.94500665138 & $ 8.33 _{ -0.16 }^{ +0.07 } $ & 4669 & $ 0.60 _{ -0.01 }^{ +0.01 }$ & $ 3728 _{ -69 }^{ +15 }$ & $\texttt{BPRP}$ $\texttt{PRI}$ \\
{\fontsize{7}{0}\selectfont 259376845\,B} & 68.46856153261 & -51.25247980639 & $ 5.28 _{ -0.13 }^{ +0.17 } $ & 1177 & $ 0.87 _{ -0.05 }^{ +0.06 }$ & $ 5457 _{ -231 }^{ +283 }$ & $\texttt{BPRP}$ $\texttt{PRI}$ \\
{\fontsize{7}{0}\selectfont 282502866\,B} & 90.09431406761 & 1.80160968503 & $ 10.41 _{ -0.08 }^{ +0.13 } $ & 3375 & $ 0.29 _{ -0.06 }^{ +0.06 }$ & $ 3066 _{ -24 }^{ +47 }$ & $\texttt{BPRP}$ $\texttt{PRI}$ \\
{\fontsize{7}{0}\selectfont 288240183\,B} & 224.78438383542 & 83.34266967111 & $ 4.58 _{ -0.17 }^{ +0.16 } $ & 9044 & $ 0.93 _{ -0.07 }^{ +0.09 }$ & $ 5866 _{ -286 }^{ +305 }$ & $\texttt{BPRP}$ $\texttt{PRI}$ \\
{\fontsize{7}{0}\selectfont 288240183\,C} & 224.78626346820 & 83.34221655772 & $ 9.88 _{ -0.25 }^{ +0.25 } $ & 8771 & $ 0.35 _{ -0.15 }^{ +0.15 }$ & $ 3478 _{ -288 }^{ +348   }$ \\
{\fontsize{7}{0}\selectfont 290596728\,B} & 306.89728308250 & -40.36459462135 & $ 7.74 _{ -0.20 }^{ +0.25 } $ & 818 & $ 0.60 _{ -0.01 }^{ +0.04 }$ & $ 4031 _{ -66 }^{ +80 }$ & $\texttt{BPRP}$ $\texttt{PRI}$ \\
{\fontsize{7}{0}\selectfont 300116105\,B} & 92.47405292459 & -39.73999660669 & $ 9.19 _{ -0.10 }^{ +0.11 } $ & 738 & $ 0.46 _{ -0.01 }^{ +0.01 }$ & $ 3623 _{ -42 }^{ +33 }$ & $\texttt{inter}$ $\texttt{BPRP}$ \\
{\fontsize{7}{0}\selectfont 308301091\,B} & 267.86777900177 & 24.81212544714 & $ 10.32 _{ -0.16 }^{ +0.18 } $ & 715 & $ 0.31 _{ -0.11 }^{ +0.14 }$ & $ 3398 _{ -280 }^{ +326   }$ \\
{\fontsize{7}{0}\selectfont 312091232\,B} & 140.31670670796 & 6.65786362756 & $ 10.00 _{ -0.14 }^{ +0.13 } $ & 2271 & $ 0.37 _{ -0.01 }^{ +0.02 }$ & $ 3422 _{ -27 }^{ +34 }$ & $\texttt{inter}$ $\texttt{EDR3}$ $\texttt{BPRP}$\\
{\fontsize{7}{0}\selectfont 326092637\,B} & 15.37238815490 & -24.93391132753 & $ 11.59 _{ -0.25 }^{ +0.10 } $ & 8137 & $ 0.21 _{ -0.01 }^{ +0.02 }$ & $ 3168 _{ -16 }^{ +37 }$ & $\texttt{inter}$ $\texttt{BPRP}$ \\
{\fontsize{7}{0}\selectfont 341411516\,B} & 119.93640614165 & -59.26953097644 & $ 13.48 _{ -0.12 }^{ +0.02 } $ & 4162 & $   _{  }^{  }$ & $   _{   }^{  }$ & $\texttt{WD}$ $\texttt{BPRP}$ \\
{\fontsize{7}{0}\selectfont 345324572\,A} & 355.48235281306 & 59.88525830586 & $ 5.20 _{ -0.15 }^{ +0.19 } $ & 1616 & $ 0.90 _{ -0.06 }^{ +0.06 }$ & $ 5475 _{ -211 }^{ +239 }$ & $\texttt{BPRP}$ $\texttt{PRI}$ \\
{\fontsize{7}{0}\selectfont 349793830\,B} & 143.44650893948 & 24.26182131498 & $ 12.70 _{ -0.08 }^{ +0.03 } $ & 2308 & $ 0.16 _{ -0.01 }^{ +0.01 }$ & $ 2862 _{ -16 }^{ +5 }$ & $\texttt{BPRP}$ \\
{\fontsize{7}{0}\selectfont 352915304\,B} & 313.79523211150 & 62.77508729271 & $ 11.92 _{ -0.13 }^{ +0.02 } $ & 427 & $ 0.19 _{ -0.01 }^{ +0.01 }$ & $ 3116 _{ -3 }^{ +21 }$ & $\texttt{inter}$ $\texttt{BPRP}$ \\
{\fontsize{7}{0}\selectfont 369376388\,B} & 59.85169681654 & -36.47623340570 & $ 11.97 _{ -0.34 }^{ +0.44 } $ & 157 & $ 0.17 _{ -0.06 }^{ +0.08 }$ & $ 3053 _{ -293 }^{ +249   }$ \\
{\fontsize{7}{0}\selectfont 369376388\,C} & 59.85740294934 & -36.47089495211 & $ 11.62 _{ -0.34 }^{ +0.44 } $ & 2292 & $   _{  }^{  }$ & $   _{   }^{  }$ & $\texttt{WD}$ $\texttt{BPRP}$ \\
{\fontsize{7}{0}\selectfont 372913337\,B} & 119.48675278787 & -60.84268804035 & $ 4.07 _{ -0.24 }^{ +0.11 } $ & 5924 & $ 1.02 _{ -0.10 }^{ +0.11 }$ & $ 6213 _{ -342 }^{ +333 }$ & $\texttt{BPRP}$ $\texttt{PRI}$ \\
{\fontsize{7}{0}\selectfont 374352402\,B} & 303.50138009888 & -69.43373843927 & $ 6.58 _{ -0.39 }^{ +0.29 } $ & 2025 & $ 0.70 _{ -0.05 }^{ +0.06 }$ & $ 4752 _{ -289 }^{ +285 }$ & $\texttt{BPRP}$ \\
{\fontsize{7}{0}\selectfont 374732772\,B} & 242.12692516072 & -38.47303765313 & $ 6.83 _{ -0.43 }^{ +0.49 } $ & 775 & $ 0.69 _{ -0.04 }^{ +0.04 }$ & $ 4457 _{ -227 }^{ +259 }$ & $\texttt{inter}$ $\texttt{BPRP}$ $\texttt{PRI}$\\
{\fontsize{7}{0}\selectfont 394721720\,B} & 20.29820671156 & -85.59341646735 & $ 10.52 _{ -0.15 }^{ +0.15 } $ & 4040 & $ 0.30 _{ -0.02 }^{ +0.02 }$ & $ 3322 _{ -28 }^{ +28 }$ & $\texttt{inter}$ $\texttt{BPRP}$ \\
{\fontsize{7}{0}\selectfont 399913539\,B} & 287.26229580166 & 48.84559355879 & $ 10.31 _{ -0.31 }^{ +0.23 } $ & 1622 & $ 0.33 _{ -0.03 }^{ +0.04 }$ & $ 3361 _{ -43 }^{ +61 }$ & $\texttt{inter}$ $\texttt{BPRP}$ \\
{\fontsize{7}{0}\selectfont 453455638\,B} & 151.96566880705 & -76.28161439253 & $ 9.51 _{ -0.17 }^{ +0.25 } $ & 1907 & $ 0.48 _{ -0.03 }^{ +0.02 }$ & $ 3370 _{ -38 }^{ +190 }$ & $\texttt{BPRP}$ $\texttt{PRI}$ \\
{\fontsize{7}{0}\selectfont 460950389\,B} & 159.14432531677 & -64.81206485569 & $ 14.31 _{ -0.15 }^{ +0.15 } $ & 8128 & $ 0.10 _{ -0.01 }^{ +0.01 }$ & $ 2876 _{ -182 }^{ +23 }$ & $\texttt{inter}$ $\texttt{BPRP}$ \\
{\fontsize{7}{0}\selectfont 467785319\,B} & 62.74680893665 & 34.15885055036 & $ 10.77 _{ -0.14 }^{ +0.14 } $ & 493 & $ 0.28 _{ -0.02 }^{ +0.02 }$ & $ 3274 _{ -25 }^{ +26 }$ & $\texttt{inter}$ $\texttt{BPRP}$ \\
{\fontsize{7}{0}\selectfont 738065944\,B} & 98.61713426810 & -41.14765885257 & $ 5.71 _{ -0.21 }^{ +0.32 } $ & 857 & $ 0.78 _{ -0.18 }^{ +0.22 }$ & $ 5311 _{ -1075 }^{ +813   }$ \\
{\fontsize{7}{0}\selectfont 901674675\,B} & 165.50033761990 & -25.69109105001 & $ 3.99 _{ -0.10 }^{ +0.23 } $ & 1127 & $ 1.03 _{ -0.12 }^{ +0.12 }$ & $ 6220 _{ -318 }^{ +393 }$ & $\texttt{BPRP}$ $\texttt{PRI}$ \\
\hline
\end{tabular}
\end{table*}

\end{document}